\documentclass{elsart}
\usepackage{epsfig}
\usepackage{graphics}
\usepackage{graphicx}
\usepackage{amsmath}
\usepackage{color}
\journal{Nuclear Physics A}
\def\lesim{{$\lower 2pt\hbox{$\scriptstyle <$}
\atop\raise4pt\hbox{$\scriptstyle\sim$}$}}     %%less than with a \sim under it
\def\grsim{{$\lower2pt\hbox{$\scriptstyle >$} \atop\raise4pt\hbox
{$\scriptstyle\sim$}$}}
\usepackage{amssymb}

\newcommand{\sqrtsNN}{\mbox{$\sqrt{\mathrm{s}_{_{\mathrm{NN}}}}$}}

\newcommand{\ks}{$\mathrm{K}^{0}_{s}$ }

\newcommand{\pt}{$p_T$ }
\def \auau      {Au+Au }

\def \pp        {p+p }

\def\pT{\mbox{$p_T$}}

\def\sqrtsNN{\mbox{$\sqrt{s_\mathrm{_{NN}}}$}}

\def\Nbinary{\mbox{$N_{bin}$}}
\def\NbinaryMean{\mbox{$\langle\Nbinary\rangle$}}

\def\pizero{\mbox{$\pi^0$}}

\def\pp{\mbox{$p+p$}}

\def\vtwo{\mbox{$v_2$}}

\def\meanpT{\mbox{$\lt\pT\gt$}}

\def\RAB{\mbox{$R_{AB}(\pT)$}}
\def\RCP{\mbox{$R_{CP}(\pT)$}}

\def\Qs{\mbox{$Q_s$}}

\def\lt{\mbox{$<$}}
\def\gt{\mbox{$>$}}

\begin{document}
\setcounter{page}{1}
\ifx\href\undefined\else\hypersetup{linktocpage=true}\fi

\begin{frontmatter}



\title{Experimental and Theoretical Challenges in the Search
for the Quark Gluon Plasma:} \title
{The STAR Collaboration's Critical
Assessment of the Evidence from RHIC Collisions}

\author[3]{J.~Adams}
\author[29]{M.M.~Aggarwal}
\author[43]{Z.~Ahammed}
\author[20]{J.~Amonett}
\author[20]{B.D.~Anderson}
\author[13]{D.~Arkhipkin}
\author[12]{G.S.~Averichev}
\author[19]{S.K.~Badyal}
\author[27]{Y.~Bai}
\author[17]{J.~Balewski}
% ...
\collab{STAR Collaboration}

% ...
\address[3]{University of Birmingham, Birmingham, United Kingdom}
% ...
\address[12]{Laboratory for High Energy (JINR), Dubna, Russia}
\address[13]{Particle Physics Laboratory (JINR), Dubna, Russia}
% ...
\address[17]{Indiana University, Bloomington, Indiana 47408, USA}
% ...
\address[19]{University of Jammu, Jammu 180001, India}
\address[20]{Kent State University, Kent, Ohio 44242, USA}
% ...
\address[27]{NIKHEF and Utrecht University, Amsterdam, The Netherlands}
% ...
\address[29]{Panjab University, Chandigarh 160014, India}
% ...
\address[43]{Variable Energy Cyclotron Centre, Kolkata 700064, India}
% ...

\begin{abstract}
% Text of abstract
We review the most important experimental results from the first
three years of nucleus-nucleus collision studies at RHIC, with
emphasis on results from the STAR experiment, and we assess their
interpretation and comparison to theory.  The theory-experiment
comparison suggests that central Au+Au collisions at RHIC produce
dense, rapidly thermalizing matter characterized by: (1) initial
energy densities above the critical values predicted by lattice
QCD for establishment of a Quark-Gluon Plasma (QGP); (2) nearly
ideal fluid flow, marked by
constituent interactions of very short mean free path,
established most probably at a stage preceding hadron formation;
and (3) opacity to jets. Many of the observations are consistent
with models incorporating QGP formation in the early collision
stages, and have not found ready
explanation in a hadronic framework.  However, the measurements
themselves do not yet establish unequivocal evidence for a
transition to this new form of matter. The theoretical treatment
of the collision evolution,
despite impressive successes, invokes a
suite of distinct models, degrees
of freedom and assumptions of as
yet unknown quantitative
consequence. We pose a set of
important open questions, and suggest additional measurements, at
least some of which should be addressed
in order to establish a
compelling basis to conclude definitively
that thermalized, deconfined
quark-gluon matter has been
produced at RHIC.
\end{abstract}

\end{frontmatter}
\newpage
\tableofcontents

% main text
\newpage
\section{Introduction}
\label{intro}

The Relativistic Heavy Ion
Collider was built to create and investigate strongly interacting
matter at energy densities unprecedented in a laboratory setting
-- matter so hot that neutrons, protons and other hadrons are
expected to ``melt". Results from the four RHIC experiments
already demonstrate that the facility has fulfilled its promise to
reach such extreme conditions during the early stages of
nucleus-nucleus collisions, forming matter that exhibits
heretofore unobserved behavior. These results are summarized in
this work and in a number of excellent recent reviews
\cite{Jacobs-Wang,Rischke,kolbheinz,GVWZ,tomasik-wiedemann}.  They
afford RHIC the exciting scientific opportunity to discover the
properties of matter under conditions believed to pertain during a
critical, though fleeting, stage of the universe's earliest
development following the Big Bang. The properties of such matter
test fundamental predictions of Quantum ChromoDynamics (QCD) in
the non-perturbative regime.

In this document we review the
results to date from RHIC experiments, with emphasis on those from
STAR, in the context of a narrower, more pointed question. The
specific prediction of QCD most often highlighted in discussions
of RHIC since its conception is that of a transition from hadronic
matter to a Quark-Gluon Plasma (QGP) phase, defined below. Recent
theoretical claims \cite{Gyulassy,McLerran-Gyulassy,RBRC} that a
type of QGP has indeed been revealed by RHIC experiments and
interest in this subject by the popular press \cite{NYT,Nature}
make it especially timely to evaluate where we are with respect to
this particular goal.  The present paper has been written in
response to a charge (see Appendix A) from the STAR Collaboration
to itself, to assess whether RHIC results yet support a compelling
discovery claim for the QGP, applying the high standards of
scientific proof merited by the importance of this issue. We began
this assessment before the end of the fourth successful RHIC
running period, and we have based our evaluation on results from
the first three RHIC runs, which are often dramatic, sometimes
unexpected, and generally in excellent agreement among the four
RHIC experiments (and we utilize results from all of the
experiments here). Since we began, some analyses of data from run
4 have progressed to yield publicly presented results that amplify
or quantify some of our conclusions in this work, but do not
contradict any of them.

In addressing our charge, it is
critical to begin by defining clearly what we mean by the QGP,
since theoretical expectations of its properties have evolved
significantly over the 20 years since the case for RHIC was first
made.  For our purposes here, we take the QGP to be \textbf{a
(locally) thermally equilibrated state of matter in which quarks
and gluons are deconfined from hadrons, so that color degrees of
freedom become manifest over {\em nuclear}, rather than merely
nucleonic, volumes}.  In concentrating on thermalization and
deconfinement, we believe our definition to be consistent with
what has been understood by the
physics community at large since
RHIC was first proposed, as
summarized by planning documents quoted in Appendix B. In
particular, thermalization is viewed as a necessary condition to
be dealing with a state of matter, whose properties can be
meaningfully compared to QCD predictions or applied to the
evolution of the early universe.  Observation of a deconfinement
transition has always been a primary goal for RHIC, in the hope of
illuminating the detailed mechanism of the normal color
confinement in QCD.  For reasons presented below, we do
significantly omit from our list of necessary conditions some
other features discussed as potentially relevant
over the years since RHIC's
conception.

\begin{itemize}

\item[$\bullet$] We do not demand that the quarks and gluons in
the produced matter be non-interacting, as has been considered in
some conceptions of the QGP.  Lattice QCD calculations suggest
that such an ideal state may be approached in static bulk QGP
matter only at temperatures very much higher than that required
for the deconfinement transition. Furthermore, attainment of
thermalization on the ultra-short timescale of a RHIC collision
must rely on frequent interactions among the constituents during
the earliest stages of the collision -- a requirement that is not
easily reconcilable with production of an ideal gas. While the
absence of interaction would allow considerable simplifications in
the calculation of thermodynamic properties of the matter, we do
not regard this as an essential feature of color-deconfined
matter. In this light, some have suggested
\cite{Gyulassy,McLerran-Gyulassy,RBRC} that we label the matter we
seek as the sQGP, for strongly-interacting Quark-Gluon Plasma.
Since we regard this as the form of QGP that should be normally
anticipated, we consider the `s' qualifier to be superfluous.

\item[$\bullet$] We do not require evidence of a first- or
second-order phase transition, even though early theoretical
conjecture \cite{Harris} often focused on possible QGP signatures
involving sharp changes in experimental observables with collision
energy density. In fact, the nature of the predicted transition
from hadron gas to QGP has only been significantly constrained by
quite recent theory.  Our definition allows for a QGP discovery in
a thermodynamic regime beyond a possible critical point.  Most
modern lattice QCD calculations indeed suggest the existence of
such a critical point at baryon densities well above those where
RHIC collisions appear to first form the matter. Nonetheless, such
calculations still predict a rapid (but unaccompanied by
discontinuities in thermodynamic observables) crossover transition
in the bulk properties of strongly interacting matter at zero
baryon density.

\item[$\bullet$] We consider that evidence for chiral symmetry
restoration would be sufficient to demonstrate a new form of
matter, but is not \emph{necessary} for a compelling QGP
discovery.  Most lattice QCD calculations do predict that this
transition will accompany deconfinement, but the question is
certainly not definitively decided theoretically. If clear
evidence for deconfinement can be provided by the experiments,
then the search for manifestations of chiral symmetry restoration
will be one of the most profound goals of further investigation of
the matter's properties, as they would provide the clearest
evidence for fundamental modifications to the QCD vacuum, with
potentially far-reaching consequences.

\end{itemize}

The above ``relaxation" of demands, in comparison to
initial expectations before
initiation of the RHIC program, makes a daunting task even more
challenging.  The possible absence
of a first- or second-order phase transition reduces hopes to
observe some well-marked changes in behavior that might serve as
an experimental ``smoking gun" for a transition to a new form of
matter. Indeed, even if there were a sharp transition as a
function of bulk matter temperature, it would be unlikely to
observe non-smooth behavior in heavy-ion collisions, which form
finite-size systems spanning some range of local temperatures even
at fixed collision energy or centrality. We thus have to rely more
heavily for evidence of QGP formation on the comparison of
experimental results with theory. But theoretical calculation of
the properties of this matter become subject to all the
complexities of strong QCD interactions, and hence to the
technical limitations of lattice gauge calculations. Even more
significantly, these QCD calculations must be supplemented by
other models to describe the complex dynamical
passage of heavy-ion collision
matter into and out of the QGP
state. Heavy ion collisions represent our best opportunity to make
this unique matter in the laboratory, but we place exceptional
demands on these collisions: they must not only produce the
matter, but then must serve ``pump and probe" functions somewhat
analogous to the modern generation of condensed matter instruments --
and they must do it all on
distance scales of femtometers and a time scale of $10^{-23}$
seconds!

There are two basic classes of probes at our disposal in heavy ion
collisions. In studying electroweak collision products, we exploit
the \emph{absence} of final-state
interactions (FSI) with the evolving strongly interacting matter,
hoping to isolate those produced during the early collision stages
and bearing the imprints of the bulk properties characterizing
those stages.  But we have to deal with the relative scarcity of
such products, and competing origins from hadron decay and
interactions during later collision stages. Most of the RHIC
results to date utilize instead the far more abundant produced
hadrons, where one exploits (but then must understand) the FSI.
It becomes critical to distinguish \emph{partonic} FSI from
\emph{hadronic} FSI, and to distinguish both from initial-state
interactions and the effects of
(so far) poorly understood parton densities at very low momentum
fraction in the entrance-channel nuclei.  Furthermore, the
formation of hadrons from a QGP involves soft processes (parton
fragmentation and recombination) that cannot be calculated from
perturbative QCD and are \emph{a priori} not well characterized
(nor even cleanly separable) inside hot strongly interacting
matter.

In light of all these complicating features, it is remarkable that
the RHIC experiments have already produced results that appear to
confirm some of the more striking, and at least semi-quantitative,
predictions made on the basis of QGP formation!  Other,
unexpected, RHIC results have stimulated new models that explain
them within a QGP-based framework.  The most exciting results
reveal phenomena not previously observed or explored at lower
center-of-mass energies, and indeed are distinct from the
observations on which a circumstantial case for QGP formation was
previously argued at CERN \cite{CERN}. In order to assess whether
a discovery claim is now justified, we must judge the robustness
of both the new experimental results and the theoretical
predictions they seem to bear out.  Do the RHIC data \emph{demand}
a QGP explanation? Can they alternatively be accounted for in a
hadronic framework? Are the theories and models used for the
predictions mutually compatible?  Are those other experimental
results that currently appear to deviate from theoretical
expectations indicative of details yet to be worked out, or rather
of fundamental problems with the QGP explanation?

We organize our discussion as follows.  In Chapter 2 we briefly
summarize the most relevant theoretical calculations and models,
their underlying assumptions, limitations and most robust
predictions.  We thereby identify the \emph{crucial} QGP features
we feel must be demonstrated experimentally to justify a
compelling discovery claim.  We divide the experimental evidence
into two broad areas in Chapters 3-4, focusing first on what we
have learned about the bulk thermodynamic properties of the early
stage collision matter from such measures as hadron spectra,
collective flow and correlations among the soft hadrons that
constitute the vast majority of outgoing particles.  We discuss
the consistency of these results with thermalization and the
exposure of new (color) degrees of freedom. Next we provide an
overview of the observations of hadron production yields and
angular correlations at high transverse momentum ($p_T \gtrsim 4$
GeV/c), and what they have taught us about the nature of FSI in
the collision matter and their bearing on deconfinement.

In Chapter 5 we focus on open questions for experiment and theory,
on important crosschecks and quantifications, on predictions not
yet borne out by experiment and experimental results not yet
accommodated by theory. Finally, we provide in Chapter 6 an
extended summary, conclusions and outlook, with emphasis on
additional measurements and theoretical improvements that we feel
are needed to strengthen the case for QGP formation.  The summary
of results in Chap. 6 is extended so that readers already familiar
with most of the theoretical and experimental background material
covered in Chaps. 2-5 can skip to the concluding section without
missing the arguments central to our assessment of the evidence.

The STAR detector and its capabilities have been described in
detail elsewhere~\cite{star:nim}, and will not be discussed.

\newpage
\section{Predicted Signatures of the QGP}
\label{signatures}

The promise, and then the delivery, of experimental results from
the AGS, SPS and RHIC have stimulated impressive and important
advances over the past decade in the theoretical treatment of the
thermodynamic and hydrodynamic properties of hot strongly
interacting matter and of the propagation of partons through such
matter.  However, the complexities of heavy-ion collisions and of
hadron formation still lead to a patchwork of theories and models
to treat the entire collision evolution, and the difficulties of
the strong interaction introduce significant quantitative
ambiguities in all aspects of this treatment. In support of a
possible compelling QGP discovery claim, we must then identify the
most striking qualitative predictions of theory, which survive the
quantitative ambiguities, and we must look for a congruence of
various observations that confirm such robust predictions.  In
this chapter, we provide a brief summary of the most important
pieces of the theoretical framework, their underlying assumptions
and quantitative limitations, and what we view as their most
robust predictions.  Some of these predictions will then be
compared with RHIC experimental results in later chapters.

\subsection{Features of the Phase Transition in Lattice QCD}

The phase diagram of bulk thermally equilibrated strongly
interacting matter should be described by QCD.  At sufficiently
high temperature one must expect hadrons to ``melt", deconfining
quarks and gluons.  The exposure of new (color) degrees of freedom
would then be manifested by a rapid increase in entropy density,
hence in pressure, with increasing temperature, and by a
consequent change in the equation of state (EOS).  In the limit
where the deconfined quarks and gluons are non-interacting, and
the quarks are massless, the (Stefan-Boltzmann) pressure $P_{SB}$
of this partonic state, as a function of temperature $T$ at zero
chemical potential (\emph{i.e.}, zero net quark density), would be
simply determined by the number of degrees of freedom
\cite{Rischke}:

\begin{equation}
\label{SB} \frac{P_{SB}}{T^4} ~=~ [2(N_c^2 - 1) + \frac{7}{2}
N_cN_f]\frac{\pi^2}{90},
\end{equation}

\noindent where $N_c$ is the number of colors, $N_f$ the number of
quark flavors, the temperature is measured in energy units
(throughout this paper), and we have taken $\hbar = c = 1$. The
two terms on the right in Eq.~\ref{SB} represent the gluon and
quark contributions, respectively. Refinements to this basic
expectation, to incorporate effects of color interactions among
the constituents, as well as of non-vanishing quark masses and
chemical potential, and to predict the location and nature of the
transition from hadronic to partonic degrees of freedom, are best
made via QCD calculations on a space-time lattice (LQCD).

In order to extract physically relevant predictions from LQCD
calculations, these need to be extrapolated to the continuum
(lattice spacing $\rightarrow 0$), chiral (actual current quark
mass) and thermodynamic (large volume) limits.
While computing power limitations
have restricted the calculations to date to numbers of lattice
points that are still considered
somewhat marginal from the
viewpoint of these extrapolations
\cite{Rischke}, enormous progress
has been made in recent years.  Within the constraints of
computing cost, there have been important initial explorations of
sensitivity to details of the calculations \cite{Rischke}:
\emph{e.g.}, the number and masses of active quark flavors
included; the technical treatment of quarks on the lattice; the
presence or absence of the $U_A(1)$ anomaly in the QGP state.
Additional numerical difficulties
have been partially overcome to
allow first calculations at nonzero chemical potential and
to improve the determination of
physical quark mass scales for a given lattice spacing
\cite{Rischke}.

\begin{figure}[bht]
%\begin{minipage}[b]{14cm}
\begin{center}
\epsfig{figure=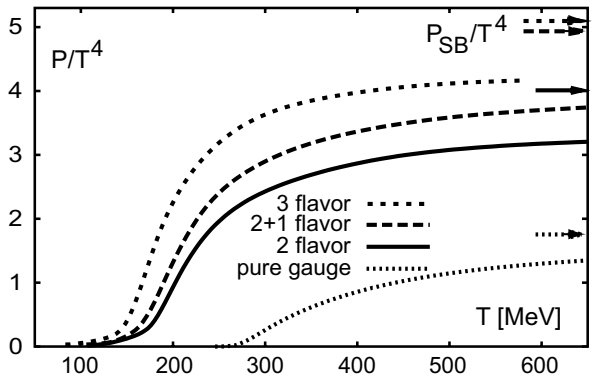,width=8.0cm}
%\end{center}
%\end{minipage}
%\vspace{6mm}
%\begin{minipage}[h]{14cm}
\caption{ {\it LQCD calculation results from Ref.~\cite{Karsch}
for the pressure divided by $T^4$ of strongly interacting matter
as a function of temperature, and for several different choices of
the number of dynamical quark flavors.  The arrows near the right
axis indicate the corresponding Stefan-Boltzmann pressures for the
same quark flavor assumptions. } }\label{LQCD-pressure}
%\end{minipage}
\end{center}
\end{figure}

Despite the technical complications, LQCD calculations have
converged on the following predictions:

\begin{figure}[thb]
%\begin{minipage}[b]{14cm}
\begin{center}
\epsfig{figure=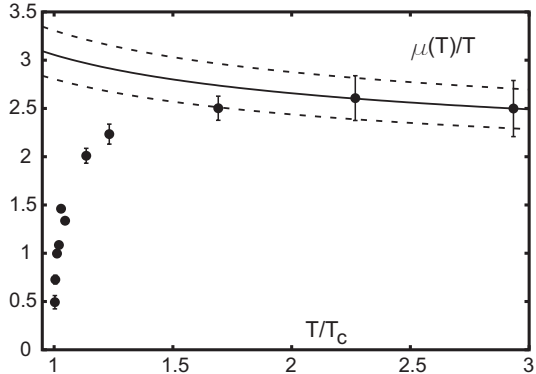,width=7.0cm}
%\end{center}
%\end{minipage}
%\vspace{6mm}
%\begin{minipage}[h]{14cm}
\caption{ {\it Temperature-dependence of the heavy-quark screening
mass (divided by temperature) as a function of temperature (in
units of the phase transition temperature), from LQCD calculations
in Ref.~\cite{Kaczmarek}. The curves represent perturbative
expectations of the temperature-dependence. } }
\label{LQCD-screening}
%\end{minipage}
\end{center}
\end{figure}

\begin{figure}[thb]
%\begin{minipage}[b]{14cm}
\begin{center}
\epsfig{figure=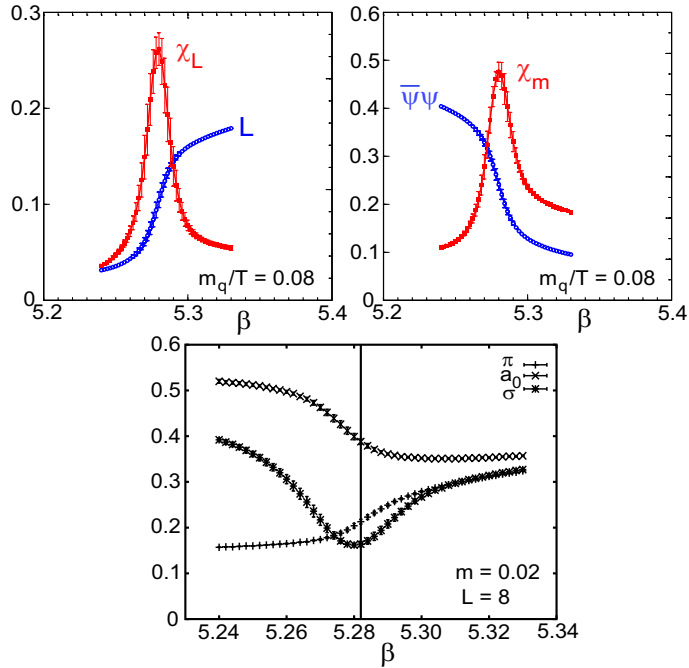,width=9.0cm}
%\end{center}
%\end{minipage}
%\vspace{6mm}
%\begin{minipage}[h]{14cm}
\caption{ {\it LQCD calculations for two dynamical quark
flavors~\cite{Karsch} showing the coincidence of the chiral
symmetry restoration (marked by the rapid decrease of chiral
condensate $\langle \overline{\psi} \psi \rangle$ in the upper
right-hand frame) and deconfinement (upper left frame) phase
transitions.  The lower plot shows that the chiral transition
leads toward a mass degeneracy of the pion with scalar meson
masses. All plots are as a function of the bare coupling strength
$\beta$ used in the calculations; increasing $\beta$ corresponds
to decreasing lattice spacing and to increasing temperature. } }
\label{LQCD-condensate}
%\end{minipage}
\end{center}
\end{figure}

\begin{figure}[bht]
%\begin{minipage}[b]{14cm}
\begin{center}
\epsfig{figure=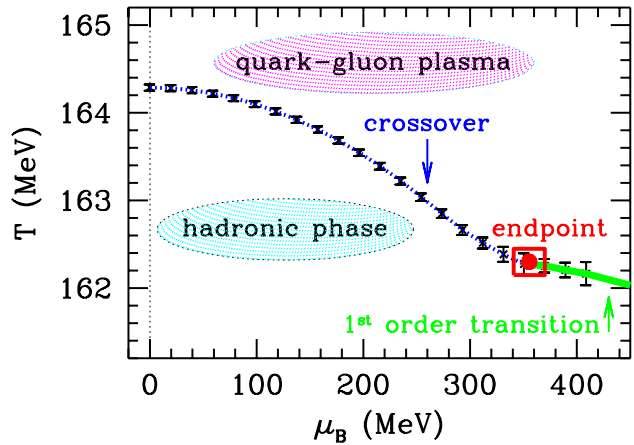,width=9.0cm}
%\end{center}
%\end{minipage}
%\vspace{6mm}
%\begin{minipage}[h]{14cm}
\caption{ {\it LQCD calculation results for non-zero chemical
potential~\cite{Fodor-2}, suggesting the existence of a critical
point well above RHIC chemical potential values. The solid line
indicates the locus of first-order phase transitions, while the
dotted curve marks crossover transitions between the hadronic and
QGP phases. } } \label{LQCD-criticalpoint}
%\end{minipage}
\end{center}
\end{figure}

\begin{itemize}

\item[$\bullet$] There is indeed a predicted transition of some
form between a hadronic and a QGP phase, occurring at a
temperature in the vicinity of $T_c \simeq 160$ MeV for zero
chemical potential.  The precise value of the transition
temperature depends on the treatment of quarks in the calculation.

\item[$\bullet$] The pressure divided by $T^4$ rises rapidly above
$T_c$, then begins to saturate by about 2$T_c$, but at values
substantially below the Stefan-Boltzmann limit (see
Fig.~\ref{LQCD-pressure}) \cite{Karsch}.  The deviation from the
SB limit indicates substantial remaining interactions among the
quarks and gluons in the QGP phase.

\item[$\bullet$] Above $T_c$, the effective potential between a
heavy quark-antiquark pair takes the form of a screened Coulomb
potential, with screening mass (or inverse screening length)
rising rapidly as temperature increases above $T_c$ (see
Fig.~\ref{LQCD-screening}) \cite{Kaczmarek}.  As seen in the
figure, the screening mass deviates strongly from perturbative QCD
expectations in the vicinity of $T_c$, indicating large
non-perturbative effects.  The increased screening mass leads to a
shortening of the range of the $q\overline{q}$ interaction, and to
an anticipated suppression of charmonium production, in relation
to open charm \cite{Matsui}.  The predicted suppression appears to
set in at substantially different temperatures for $J/\psi$
($1.5-2.0 T_c$) and $\psi^\prime$ ($\sim 1.0 T_c$) \cite{Asakawa}.

\item[$\bullet$] In most calculations, the deconfinement
transition is also accompanied by a chiral symmetry restoration
transition, as seen in Fig.~\ref{LQCD-condensate} \cite{Karsch}.
The reduction in the chiral condensate leads to significant
predicted variations in in-medium meson masses.  These are also
affected by the restoration of $U_A(1)$ symmetry, which occurs at
higher temperature than chiral symmetry restoration in the
calculation of Fig.~\ref{LQCD-condensate}.

\item[$\bullet$] The nature of the transition from hadronic to QGP
phase is highly sensitive to the number of dynamical quark flavors
included in the calculation and to the quark masses
\cite{Laermann}.  For the most realistic calculations,
incorporating two light ($u,d$) and one heavier ($s$) quark flavor
relevant on the scale of $T_c$, the transition is most likely of
the crossover type (with no discontinuities in thermodynamic
observables -- as opposed to first- or second-order phase
transitions) at zero chemical potential, although the ambiguities
in tying down the precise values of quark masses corresponding to
given lattice spacings still permit some doubt.

\item[$\bullet$] Calculations at non-zero chemical potential,
though not yet mature, suggest the existence of a critical point
such as that illustrated in Fig.~\ref{LQCD-criticalpoint}
\cite{Fodor-2}.  The numerical
challenges in such calculations leave considerable ambiguity about
the value of $\mu_B$ at which the critical point occurs
(\emph{e.g.}, it changes from
$\mu_B \approx 700$ to 350 MeV between Refs.~\cite{Fodor-1} and
\cite{Fodor-2}), but it is most likely above the value at which
RHIC collision matter is formed, consistent with the crossover
nature of the transition anticipated at RHIC.

\item[$\bullet$] Even for crossover transitions, the LQCD
calculations still predict a rapid temperature-dependence of the
thermodynamic properties, as revealed in all of the figures
considered above.  However, in
basing experimental expectations on this feature, it must be kept
in mind that the early collision temperature varies slowly with
collision energy and is not directly measured by any of the probes
studied most extensively to date.

\end{itemize}

\subsection{Hydrodynamic Signatures}

In order to determine how the properties of bulk QGP matter, as
determined in LQCD calculations, may influence observable particle
production spectra from RHIC collisions, one needs to model the
time evolution of the collision ``fireball".  To the extent that
the initial interactions among the constituents are sufficiently
strong to establish local thermal equilibrium rapidly, and then to
maintain it over a significant evolution time, the resulting
matter may be treated as a relativistic fluid undergoing
collective, hydrodynamic flow~\cite{kolbheinz}. The application of
hydrodynamics for the description of hadronic fireballs has a long
history~\cite{landau,shuryak0312}. Relativistic hydrodynamics has
been extensively applied to heavy ion collisions from BEVALAC to
RHIC~\cite{kolbheinz,shuryak0312,stoeck}, but with the most
striking successes at RHIC.  The applicability of hydrodynamics at
RHIC may provide the clearest evidence for the attainment of local
thermal equilibrium at an early stage in these collisions.
(On the other hand, there are
alternative, non-equilibrium treatments of the fireball evolution
that have also been compared to RHIC data \cite{biro04}.) The
details of the hydrodynamic evolution are clearly sensitive to the
EOS of the flowing matter, and hence to the possible crossing of a
phase or crossover transition
during the system expansion and cooling.  It is critical to
understand the relative sensitivity to the EOS as compared with that to other
assumptions and parameters of the hydrodynamic treatment.

Traditional hydrodynamics calculations cannot be applied to matter
not in local thermal equilibrium, hence they must be supplemented
by more phenomenological treatments of the early and late stages
of the system evolution. These parameterize the initial conditions
for the hydrodynamic flow and the transition to freezeout, where
the structureless matter flow is converted into final hadron
spectra. Since longitudinal flow is especially sensitive to
initial conditions beyond the scope of the theory, most
calculations to date have concentrated on \emph{transverse} flow,
and have assumed longitudinal boost-invariance of the
predictions~\cite{kolbheinz}. Furthermore, it is anticipated that
hadrons produced at sufficiently high transverse momentum in
initial partonic collisions will not have undergone sufficient
rescatterings to come to thermal equilibrium with the surrounding
matter, so that hydrodynamics will be applicable at best only for
the softer features of observed spectra. Within the time range and
momentum range of its applicability, most hydrodynamics
calculations to date have treated the matter as an \emph{ideal},
non-viscous fluid.  The motion of this fluid is completely
determined given the three components of the fluid velocity ${\vec
v}$, the pressure ($P$) and the energy and baryon densities ($e$
and $n_B$). The hydrodynamic equations of motion for an ideal
fluid are derived from the exact local conservation laws for
energy, momentum, and baryon number by assuming an ideal-fluid
form for the energy-momentum tensor and baryon number current;
they are closed by an equation of state $P(e,n_B)$~\cite{landau}.

\begin{figure}[thb]
%\begin{minipage}[b]{14cm}
\begin{center}
\epsfig{figure=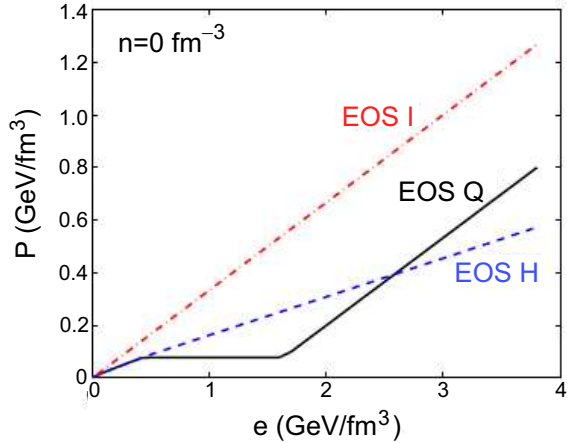,width=9.0cm}
%\end{center}
%\end{minipage}
%\vspace{6mm}
%\begin{minipage}[h]{14cm}
\caption{ {\it Pressure as a function of energy density at
vanishing net baryon density for three different equations of
state of strongly interacting matter:  a Hagedorn resonance gas
(EOS H), an ideal gas of massless partons (EOS I) and a connection
of the two via a first-order phase transition at $T_c$=164 MeV
(EOS Q).  These EOS are used in hydrodynamics calculations in
Ref.~\cite{kolbheinz}, from which the figure is taken. } }
\label{hydro-EOS}
%\end{minipage}
\end{center}
\end{figure}

The EOS in hydrodynamics calculations for RHIC has been
implemented using simplified models inspired by LQCD results,
though not reproducing their details.  One example is illustrated
by the solid curve in Fig.~\ref{hydro-EOS}, connecting an ideal
gas of massless partons at high temperature to a Hagedorn hadron
resonance gas~\cite{hagedorn} at low temperatures, via a
first-order phase transition chosen to ensure consistency with
($\mu_B = 0$) LQCD results for critical temperature and net
increase in entropy density across the
transition~\cite{kolbheinz}. In this implementation, the slope
$\partial{P}/\partial{e}$ (giving the square of the velocity of
sound in the matter) exhibits high values for the hadron gas and,
especially, the QGP phases, but has a soft point at the mixed
phase~\cite{kolbheinz,shuryak0312}. This generic softness of the
EOS during the assumed phase transition has predictable
consequences for the system evolution.

\begin{figure}[thb]
%\begin{minipage}[b]{14cm}
\begin{center}
\epsfig{figure=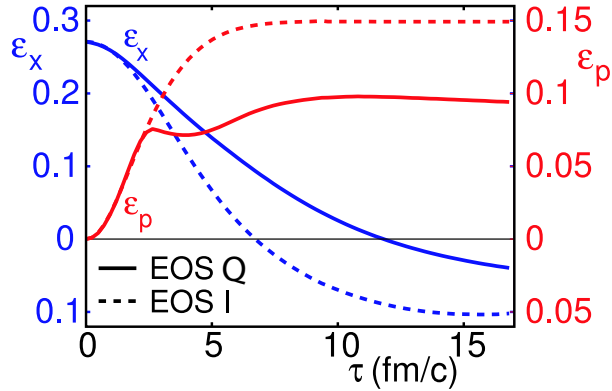,width=8.0cm}
%\end{center}
%\end{minipage}
%\vspace{6mm}
%\begin{minipage}[h]{14cm}
\caption{ {\it Hydrodynamics calculations for the time evolution
of the spatial eccentricity $\epsilon_x$ and the momentum
anisotropy $\epsilon_p$ for non-central (7 fm impact parameter)
Au+Au collisions at RHIC~\cite{kolbheinz}.  The solid and dashed
curves result, respectively, from use of EOS Q and EOS I from
Fig.~\ref{hydro-EOS}.  The gradual removal of the initial spatial
eccentricity by the pressure gradients that lead to the buildup of
$\epsilon_p$ reflects the self-quenching aspect of elliptic flow.
The time scale runs from initial attainment of local thermal
equilibrium through freezeout in this calculation. } }
\label{hydro-anisotropy}
%\end{minipage}
\end{center}
\end{figure}

In heavy ion collisions, the measurable quantities are the momenta
of the produced particles at the final state and their
correlations. Transverse flow measures are key observables to
compare quantitatively with model predictions in studying the EOS
of the hot, dense matter. In non-central collisions, the reaction
zone has an almond shape, resulting in azimuthally anisotropic
pressure gradients, and therefore a nontrivial elliptic flow
pattern.  Experimentally, this elliptic flow pattern is usually
measured using a Fourier decomposition of momentum spectra
relative to the event-by-event reaction plane, in which the second Fourier
component $v_2$ is the dominant contribution.
The important feature of elliptic flow is that it is
``self-quenching"~\cite{sorge97,sorge99},
because the pressure-driven expansion tends to reduce the spatial
anisotropy that causes the azimuthally anisotropic pressure
gradient in the first place. This robust feature is illustrated in
Fig.~\ref{hydro-anisotropy}, which compares predictions for the
spatial and resulting momentum eccentricities as a function of
time during the system's hydrodynamic evolution, for two different
choices of EOS~\cite{kolbheinz}. The self-quenching makes the
elliptic flow particularly sensitive to earlier collision stages,
when the spatial anisotropy and pressure gradient are the
greatest.  In contrast, hadronic interactions at later stages
may contribute
significantly to the radial
flow~\cite{bass,teaney}.

\begin{figure}[thb]
%\begin{minipage}[b]{14cm}
\begin{center}
\epsfig{figure=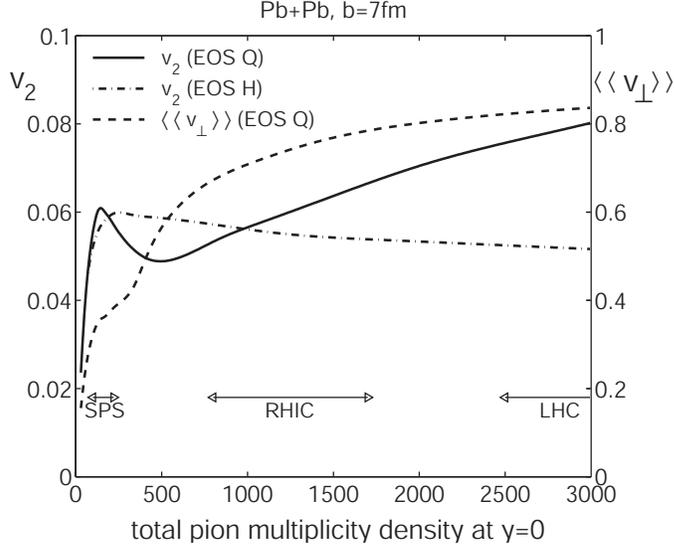,width=9.0cm}
%\end{center}
%\end{minipage}
%\vspace{6mm}
%\begin{minipage}[h]{14cm}
\caption{ {\it Predicted hydrodynamic excitation function of
$p_T$-integrated elliptic ($v_2$, solid curve, left axis) and
radial ($\langle \langle v_\perp \rangle \rangle$, dashed, right
axis) flow for non-central Pb+Pb collisions~\cite{kolbsollfrank}.
The calculations assume a sharp onset for freezeout along a
surface of constant energy density corresponding to temperature
$\approx 120$ MeV. The soft phase transition stage in EOS Q gives
rise to a dip in the elliptic flow.  The horizontal arrows at the
bottom reflect early projections of particle multiplicity for
different facilities, but we now know that RHIC collisions produce
multiplicities in the vicinity of the predicted dip.} }
\label{hydro-excitation}
%\end{minipage}
\end{center}
\end{figure}

The solid momentum anisotropy curve in Fig.~\ref{hydro-anisotropy}
also illustrates that entry into the soft EOS mixed phase during a
transition from QGP to hadronic matter stalls the buildup of
momentum anisotropy in the flowing matter.  An even more dramatic
predicted manifestation of this stall is shown by the dependence
of $p_T$-integrated elliptic flow on produced hadron multiplicity
in Fig.~\ref{hydro-excitation}, where a dip is seen under
conditions where the phase transition occupies most of the early
collision stage. Since the calculations are carried out for a
fixed impact parameter, measurements to confirm such a dip would
have to be performed as a function of collision energy.  In
contrast to early (non-hydrodynamic) projections of particle
multiplicities at RHIC (represented by horizontal arrows in
Fig.~\ref{hydro-excitation}), we now know that the multiplicity at
the predicted dip is approximately achieved for appropriate
centrality in RHIC Au+Au collisions at full energy. However,
comparisons of predicted with measured excitation functions for
elliptic flow are subject to an overriding ambiguity concerning
where and when appropriate conditions of initial local thermal
equilibrium for hydrodynamic applicability are actually achieved.
Hydrodynamics itself has nothing to say concerning this issue.

\begin{figure}[thb]
%\begin{minipage}[b]{14cm}
\begin{center}
\epsfig{figure=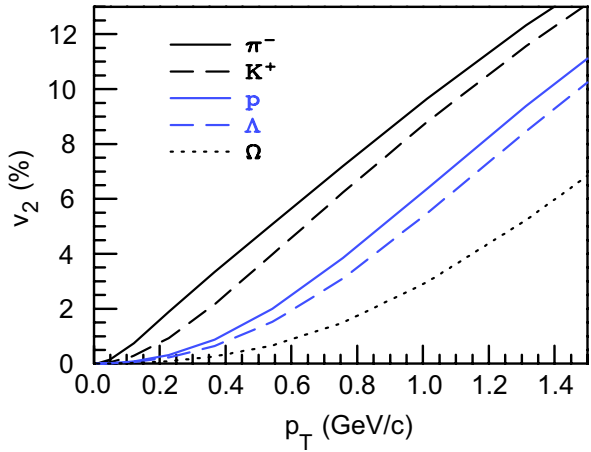,width=8.0cm}
%\end{center}
%\end{minipage}
%\vspace{6mm}
%\begin{minipage}[h]{14cm}
\caption{ {\it Hydrodynamics predictions~\cite{Huovinen} of the
$p_T$ and mass-dependences of the elliptic flow parameter $v_2$
for identified final hadrons from Au+Au collisions at
$\sqrt{s_{NN}}$=130 GeV. The calculations employ EOS Q (see
Fig.~\ref{hydro-EOS}) and freezeout near 120 MeV. } }
\label{hydro-massdep}
%\end{minipage}
\end{center}
\end{figure}

One can alternatively attain sensitivity to the EOS in
measurements for given collision energy and centrality by
comparing to the predicted dependence of elliptic flow strength on
hadron $p_T$ and mass (see Fig.~\ref{hydro-massdep}). The
mass-dependence is of simple kinematic origin~\cite{kolbheinz},
and is thus a robust feature of hydrodynamics, but its
quantitative extent, along with the magnitude of the flow itself,
depends on the EOS~\cite{kolbheinz}.

\begin{figure}[thb]
%\begin{minipage}[b]{14cm}
\begin{center}
\epsfig{figure=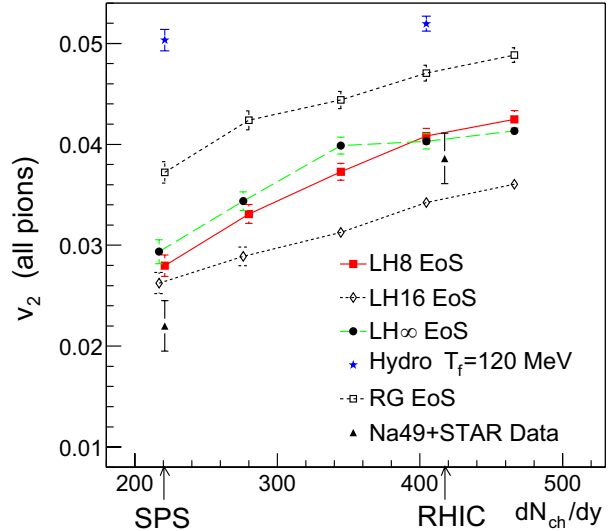,width=8.0cm}
%\end{center}
%\end{minipage}
%\vspace{6mm}
%\begin{minipage}[h]{14cm}
\caption{ {\it Predictions~\cite{teaney} from a hybrid
hydrodynamics-RQMD approach for the elliptic flow as a function of
charged particle multiplicity in Pb+Pb collisions at an impact
parameter $b=6$ fm.  Curves for different choices of EOS (LH8 is
most similar to EOS Q in Fig.~\ref{hydro-excitation}) are compared
to experimental results derived \cite{teaney} from SPS and RHIC
measurements.  The replacement of a simplified freezeout model for
all hadron species and of the
assumption of ideal hadronic fluid flow with the RQMD hadron
cascade appears to remove any dip in $v_2$ values, such as seen in
Fig.~\ref{hydro-excitation}. } } \label{hydro-RQMD}
%\end{minipage}
\end{center}
\end{figure}

Of course, the energy- and mass-dependence of $v_2$ can also be
affected by species-specific hadronic FSI at and close to the
freezeout where the particles decouple from the system, and
hydrodynamics is no longer applicable~\cite{bass,teaney}. A
combination of macroscopic and microscopic models, with
hydrodynamics applied at the early partonic and mixed-phase stages
and a hadronic transport model such as RQMD \cite{RQMD} at the
later hadronic stage, may offer a more realistic description of
the whole evolution than that achieved with a simplified sharp
freezeout treatment in
Figs.~\ref{hydro-anisotropy},\ref{hydro-excitation},\ref{hydro-massdep}.
The combination of hydrodynamics with RQMD \cite{teaney} has, for
example, led to predictions of a substantially different, and
monotonic, energy-dependence of elliptic flow, as can be seen by
comparing Fig.~\ref{hydro-RQMD} to Fig.~\ref{hydro-excitation}.
The difference between the two calculations may result primarily
\cite{Gyulassy} from the elimination in \cite{teaney} of the
assumption of ideal fluid expansion even in the hadronic phase. In
any case, this comparison suggests that the energy dependence of
elliptic flow in the quark-hadron transition region is at least as
sensitive to the late hadronic interaction details as to the
softening of the EOS in the mixed-phase region. Flow for
multi-strange and charmed particles with small hadronic
interaction cross sections may provide more selective sensitivity
to the properties of the partonic and mixed
phases~\cite{teaney,sorge,xu2}.  There may be non-negligible
sensitivity as well to the addition of such other complicating
features as viscosity \cite{viscous} and deviations from
longitudinal boost-invariance, studies of the latter effect
requiring computationally challenging (3+1)-dimensional
hydrodynamics calculations \cite{3Dhydro}. Certainly, the relative
sensitivities to EOS variations vs. treatments of viscosity,
boost-invariance, and the evolution of the hadronic stage must be
clearly understood in order to interpret agreement between
hydrodynamics calculations and measured flow.

In addition to predicting one-body hadron momentum spectra as a
function of many kinematic variables, hydrodynamic evolution of
the matter is also relevant for understanding two-hadron
Hanbury-Brown-Twiss (HBT) quantum correlation
functions~\cite{tomasik-wiedemann}.  From these correlation
measurements one can extract information concerning the size and
shape of the emitting surface at freezeout, \emph{i.e.}, at the
end of the space-time evolution stage treated by hydrodynamics.
While the detailed comparison certainly depends on improving
models of the freezeout stage, it is reasonable to demand that
hydrodynamics calculations consistent with the one-body hadron
measurements be also at least roughly consistent with HBT results.

\subsection{Statistical Models}

The aim of statistical models is to derive the equilibrium
properties of a macroscopic system from the measured yields of the
constituent particles~\cite{pbm,huang}. Statistical models,
however, do not describe how a system approaches
equilibrium~\cite{huang}.  Hagedorn~\cite{hagedorn} and
Fermi~\cite{fermi} pioneered their application to computing
particle production yield ratios in high energy collisions, where
conserved quantities such as baryon number and strangeness play
important roles~\cite{redlich}. Statistical methods have become an
important tool to study the properties of the fireball created in
high energy heavy ion collisions~\cite{pbm,nxu}, where they
succeed admirably in reproducing measured yield ratios.  Can this
success be taken as evidence that the matter produced in these
collisions has reached thermal and
chemical equilibrium before hadronization?
Can the temperature and chemical potential
values extracted from such statistical model fits be interpreted
as the equilibrium properties of the collision matter?

The answer to both of the above questions is ``not on the basis of
fits to integrated yields alone." The essential condition for
applicability of statistical models is phase-space dominance in
determining the distribution of a system with many degrees of
freedom among relatively few observables~\cite{fermi,koch}, and
this does not necessarily reflect a process of thermodynamic
equilibration via interactions of the constituents. Indeed,
statistical model fits can describe the observed hadron abundances
well (albeit, only by including a strangeness undersaturation
factor, $\gamma_s < 1$) in p+p, $e^+ + e^-$ and p+A collisions,
where thermal and chemical equilibrium are thought not to be
achieved~\cite{pbm} . It is thus desirable to distinguish a system
driven towards thermodynamic equilibrium from one born at
hadronization with statistical phase space distributions, where
``temperature" and ``chemical potential" are simply Lagrange
multipliers~\cite{heinz-qm99}.  In order to make this distinction,
it is necessary and sufficient to measure the extensive
interactions among particles and to observe the change from
canonical ensemble in a small system with the size of a nucleon
(p+p, $e^+ + e^-$) and tens of produced particles, to grand
canonical ensemble in a large system with extended volume and
thousands of produced particles (central
Au+Au)~\cite{pbm,redlich}.

The evolution of the system from canonical to grand canonical
ensemble can be observed, for example, via multi-particle
correlations (especially of particles constrained by conservation
laws~\cite{koch}) or by the centrality dependence of the
strangeness suppression factor $\gamma_s$.  The interactions among
constituent particles, necessary to attainment of \emph{thermal}
equilibrium, can be measured by collective flow of many identified
particles~\cite{teaney,art} and by resonance
yields~\cite{rafelski1} that follow their hadronic rescattering
cross sections.  (Collective flow and resonance formation could,
in principle, proceed via the dominant hadronic interactions that
do not change hadron species, and hence are not strictly
sufficient to establish \emph{chemical} equilibration among
hadrons, which would have to rely on relatively weak
\emph{inelastic} processes \cite{heinz-qm99}.)

If other measurements confirm the applicability of a grand
canonical ensemble, then the hadron yield ratios can be used to
extract the temperature and chemical potential of the
system~\cite{pbm} at chemical freezeout.  The latter is defined as
the stage where hadrons have been created and the net numbers of
stable particles of each type no longer change in further system
evolution.  These values place constraints on, but do not directly
determine, the properties of the matter when thermal equilibrium
was first attained in the wake of the collision.  Direct
measurement of the temperature at this early stage requires
characterization of the yields of particles such as photons that
are produced early but do not significantly interact on their way
out of the collision zone.

%\section{Blast Wave Model}

\subsection{Jet Quenching and Parton Energy Loss}

Partons from the colliding nuclei that undergo a hard scattering
in the initial stage of the collision provide colored probes for
the colored bulk matter that may be formed in the collision's
wake.  It was Bjorken \cite{Bjorken} who first suggested that
partons traversing bulk partonic matter might undergo significant
energy loss, with observable consequences on the parton's
subsequent fragmentation into hadrons.  More recent theoretical
studies have demonstrated that the elastic parton scattering
contribution to energy loss first contemplated by Bjorken is
likely to be quite small, but that gluon radiation induced by
passage through the matter may be quite sizable \cite{GVWZ}.  Such
induced gluon radiation would be manifested by a significant
softening and broadening of the jets resulting from the
fragmentation of partons that traverse substantial lengths of
matter containing a high density of partons -- a phenomenon called
``jet quenching".  As will be documented in later chapters, some
of the most exciting of the RHIC results reveal jet quenching
features quite strikingly.  It is thus important to understand
what features of this phenomenon may distinguish parton energy
loss through a QGP from other possible sources of jet softening
and broadening.

Several different theoretical
evaluations of the non-Abelian
radiative energy loss of partons in dense, but finite, QCD matter
have been developed \cite{{GLV},{WW},{BDMS},{Wied}}.  They give essentially
consistent results, including the non-intuitive prediction
 that the energy loss
varies with the square ($L^2$) of the thickness traversed through
static matter, as a consequence of destructive interference
effects in the coherent system of the leading quark and its first
radiated gluon as they propagate through the matter. The overall
energy loss is reduced, and the $L$-dependence shifted toward
linearity, by the expansion of the matter resulting from heavy ion
collisions. The significant deformation of the collision zone for
non-central collisions, responsible for the observed elliptic flow
(hence also for an azimuthal dependence of the rate of matter
expansion), should give rise to a significant variation of the
energy loss with angle with respect to the impact parameter plane.
The scale of the net energy loss depends on factors that can all
be related to the rapidity density of gluons ($dN_g/dy$) in the
matter traversed.

The energy loss calculated via any of these approaches is then
embedded in a perturbative QCD (pQCD) treatment of the hard parton
scattering. The latter treatment makes the standard factorization
assumption (untested in the many-nucleon environment) that the
cross section for producing a given final-state high-$p_T$ hadron can
be written as the product of suitable initial-state parton
densities, pQCD hard-scattering cross section,
and final-state
fragmentation functions for the scattered partons. Nuclear
modifications must be expected for the initial parton densities as
well as for the fragmentation functions. Entrance-channel
modifications -- including both nuclear shadowing of parton
densities and the introduction by multiple scattering of
additional transverse momentum to the colliding partons -- are
capable of producing some broadening and softening of the
final-state jets.  But these effects can, in principle, be
calibrated by complementing RHIC A+A collision studies with p+A or
d+A, where QGP formation is not anticipated.

The existing theoretical treatments of the final-state
modifications attribute the changes in effective fragmentation
functions to the parton energy loss. That is, they assume vacuum
fragmentation (as characterized phenomenologically from jet
studies in more elementary systems) of the degraded parton and its
spawned gluons \cite{GVWZ}.  This assumption may be valid in the
high-energy limit, when the dilated fragmentation time should
exceed the traversal time of the leading parton through the
surrounding matter.  However, its justification seems questionable
for the soft radiated gluons and over the leading-parton momentum
ranges to which it has been applied so far for RHIC collisions. In
these cases, one might expect hadronization to be aided by the
pickup of other partons from the surrounding QGP, and not to rely
solely on the production of $q{\overline q}$ pairs from the
vacuum. Indeed, RHIC experimental results to be described later in
this document hint that the distinction between such recombination
processes and parton fragmentation in the nuclear environment may
not be clean. Furthermore, one of the developed models of parton
energy loss \cite{WW} explicitly includes energy \emph{gain} via
absorption of gluons from the surrounding thermal QGP bath.

The assumption of vacuum fragmentation also implies a neglect of
FSI effects for the hadronic fragmentation products, which might
further contribute to jet broadening and softening. Models that
attempt to account for \emph{all} of the observed jet quenching
via the alternative description of hadron energy loss in a
hadronic gas environment are at this time still incomplete
\cite{Greiner}.  They must contend with the initial expectation of
\emph{color transparency} \cite{CT}, \emph{i.e.}, that high
momentum hadrons formed in strongly interacting matter begin their
existence as point-like color-neutral particles with very small
color dipole moments, hence weak interactions with surrounding
nuclear matter.  In order to produce energy loss consistent with
RHIC measurements, these models must then introduce \emph{ad hoc}
assumptions about the rate of growth of these ``pre-hadron"
interaction cross sections during traversal of the surrounding
matter \cite{Greiner}.

The above caveats concerning assumptions of the parton energy loss
models may call into question some of their quantitative
conclusions, but are unlikely to alter the basic qualitative
prediction that substantial jet quenching is a \emph{necessary}
result of QGP formation.  The more difficult question is whether
the observation of jet quenching can also be taken as a
\emph{sufficient} condition for a QGP discovery claim?  Partonic
traversal of matter can, in principle, be distinguished from
effects of \emph{hadronic} traversal by detailed dependences of
the energy loss, \emph{e.g.}, on azimuthal angle and system size
(reflecting the nearly quadratic length-dependence characteristic
of gluon radiation), on $p_T$ (since hadron formation times should
increase with increasing partonic momentum~\cite{Wang}), or on
type of detected hadron (since hadronic energy losses should
depend on particle type and size, while partonic energy loss
should be considerably reduced for heavy quarks
\cite{Wang,Djordjevic}). However, the energy loss calculations do
not (with the exception of the small quantitative effect of
\emph{absorption} of thermal gluons \cite{WW}) distinguish
confined from deconfined quarks and gluons in the surrounding
matter.  Indeed, the same approaches have been applied to
experimental results from semi-inclusive deep inelastic scattering
\cite{HERMES} or Drell-Yan dilepton production \cite{E866}
experiments on nuclear targets to infer quark energy losses in
\emph{cold}, confined nuclear matter \cite{Zhang-Wang}.
Baier, \emph{et al.}~\cite{Baier}
have shown that the energy loss is expected to vary smoothly with
energy density from cold hadronic to hot QGP matter, casting doubt
on optimistic speculations \cite{Wang} that the QGP transition
might be accompanied by a rapid change in the extent of jet
quenching with collision energy. Thus, the relevance of the QGP
can only be inferred indirectly, from the magnitude of the gluon
density $dN_g/dy$ needed to reproduce jet quenching in RHIC
collision matter, vis-a-vis that needed to explain the energy loss
in cold nuclei. Is the extracted gluon density consistent with
what one might expect for a QGP formed from RHIC collisions? To
address this critical question, one must introduce new theoretical
considerations of the initial state for RHIC collisions.

\subsection{Saturation of Gluon Densities}

In a partonic view, the initial conditions for the expanding
matter formed in a RHIC collision are dominated by the scattering
of gluons carrying small momentum fractions (Bjorken $x$) in the
nucleons of the colliding nuclei.  Gluon densities in the proton
have been mapped down to quite small values of $x \sim 10^{-4}$ in
deep inelastic scattering experiments at HERA~\cite{HERA-glue}.
When the measurements are made with high resolving power
(\emph{i.e.}, with large 4-momentum transfer $Q^2$), the extracted
gluon density $xg(x,Q^2)$ continues to grow rapidly down to the
lowest $x$ values measured.  However, at moderate $Q^2 \sim$ few
(GeV)$^2$, there are indications from the HERA data that
$xg(x,Q^2)$ begins to saturate, as might be expected from the
competition between gluon fusion ($g+g \rightarrow g$) and gluon
splitting ($g \rightarrow g+g$) processes.  It has been
conjectured~\cite{mueller,McLerran1,McLerran2,raju} that the onset
of this saturation moves to considerably higher $x$ values (for
given $Q^2$) in a nuclear target, compared to a proton, and that a
QGP state formed in RHIC collisions may begin with a saturated
density of gluons.  Indeed, birth within this saturated state
might provide a natural mechanism for the rapid achievement of
thermal equilibrium in such collisions \cite{mueller}.

\begin{figure}[thb]
%\begin{minipage}[b]{14cm}
\begin{center}
\epsfig{figure=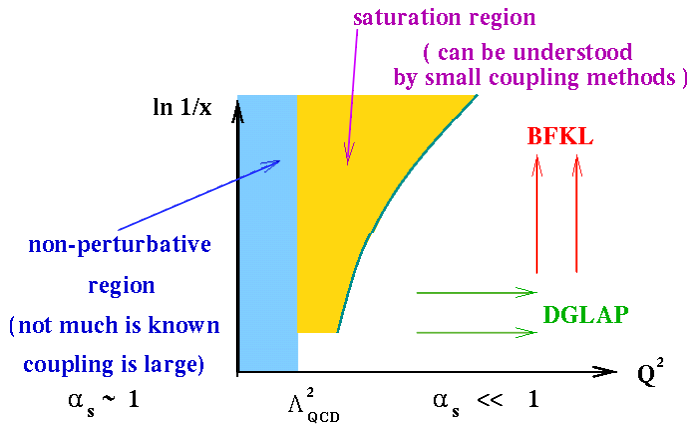,width=9.0cm}
%\end{center}
%\end{minipage}
%\vspace{6mm}
%\begin{minipage}[h]{14cm}
\caption{ {\it Schematic layout of the QCD landscape in $x-Q^2$
space.  The region at the right is the perturbative region, marked
by applicability of the linear DGLAP\protect\cite{DGLAP} and
BFKL\protect\cite{BFKL} evolution equations for the $Q^2$- and
$x$-dependence, respectively, of the parton distribution
functions. At $Q^2 < \Lambda_{QCD}^2$, the coupling constant is
large and non-perturbative methods must be used to treat strongly
interacting systems.  The matter in RHIC collisions may be formed
in the intermediate region, where gluon densities saturate, the
coupling is still weak, but very strong color fields lead to
non-linear behavior describable by classical field methods.  The
curve separating the saturation and perturbative regimes sets the
saturation scale. Figure courtesy of Y. Kovchegov. } }
\label{CGC-landscape}
%\end{minipage}
\end{center}
\end{figure}

The onset of saturation occurs when the product of the cross
section for a QCD process (such as gluon fusion) of interest
($\sigma \sim \pi \alpha_s(Q^2)/Q^2$) and the areal density of
partons ($\rho$) available to participate exceeds
unity~\cite{Kharzeev}. In this so-called Color Glass Condensate
region (see Fig.~\ref{CGC-landscape}), QCD becomes highly
non-linear, but amenable to classical field treatments, because
the coupling strength remains weak ($\alpha_s << 1$) while the
field strength is large~\cite{mueller,McLerran1,McLerran2,raju}.
The borderline of the CGC region is denoted by the ``saturation
scale" $Q_s^2(x,A)$. It depends on both $x$ and target mass number
$A$, because the target gluon density depends on both factors.  In
particular, at sufficiently low $x$ and moderate $Q^2$, $\rho$ is
enhanced for a nucleus compared to a nucleon by a factor $\sim
A^{1/3}$: the target sees the probe as having a longitudinal
coherence length ($\ell_c \sim 1/m_Nx$) much greater, but a
transverse size ($\sim 1/Q^2$) much smaller, than the nuclear
diameter. The probe thus interacts coherently with all the target
gluons within a small diameter cylindrical ``core" of the nucleus.
The HERA data~\cite{HERA-glue} suggest a rather slow variation --
$xg(x) \propto x^{-\lambda}$, with $\lambda \sim 0.3$ at $Q^2
\sim$ few (GeV)$^2$ -- of gluon densities with $x$ at low $x$.
Consequently, one would have to probe a proton at roughly two
orders of magnitude lower $x$ than a Au nucleus to gain the same
factor growth in gluon densities as is provided by $A^{1/3}$.

Under the assumption that QGP formation in a RHIC collision is
dominated by gluon-gluon interactions below the saturation scale,
saturation models predict the density of gluons produced per unit
area and unit rapidity~\cite{mueller}:
\begin{equation}
\label{glue-density} {d^2 N \over d^2b dy} ~=~ C {{N_c^2 - 1} \over
{4 \pi^2 \alpha_s(Q_s^2)N_c}} Q_s^2(x,N_{part}),
\end{equation}
where A has been replaced by $N_{part}$, the number of nucleons
participating in an A+A collision at given impact parameter $b$,
and $\hbar=c=1$.
The $x$-dependence of the saturation scale is taken from the HERA
data,
\begin{equation}
\label{sat-scale} Q_s^2(x) ~=~ Q_0^2 ({x_0 \over x})^\lambda,
\end{equation}
and the same values of $\lambda \sim 0.2-0.3$ are generally
assumed to be valid inside the nucleus as well.
However, the multiplicative factor
$C$ above, parameterizing the number of outgoing hadrons per
initially present gluon, is typically adjusted to fit observed
outgoing hadron multiplicities from RHIC collisions. (Variations
in $C$ are clearly not distinguishable, in the context of
Eqn.~(\ref{glue-density}), from changes to the overall saturation
scale $Q_0^2$.) Once this parameter is fixed, gluon saturation
models should be capable of predicting the dependence of hadron
multiplicity on collision energy, rapidity, centrality and mass
number. Furthermore, the initial QGP gluon densities extracted can
be compared with the independent values obtained from parton
energy loss model fits to jet quenching observations
or from hydrodynamics calculations
of elliptic flow.

While it is predictable within the QCD framework that gluon
saturation should occur under appropriate conditions, and the
theoretical treatment of the CGC state is highly evolved
~\cite{mueller,McLerran1,McLerran2,raju}, the dependences of the
saturation scale are not yet fully exposed by supporting data.
Eventual confirmation of the existence of such a scale must come
from comparing results for a wide range of high energy experiments
from Deep Inelastic Scattering in ep and eA (HERA, eRHIC) to pA
and AA (RHIC, LHC) collisions.

\subsection{Manifestations of Quark Recombination}\label{sec:theoryReco}

The concept of quark recombination was introduced to describe
hadron production in the forward region in p+p
collisions~\cite{recom77}.  At forward rapidity, this mechanism
allows a fast quark resulting from a hard parton scattering to
recombine with a slow anti-quark, which could be one in the
original sea of the incident hadron, or one excited by a
gluon~\cite{recom77}. If a QGP is formed in relativistic heavy ion
collisions, then one might expect recombination of a different
sort, namely, coalescence of the abundant thermal partons, to
provide another important hadron production mechanism, active over
a wide range of rapidity and transverse
momentum~\cite{coalescence95}. In particular, at moderate $p_T$
values (above the realm of hydrodynamics applicability), this
hadron production ``from below" (recombination of lower $p_T$
partons from the thermal bath) has been predicted~\cite{recom03}
to be competitive with production ``from above" (fragmentation of
higher $p_T$ scattered partons). It has been
suggested~\cite{MuellerRBRC} that the need for substantial
recombination to explain observed hadron yields and flow may be
taken as a signature of QGP formation.

In order to explain observed features of RHIC collisions, the
recombination models~\cite{coalescence95,recom03} make the central
assumption that coalescence proceeds via \emph{constituent}
quarks, whose number in a given hadron determines its production
rate. The constituent quarks are presumed to follow a thermal
(exponential) momentum spectrum and to carry a collective
transverse velocity distribution. This picture leads to clear
predicted effects on baryon and meson production rates, with the
former depending on the spectrum of thermal constituent quarks and
antiquarks at roughly one-third the baryon $p_T$, and the latter
determined by the spectrum at one-half the meson $p_T$. Indeed,
the recombination model was recently re-introduced in the RHIC
context, precisely to explain an anomalous abundance of baryons
vs. mesons observed at moderate $p_T$ values~\cite{recom03}.  If
the observed (saturated) hadronic elliptic flow values in this
momentum range result from coalescence of collectively flowing
constituent quarks, then one can expect a similarly simple baryon
vs. meson relationship~\cite{recom03}: the baryon (meson) flow
would be 3 (2) times the quark flow at roughly one-third
(one-half) the baryon $p_T$.

As will be discussed in later
chapters, RHIC experimental results showing just such simple
predicted baryon vs. meson features would appear to provide strong
evidence for QGP formation.  However, the models do not spell out
the connection between the inferred spectrum and flow of
constituent quarks and the properties of the essentially massless
partons (predominantly gluons) in a chirally restored QGP, where
the chiral condensate (hence most of the constituent quark mass)
has vanished. One may guess that the constituent quarks themselves
arise from an earlier coalescence of gluons and \emph{current}
quarks during the chiral symmetry breaking transition back to
hadronic matter, and that the constituent quark flow is carried
over from the partonic phase.

However, alternative guesses
concerning the relation of partons to the recombination degrees of
freedom are also conceivable. Perhaps it is valence current,
rather than constituent, quarks that recombine to determine hadron
flow and momentum in this moderate-$p_T$ range. In that case,
hadronization might proceed through the formation of
``pre-hadrons" (\emph{e.g.}, the pointlike color singlet objects
discussed in connection with color transparency \cite{CT}) from
the leading Fock (valence quark only) configurations, giving rise
to the same 3-to-2 baryon-to-meson ratios as for constituent
quarks.  The internal pre-hadron wave functions would
then subsequently evolve toward
those of ordinary hadrons on their way out of the collision zone,
so that the little-modified hadron momentum would in the end be
shared substantially among sea quarks and gluons, as well as the
progenitor valence quarks.  Either
of the above speculative (and quite possibly not orthogonal)
interpretations of recombination would suggest that the hadron
flow originates in, but is two steps removed from, \emph{partonic}
collectivity. But it is difficult to draw firm conclusions in
light of the present ambiguity in connecting the
effective degrees of freedom in
coalescence models to the quarks and gluons treated by LQCD.

In addition, it is yet to be demonstrated that the coexistence of
coalescence and fragmentation processes is quantitatively
consistent with hadron angular correlations observed over $p_T$
ranges where coalescence is predicted to dominate. These
correlations exhibit prominent
(near-side) peaks with angular
widths (at least in azimuthal difference between two
moderate-$p_T$ hadrons) and charge sign ordering characteristic of
jets from vacuum fragmentation of hard
partons~\cite{star:highpTbtob}. The coalescence yield might simply
contribute to the background underlying these peaks, but one
should also expect contributions from the ``fast-slow"
recombination (hard scattered parton with QGP bath partons)
\cite{Hwa} for which the model was first introduced, and these
could produce charge sign ordering. The latter effects -- part of
in-medium, as opposed to vacuum, fragmentation -- complicate the
interpretation of the baryon/meson comparisons and, indeed, muddy
the distinction between fragmentation and recombination processes.

Finally, the picture provided by recombination is distinctly
different from ideal hydrodynamics
at a hadronic level, where
velocity (mass) of a hadron is
the crucial factor determining flow, rather than the number of
constituent (or valence) quarks.
At low momentum, energy and entropy conservations become a serious
problem for quark coalescence, placing an effective lower limit on
the $p_T$ range over which the models can be credibly applied. The
solution of this problem would require a dynamical, rather than
purely kinematic treatment of the recombination
process~\cite{recom03}. Such parton dynamics at low momentum
might account for the
thermodynamic properties of the macroscopic system discussed
earlier, but we do not yet have a unified partonic theoretical
framework.

\newpage
%--=================================================================
\section{Bulk properties}

The multiplicities, yields, momentum spectra and correlations of
hadrons emerging from heavy-ion collisions, especially in the soft
sector comprising particles at transverse momenta $p_T \lesssim
1.5$ GeV/c, reflect the properties of the bulk of the matter
produced in the collision. In particular, we hope to infer
constraints on its initial conditions, its degree of
thermalization and its Equation of State from measurements of soft
hadrons.

The measured hadron spectra reflect the properties of the bulk of
the matter at kinetic freezeout, after elastic collisions among
the hadrons have ceased. At this stage the system is already
relatively dilute and ``cold''. However from the detailed
properties of the hadron spectra at kinetic freezeout, information
about the earlier hotter and denser stage can be obtained.
Somewhat more direct information on an earlier stage can be
deduced from the integrated yields of the different hadron
species, which change only via \emph{inelastic} collisions. These
inelastic collisions cease already (at so-called chemical
freezeout) before kinetic freezeout.

The transverse momentum distributions of the different particles
reflect a random and a collective component. The random component
can be identified with the temperature of the system at kinetic
freezeout. The collective component arises from the matter density
gradient from the center to the boundary of the fireball created
in high-energy nuclear collisions. Interactions among constituents
push matter outwards; frequent interactions lead to a common
constituent velocity distribution. This so-called \emph{collective
flow} is therefore sensitive to the strength of the interactions.
The collective flow is additive and thus accumulated over the
whole system evolution, making it potentially sensitive to the
Equation of State of the expanding matter. At lower energies the
collective flow reflects the properties of dense hadronic
matter~\cite{ritter97}, while at RHIC energies a contribution from
a pre-hadronic phase is anticipated.

In non-central heavy-ion collisions the initial transverse density
gradient has an azimuthal anisotropy that leads to an azimuthal
variation of the collective transverse flow velocity with respect
to the impact parameter plane for the event. This azimuthal
variation of flow is expected to be self-quenching (see Sec. 2.2),
hence, especially sensitive to the interactions among constituents
in the \emph{early} stage of the
collision~\cite{sorge97,ollitrault92}, when the system at RHIC
energies is anticipated to be well above the critical temperature
for QGP formation.

Study of quantum (boson) correlations among pairs of emerging
hadrons utilizes the Hanbury-Brown-Twiss effect to complement
measurements of momentum spectra with information on the spatial
size and shape of the emitting system.  The measurement of more
general two-particle correlations and of event-wise fluctuations
can illuminate the degree of equilibration attained in the final
hadronic system, as well as the dynamical origin of any observed
non-equilibrium structures.  Such dynamical correlations are
prevalent in high-energy collisions of more elementary particles
-- where even relatively soft hadrons originate in large part from
the fragmentation of partons -- but are expected to be washed out
by thermalization processes that produce phase space dominance of
the final distribution probabilities.

In this section, we review the most important implications and
questions arising from RHIC's vast
body of data on soft hadrons.  We also discuss some features of
the transition region ($1.5 \lesssim p_T \lesssim 6$ GeV/c), where
the spectra gradually evolve toward the characteristic behavior of
the hard parton fragmentation regime.
In the process of going through
the measured features of hadron spectra in the logical sequence
outlined above, we devote special attention to a few critical
features observed \emph{for the first time} for central and
near-central Au+Au collisions at STAR, that bear directly on the
case for the QGP:

\begin{itemize}
\item[$\bullet$] hadron yields
suggestive of chemical equilibration across the $u,~d$ and $s$
quark sectors;

\item[$\bullet$] elliptic flow of
soft hadrons attaining the strength expected for an ideal
relativistic fluid thermalized very shortly after the collision;

\item[$\bullet$] elliptic flow
results at intermediate $p_T$ that appear to arise from the flow
of quarks in a pre-hadronic stage of the matter.

\end{itemize}

%--=================================================================
 \subsection{Rapidity Densities}

%--=================================================================

\begin{figure} [bht]
\begin{center}
\includegraphics[width=0.7\textwidth]{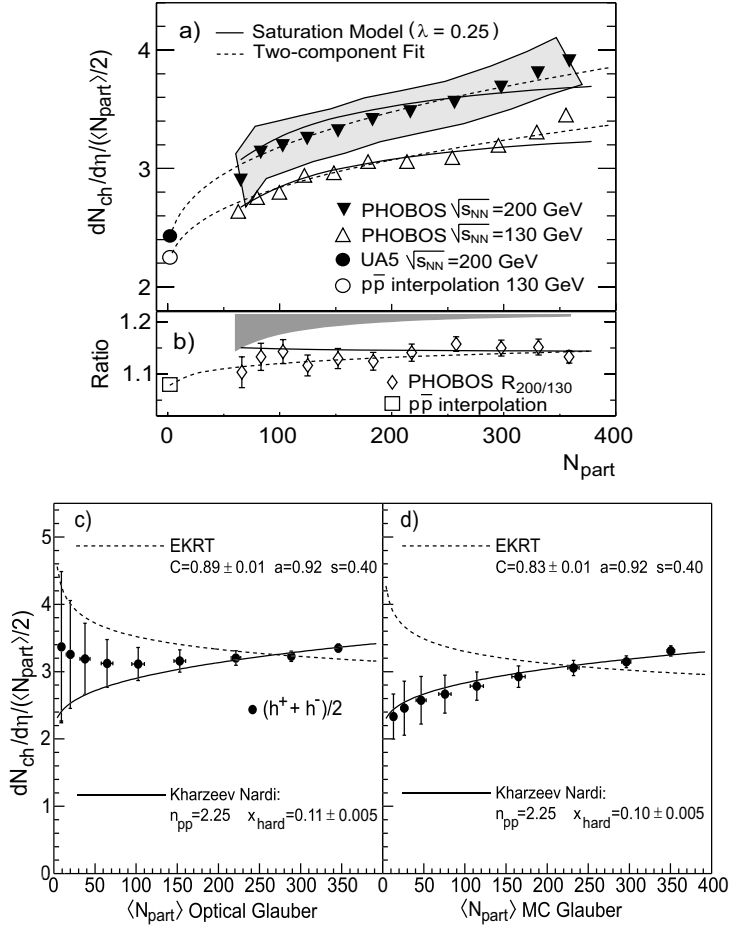}

\caption{ {\it Measured and calculated pseudorapidity densities
$dN_{ch}/d\eta_{|\eta|\le 1}/(\langle N_{part}\rangle /2)$ of
charged particles from RHIC \auau collisions as a function of
$N_{part}$. PHOBOS data \cite{phobos02} at \sqrtsNN = 130 GeV
(open triangles) and 200 GeV (closed triangles) are shown in frame
(a), and their ratio is plotted in (b).  The open and solid
circles in (a) are $\overline{p}p$ collision results.  STAR data
for 130 GeV \cite{Laue-spectra} are shown in frames (c) and (d),
plotted with two different Glauber model treatments to deduce
$\langle N_{part} \rangle$. The data plotted in frames (a), (b)
and (d) utilize the preferred Monte-Carlo Glauber approach.
However, the initial-state gluon saturation model calculations
\cite{Kharzeev} shown as solid curves in all frames have been
carried out utilizing the questionable optical approximation to
the Glauber treatment, which is applied as well to the
experimental results only in frame (c). The dashed curves in
frames (a) and (b) represent two-component fits to the data
\cite{phobos02} and in frames (c) and (d) represent an alternative
model \cite{EKRT} assuming saturation of final-state mini-jet
production.}} \label{mult}
\end{center}
\end{figure}

Much has been made of the fact that predictions of hadron
multiplicities in RHIC collisions before the year 2000 spanned a
wide range of values, so that even the earliest RHIC measurements
had significant discriminating power
\cite{McLerran-Gyulassy,Wang-Gyulassy}. Mid-rapidity charged
hadron densities measured in PHOBOS \cite{phobos02} and in STAR
\cite{Laue-spectra} are plotted in Fig.~\ref{mult} as a function
of collision centrality, as characterized by the number of
participating nucleons, $N_{part}$, inferred from the fraction of
the total geometric cross section accounted for in each analyzed
bin.  The solid curves in the figure represent calculations within
a gluon saturation model \cite{Kharzeev}, while the dashed curves
in frames (a) and (b) represent two-component fits to the data
\cite{phobos02} and in frames (c) and (d) represent an alternative
model \cite{EKRT} assuming saturation of final-state mini-jet
production.  The apparent logarithmic dependence of the measured
pseudorapidity densities on $\langle N_{part} \rangle$ is a
characteristic feature of the gluon saturation model
\cite{Kharzeev}. Consequently, the model's ability to reproduce
the measured centrality and energy dependences have been presented
as evidence for the relevance of the Color Glass Condensate to the
initial state for RHIC collisions, and used to constrain the
saturation scale for initial gluon densities.  This scale is in
fair agreement with the scale extrapolated from HERA e-p
measurements at low Bjorken $x$ \cite{McLerran-Gyulassy}.

%--============================================================

However, these arguments are compromised because the particle
multiplicity appears not to have strong discriminating power once
one allows for adjustment of theoretical
parameters.  Furthermore,
$\langle N_{part} \rangle$, which affects the scale on both axes
in Fig.~\ref{mult}, is not a direct experimental observable.
Glauber model calculations to associate values of $\langle
N_{part} \rangle$ with given slices of the geometric cross section
distribution have been carried out in two different ways,
leading to an inconclusive theory
\emph{vs.} experiment comparison in Figs.~\ref{mult}(a) and (b).
The preferred method for evaluating Glauber model cross sections
in nucleus-nucleus collisions \cite{Czyz} invokes a Monte Carlo
approach for integrating over all nucleon configurations, and has
been used for the experimental results in frames (a), (b) and (d).
However, the gluon saturation model calculations in \emph{all}
frames of Fig.~\ref{mult} have employed the optical approximation,
which ignores non-negligible correlation effects \cite{Czyz}.
Comparison of the \emph{experimental} results in frames (c) and
(d), where the same STAR data have been plotted using these two
Glauber prescriptions, illustrates the significant sensitivity to
the use of the optical approximation.  The ``apples-to-apples"
comparison of experiment and theory in frame (c) does not argue
strongly in favor of initial-state gluon saturation, although an
analogous ``apples-to-apples" comparison within the Monte Carlo
Glauber framework is clearly desirable.

Furthermore, over a much broader energy range, the charged
particle multiplicity is found to vary quite smoothly from AGS
energies (\sqrtsNN ~$\approx$ few GeV) to the top RHIC energy
(\sqrtsNN=200 GeV)\cite{phobos03} (see Fig.~\ref{dNdeta}).  One
would not expect CGC conditions to be dominant in collisions over
this entire range \cite{McLerran-Gyulassy}, so the apparent
success of CGC arguments for RHIC hadron multiplicities is less
than compelling.  Other evidence more directly relevant to CGC
predictions will be discussed in Chap. 4.

Whatever physics ultimately governs the smooth increase in
produced particle multiplicity with increasing collision energy
and centrality seems also to govern the growth in total transverse
energy per unit pseudorapidity ($dE_T/d\eta$). PHENIX measurements
at \sqrtsNN\ = 130 GeV \cite{phenix:et130}
first revealed that RHIC collisions generate
$\approx 0.8$ GeV of transverse energy per produced charged
particle near mid-rapidity, independent of centrality --
essentially the same value that is observed also in SPS collisions
at an order of magnitude lower center-of-mass energy
\cite{WA98:dETdeta}.
This trend persists to
\sqrtsNN\ = 200 GeV \cite{star:et200}.
For RHIC central Au+Au collisions, this
translates to the conversion of nearly 700 GeV per unit rapidity
($dE_T/dy$) from initial-state longitudinal to final-state
transverse energy \cite{phenix:et130}.  Under simplifying
assumptions (longitudinal boost-invariance, free-streaming
expansion in which the matter does no work) first suggested by
Bjorken \cite{BjorkenHydro}, one can extract from this observation
a crude estimate of the initial spatial energy density of the bulk
matter at the start of its transverse expansion:

\begin{equation}
\label{epsilon} e_{Bj} ~=~ \frac{dE_T}{dy} \frac{1}{\tau_0 \pi
R^2},
\end{equation}

\noindent where $\tau_0$ is the formation time and $R$ the initial
radius of the expanding system.  With reasonable guesses for these
parameter values ($\tau_0 \approx 1$ fm/c, $R \approx 1.2A^{1/3}$
fm), the PHENIX $dE_T/d\eta$ measurements suggest an initial
energy density $\sim 5$ GeV/fm$^3$ for central Au+Au collisions at
RHIC, well above the critical
energy density $\sim 1$ GeV/fm$^3$ expected from LQCD for the
transition to the QGP phase.  This estimate of the
 initial energy density is larger
than that in SPS collisions, since the particle multiplicity grows
at RHIC, but by a modest factor ($\approx 1.6$ \cite{phenix:et130}
at \sqrtsNN\ = 130 GeV).

%--=================================================================
\subsection{Hadron Yields and Spectra}

\begin{figure} [thb]
\begin{center}
\includegraphics[width=0.70\textwidth]{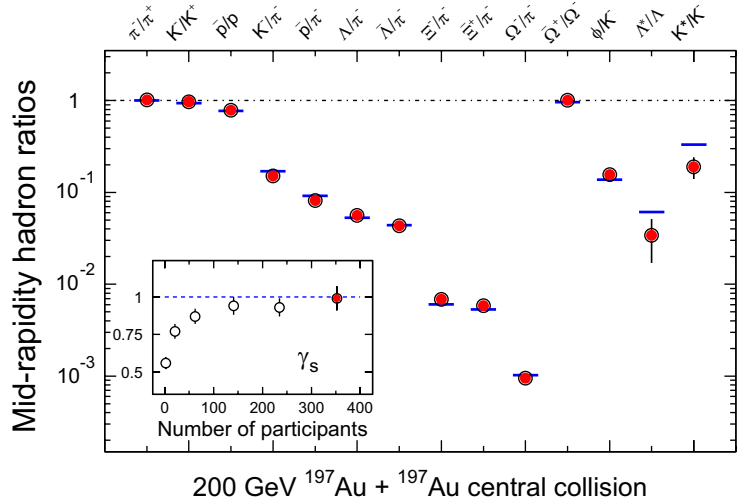}

\caption{ {\it Ratios of $p_T$-integrated mid-rapidity yields for
different hadron species measured in STAR for central Au+Au
collisions at \sqrtsNN\ = 200 GeV. The horizontal bars represent
statistical model fits to the measured yield ratios for stable and
long-lived hadrons.  The fit parameters are $T_{ch}=163 \pm 4$
MeV, $\mu_B = 24 \pm 4$ MeV, $\gamma_s = 0.99 \pm 0.07$
\cite{barranikQM04}. The variation of $\gamma_s$ with centrality
is shown in the inset, including the value (leftmost point) from
fits to yield ratios measured by STAR for 200 GeV p+p collisions.
} }\label{ratio}
\end{center}
\end{figure}

Figure~\ref{ratio} compares STAR measurements of integrated hadron
yield ratios for central Au+Au collisions to statistical model
fits.  In comparison to results from p+p collisions at similar
energies, the relative yield of multi-strange baryons $\Xi$ and
$\Omega$ is considerably enhanced in RHIC Au+Au collisions
\cite{star:multi-strange,barranikQM04}. The measured ratios are
used to constrain the values of system temperature and baryon
chemical potential at chemical freezeout, under the statistical
model assumption that the system is in thermal and chemical
equilibrium at that stage.  The excellent fit obtained to the
ratios in the figure, including stable and long-lived hadrons
through multi-strange baryons, is consistent with the light
flavors, $u, d,$ and $s$, having reached chemical equilibrium (for
central and near-central collisions only) at $T_{ch} = 163 \pm 5$
MeV \cite{pbm,star:multi-strange,barranikQM04,star_pikp}. (The
deviations of the short-lived resonance yields, such as those for
$\Lambda ^*$ and K$^*$ collected near the right side of
Fig.~\ref{ratio}, from the statistical model fits, presumably
result from hadronic rescattering after the chemical freezeout.)

%--=================================================================

Although the success of the statistical model in Fig.~\ref{ratio}
might, in isolation, indicate hadron production mechanisms
dominated by kinematic phase space in elementary collisions (see
Sec. 2.3), other measurements to be discussed below suggest that
true thermal and chemical equilibration is at least approximately
achieved in heavy-ion collisions at RHIC by interactions among the
system's constituents. The saturation of the strange sector
yields, attained for the first time in near-central RHIC
collisions, is particularly significant. The saturation is
indicated quantitatively by the value obtained for the
non-equilibrium parameter $\gamma_s$ for the strange sector
\cite{Xu-Kaneta}, included as a free parameter in the statistical
model fits.  As seen in the inset of Fig.~\ref{ratio}, $\gamma_s$
rises from $\approx 0.75$ in peripheral Au+Au collisions to values
statistically consistent with unity
~\cite{star:multi-strange,barranikQM04} for central collisions.
The temperature deduced from the fits is essentially equal to the
critical value for a QGP-to-hadron-gas transition predicted by
LQCD \cite{Rischke,Karsch}, but is also close to the Hagedorn
limit for a hadron resonance gas, predicted without any
consideration of quark and gluon degrees of
freedom~\cite{hagedorn}.\footnote{Note that Hagedorn himself
considered the Hagedorn temperature and the LQCD critical
temperature to be identical \protect\cite{Hagedorn:1984hz}.}  If
thermalization is indeed achieved by the bulk matter \emph{prior}
to chemical freezeout, then the deduced value of $T_{ch}$
represents a lower limit on that thermalization temperature.

\begin{figure} [bht]
\begin{center}
\includegraphics[width=0.80\textwidth]{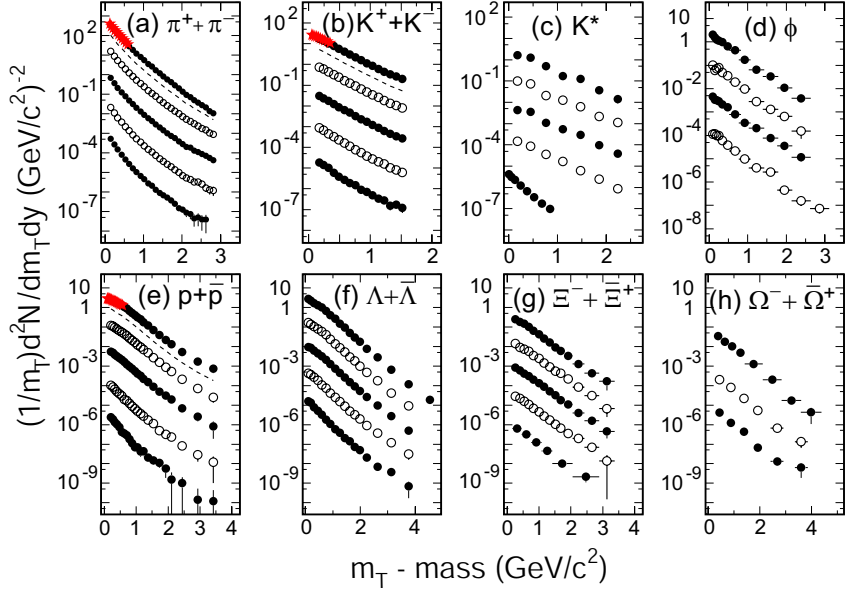}

\caption{ {\it Mid-rapidity hadron spectra from \sqrtsNN = 200 GeV
\auau collisions, as reported in
Refs.~\cite{star_pikp,phenixspectra,starphi,kaiqm04}. The spectra
are displayed for steadily decreasing centrality from the top
downwards within each frame, with appropriate scaling factors
applied to aid visual comparison of the results for different
centralities. For K$^*$ only, the lowest spectrum shown is for 200
GeV \pp\ collisions.  The dashed curves in frames (a), (b) and (e)
represent spectra from minimum-bias collisions. The invariant
spectra are plotted as a function of $m_T - {\rm mass} \equiv
\sqrt{p_T^2/c^2+{\rm mass}^2}-{\rm mass}$.} } \label{figspectra}
\end{center}
\end{figure}

\begin{figure}[thb]
\begin{center}
\includegraphics[width=0.7\textwidth]{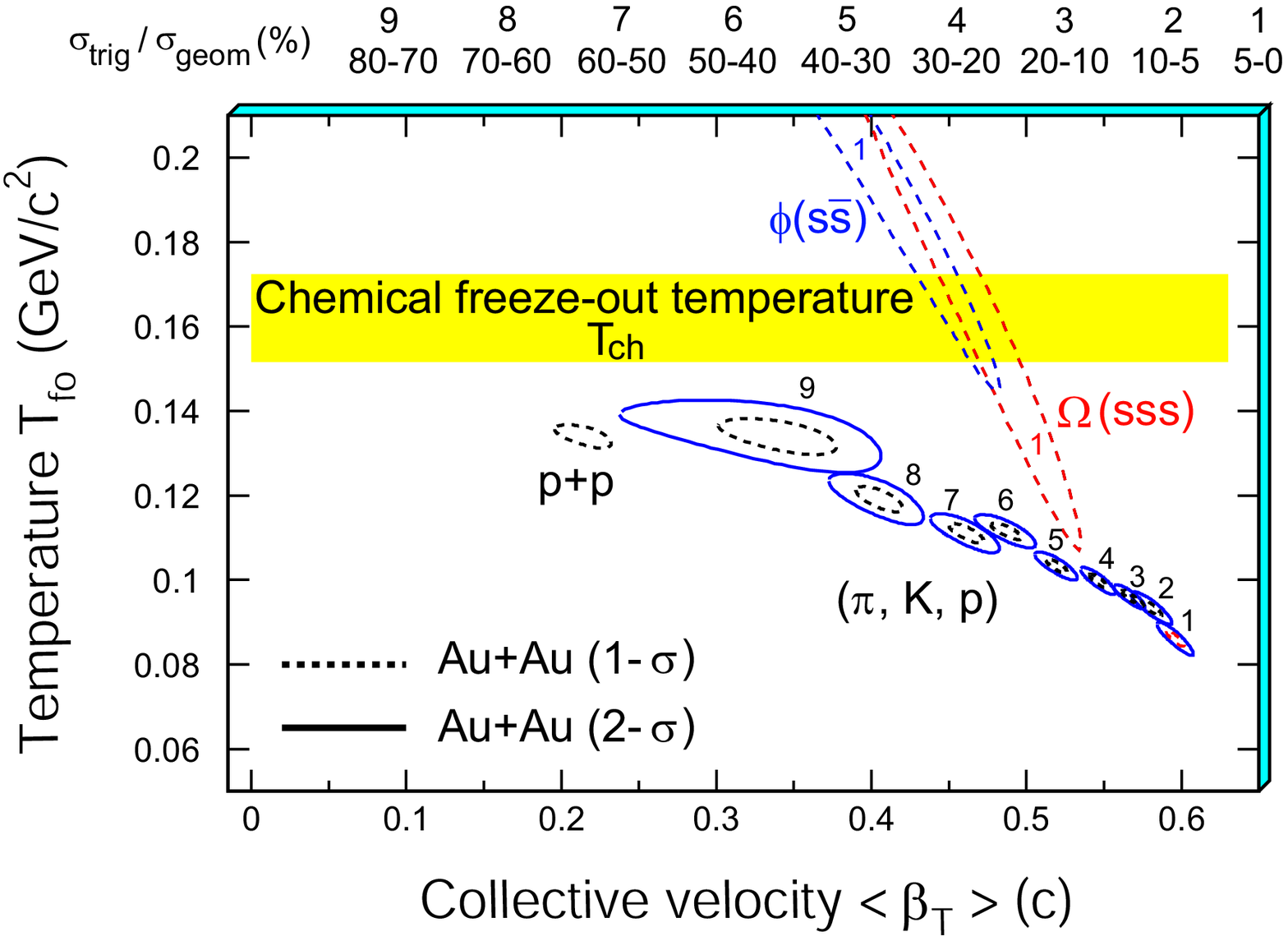}
\caption{ {\it The $\chi^2$ contours, extracted from thermal +
radial flow fits (without
allowance for resonance feed-down), for copiously produced
hadrons $\pi, K$ and $p$ and multi-strange hadrons $\phi$ and
$\Omega$. On the top of the plot, the numerical labels indicate
the centrality selection. For $\pi, K$ and $p$, 9 centrality bins
(from top 5\% to 70-80\%) were used for $\sqrt{s_{NN}}=200$ GeV
\auau collisions \cite{star_pikp}. The results from \pp\
collisions are also shown. For $\phi$ and $\Omega$, only the most
central results \cite{barranikQM04} are presented. Dashed and
solid lines are the 1-$\sigma$ and 2-$\sigma$ contours,
respectively.} } \label{figtbeta}
\end{center}
\end{figure}

The characteristics of the system at \emph{kinetic} freezeout can
be explored by analysis of the transverse momentum distributions
for various hadron species, some of which are shown in
Fig.~\ref{figspectra}. In order to characterize the transverse
motion, hydrodynamics-motivated fits \cite{heinz93} have been made
to the measured spectra, permitting extraction of model parameters
characterizing the random (generally interpreted as a kinetic
freezeout temperature $T_{fo}$) and collective (radial flow
velocity $\langle \beta_T \rangle$) aspects. Results for these
parameters are shown for different centrality bins and different
hadron species in Fig.~\ref{figtbeta}. (While theoretical studies
\cite{heinz93} suggest caution in interpreting spectrum fits made
without correction for resonance feed-down, as is the case in
Fig.~\ref{figtbeta}, auxiliary STAR analyses show little
quantitative effect of the feed-down within the STAR $p_T$
coverage.)

%--=================================================================

%--======================= T versus beta plot

As the collisions become more and more central, the bulk of the
system, dominated by the yields of $\pi, K, p$, appears from
Fig.~\ref{figtbeta} to grow cooler at kinetic freezeout and to
develop stronger collective flow. These results may indicate a
more rapid expansion after chemical freezeout with increasing
collision centrality.  On the other hand, even for the most
central collisions, the spectra for multi-strange particles $\phi$
and $\Omega$ appear, albeit with still large uncertainties, to
reflect a higher temperature \cite{barranikQM04}. The $\phi$ and
$\Omega$ results suggest diminished hadronic interactions with the
expanding bulk matter after chemical freezeout
\cite{star:multi-strange,barranikQM04,pbm9596,bass99}, as
predicted \cite{bass,sorge,dumitru} for hadrons containing no
valence $u$ or $d$ quarks. If this interpretation is correct, the
substantial radial flow velocity inferred for $\phi$ and $\Omega$
would have to be accumulated prior to chemical freezeout, giving
the multi-strange hadrons perhaps greater sensitivity to
collective behavior during earlier partonic stages of the system
evolution.

\begin{figure} [bht]
\begin{center}
\includegraphics[width=0.7\textwidth]{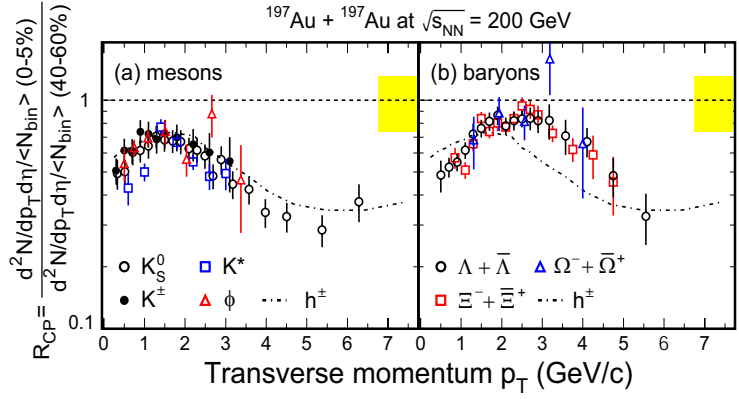}

\caption{ {\it STAR results \cite{kaiqm04} from \sqrtsNN = 200 GeV
Au+Au collisions for the ratio of mid-rapidity hadron yields
$R_{CP}$ in a central (0-5\%) over a peripheral (40-60\%) bin,
plotted vs. \pt for mesons (a) and baryons (b). The yields are
scaled in each centrality region by the calculated mean number
$\langle N_{bin} \rangle$ of binary contributing nucleon-nucleon
collisions, calculated within a Monte Carlo Glauber model
framework. The width of the shaded band around the line at unity
represents the systematic uncertainty in model calculations of the
centrality dependence of $\langle N_{bin} \rangle$. $R_{CP}$ for
the sample of all charged hadrons is also shown by dot-dashed
curves in both plots. The error bars on the measured ratios
include both statistical and systematic uncertainties.} }
\label{rcp}
\end{center}
\end{figure}

As one moves beyond the soft sector, the \pt- and centrality
dependences of the observed hadron spectra develop a systematic
difference between mesons and baryons, distinct from the
mass-dependence observed at lower $p_T$.  This difference is
illustrated in Fig.~\ref{rcp} by the binary-scaled ratio $R_{CP}$
of hadron yields for the most central vs. a peripheral bin,
corrected by the expected ratio of contributing binary
nucleon-nucleon collisions in the two centrality bins
\cite{kaiqm04}. The results are plotted as a function of \pt for
mesons and baryons separately in panels (a) and (b), respectively,
with the ratio of binary collision-scaled yields of all charged
hadrons indicated in both panels by a dot-dashed curve to aid
comparison. If the centrality-dependence simply followed the
number of binary collisions, one would expect $R_{CP}=1$. This
condition is nearly achieved for baryons near $p_T \approx 2.5$
GeV/c, but is never reached for mesons.  The initial results for
$\phi$-mesons and $\Omega$-baryons included in Fig.~\ref{rcp}
suggest that the difference is not very sensitive to the mass of
the hadron, but rather depends primarily on the number of valence
quarks contained within it.  The meson and baryon values appear to
merge by $p_T \approx 5$ GeV/c, by which point $R_{CP} \approx
0.3$.

The origin of this significant shortfall in central high-\pt
hadron production will be discussed at length in Sec. 4. Here, we
want simply to note that the clear difference seen in the
centrality dependence of baryon vs. meson production is one of the
defining features of the intermediate \pt range from $\sim 1.5$ to
$\sim 6$ GeV/c in RHIC heavy-ion collisions, and it cannot be
understood from \pp\ collision results \cite{phenix:meson-baryon}.
Another defining feature of this medium \pt range, to be discussed
further below, is a similar meson-baryon difference in elliptic
flow. Both facets of the meson-baryon differences can be explained
naturally in quark recombination models for hadron formation
\cite{recom03}.

%--=================================================================

%--==============================================================
\subsection{Hadron yields versus the reaction plane}

In non-central heavy-ion collisions, the beam direction and the
impact parameter define a reaction plane for each event, and hence
a preferred azimuthal orientation. The orientation of this plane
can be estimated experimentally by various methods, \emph{e.g.},
using 2- or 4-particle correlations \cite{Poskanzer,Borghini},
with different sensitivities to azimuthal anisotropies not
associated with collective flow. The observed particle yield
versus azimuthal angle with respect to the event-by-event reaction
plane promises information on the early collision
dynamics~\cite{sorge99,ollitrault92}. The anisotropy of the
particle yield versus the reaction plane can be characterized in a
Fourier expansion. Due to the geometry of the collision overlap
region the second coefficient of this Fourier series -- $v_2$,
often referred to as the elliptic flow -- is expected to be the
dominant contribution.

\begin{figure} [thb]
\begin{center}
\includegraphics[width=0.6\textwidth]{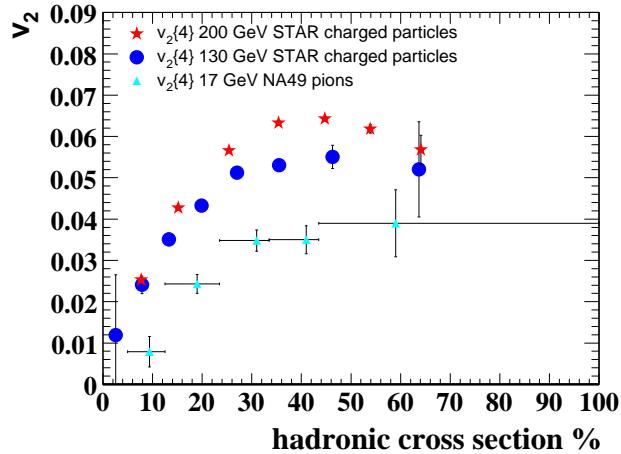}
\caption{ {\it Centrality dependence of $v_2$, integrated over
$p_T$. The triangles are the NA49 measurements for pions at
$\sqrt{s_{_{NN}}}$ = 17 GeV \cite{NA49flow}. The circles and
crosses are STAR measurements for charged particles at
$\sqrt{s_{_{NN}}}$ = 130 GeV \cite{star:flow4part130} and 200 GeV
\cite{Tang:2003kz}, respectively. The 4-particle cumulant method
has been used to determine $v_2$ in each case. } } \label{v2int}
\end{center}
\end{figure}

Figure~\ref{v2int} shows the mid-rapidity elliptic flow
measurements, integrated over transverse momentum, as a function
of collision centrality for one SPS \cite{NA49flow} and two RHIC
\cite{star:flow4part130,Tang:2003kz} energies. One clearly
observes a characteristic centrality dependence that reflects the
increase of the initial spatial eccentricity of the collision
overlap geometry with increasing impact parameter. The integrated
elliptic flow value for produced particles increases about 70\%
from the top SPS energy to the top RHIC energy, and it appears to
do so smoothly as a function of energy (see
Fig.~\ref{v2-vs-energy}), so far exhibiting no obvious ``dip" of
the sort predicted
\cite{kolbsollfrank} by ideal
hydrodynamics in Fig.~\ref{hydro-excitation}.

\begin{figure} [bht]
\begin{center}
\includegraphics[width=0.7\textwidth]{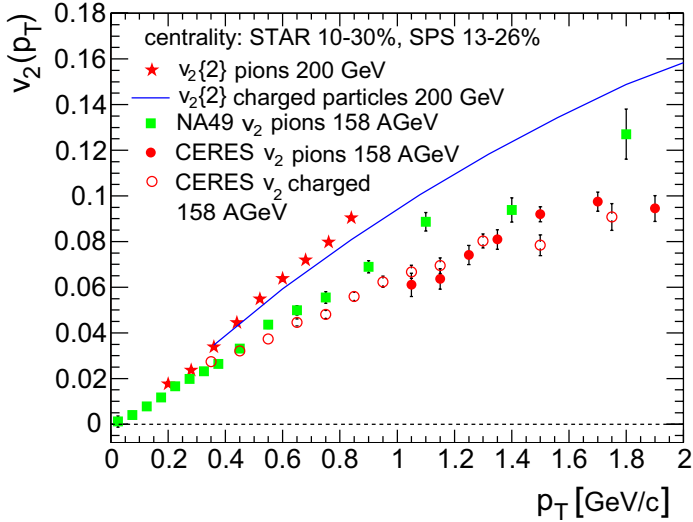}
\caption{ {\it $v_2$($p_t$) for one centrality (10-30\%) range.
The circles and squares are the CERES \cite{CERESAwaySide} and NA49
\cite{NA49flow} measurements, respectively, at $\sqrt{s_{_{NN}}}$
= 17 GeV. The stars and the solid line are STAR measurements
\cite{Tang:2003kz} for pions and for all charged particles,
respectively, at $\sqrt{s_{_{NN}}}$ = 200 GeV (evaluated here by
the 2-particle correlation method). }
 }
 \label{v2ptspsrhic}
 \end{center}
\end{figure}

The origin of the energy dependence can be discerned by examining
the differential $v_2$($p_t$), shown for the centrality selection
10--30\% in Fig.~\ref{v2ptspsrhic}. The comparison of the results
for pions at $\sqrt{s_{_{NN}}}$ = 200 GeV and at the top SPS
energy clearly reveals an increase in slope vs. \pt that accounts
for part of the increase in
\pt-integrated $v_2$ from SPS to RHIC. The remaining
part of the change is due to the
increase in $\langle p_t \rangle$.  As measurements become
available at more collision energies, it will
 be important to
remove kinematic effects, such as the increase in $\langle p_t
\rangle$, from comparisons of results, as they might mask finer,
but still significant, deviations from smooth energy dependence.

\begin{figure} [thb]
\begin{center}
 \includegraphics[width=0.9\textwidth]{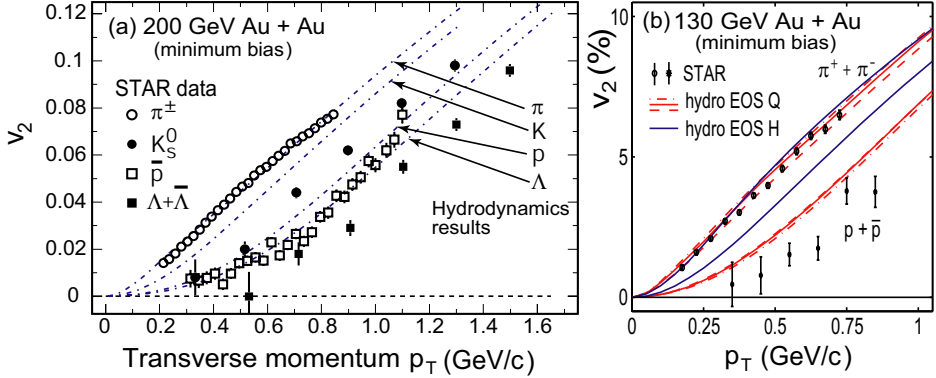}

\caption{{\it (a) STAR experimental results of the transverse
momentum dependence of the elliptic flow parameter in 200 GeV
\auau collisions for charged $\pi^+ + \pi^-$, \ks, $\overline{p}$,
and $\Lambda$ \cite{starklv2}. Hydrodynamics calculations
\cite{Huovinen,pasi03} assuming early thermalization, ideal fluid
expansion, an equation of state consistent with LQCD calculations
including a phase transition at $T_c$=165 MeV (EOS Q in
\cite{Huovinen} and Fig.~\ref{hydro-EOS}), and a sharp kinetic
freezeout at a temperature of 130 MeV, are shown as dot-dashed
lines. Only the lower \pt portion (\pt $\le 1.5$ GeV/c) of the
distributions is shown. (b) Hydrodynamics calculations of the same
sort as in (a), now for a hadron gas (EOS H) vs. QGP (EOS Q)
equation of state (both defined in Fig.~\ref{hydro-EOS})
\cite{kolbheinz,Huovinen}, compared to STAR $v_2$ measurements for
pions and protons in minimum bias 130 GeV \auau collisions
\cite{star130flow}. Predictions with EOS Q are shown for a wider
variety of hadron species in Fig.~\ref{hydro-massdep}. }}
\label{v2low}
\end{center}
\end{figure}

Collective motion leads to predictable behavior of the shape of
the momentum spectra as a function of particle mass, as reflected
in the single inclusive spectra in Fig.~\ref{figspectra}. It is
even more obvious in the dependence of $v_2$($p_t$) for the
different mass particles. Figure~\ref{v2low} shows the measured
low-\pt $v_2$ distributions from 200 and 130 GeV \auau minimum
bias collisions.  Shown are the measurements for charged pions,
$K_S^0$, antiprotons and $\Lambda$\cite{starklv2,star130flow}. The
clear, systematic mass-dependence of $v_2$ shown by the data is a
strong indicator that a common transverse velocity field underlies
the observations.  This mass-dependence, as well as the absolute
magnitude of $v_2$, is reproduced reasonably well (\emph{i.e.}, at
the $\pm 30\%$ level) by the hydrodynamics calculations shown in
Fig.~\ref{v2low}. Parameters of these calculations have been tuned
to achieve good agreement with the measured spectra for different
particles, implying that they account for the observed radial flow
and elliptic flow simultaneously.
In particular, since the
parameters are tuned for zero impact parameter, the
theory-experiment comparison for $v_2$ as a function of centrality
represents a significant test of these hydrodynamics calculations.

The agreement of these hydrodynamics calculations, which assume
\emph{ideal} relativistic fluid flow, with RHIC spectra and $v_2$
results is one of the centerpieces of recent QGP discovery claims
\cite{Gyulassy,McLerran-Gyulassy,RBRC}.  The agreement appears to
be optimized (though still with some quantitative differences, see
Fig.~\ref{v2low}) when it is assumed that local thermal
equilibrium is attained very early ($\tau < 1$ fm/c) during the
collision, and that the hydrodynamic expansion is characterized by
an EOS (labeled Q in Fig.~\ref{v2low}) containing a soft point
roughly consistent with the LQCD-predicted phase transition from
QGP to hadron gas \cite{kolbheinz,teaney,Huovinen}.  When the
expanding matter is treated as a pure hadron gas (EOS H in
Fig.~\ref{v2low}(b)), the mass-dependence of $v_2$ is
significantly underpredicted.  The inferred early thermalization
suggests that the collision's early stages are dominated by very
strongly interacting matter with very short constituent mean free
paths -- essentially a ``perfect liquid"
\cite{Shuryak}, free of
viscosity. Similar QGP-based calculations
that invoke ideal hydrodynamics up
to freezeout overpredict the elliptic flow for more peripheral
RHIC collisions and for lower
energies. One possible
interpretation of this observation is that thermalized, strongly
interacting QGP matter dominates
near-central Au+Au collisions at or near the full RHIC energy.

In assessing these claims, it is critical to ask how unique and
robust the hydrodynamics account
is in detail for the near-central RHIC
collision
flow measurements (radial and elliptic).  Might the observed $v_2$ result
alternatively from a harder EOS (such as EOS H) combined with
later achievement of thermalization
or with higher viscosity
\cite{viscous} (both conditions impeding the development of
collective flow)? How does the sensitivity to the EOS in the
calculations compare quantitatively with the sensitivity to other
ambiguities or questionable assumptions in the hydrodynamics
treatments?  For example, the particular calculations in
Fig.~\ref{v2low} \cite{Huovinen,pasi03} invoke a simplified
treatment with a sharp onset of kinetic freezeout along a surface
of constant energy density corresponding to $T_{fo} \approx
130$ MeV. The sensitivity to the
assumed value of $T_{fo}$, if it is kept within the range spanned
by the measurements in Fig.~\ref{figtbeta}, is relatively weak
\cite{Huovinen}. However, alternative approaches combining
ideal hydrodynamics for the
partonic stage with a hadron transport (RQMD) treatment of the
presumably viscous hadronic stage
\cite{teaney} yield similar success in accounting for RHIC
results, but certainly predict a dependence of $v_2$ on collision
energy differing significantly from the sharp-freezeout
predictions (compare Fig.~\ref{hydro-excitation} and
Fig.~\ref{hydro-RQMD}). While the
combination of partonic
hydrodynamics and hadron transport offers the promise of a
reasonable QGP-based account for
the observed smooth energy dependence of \pt-integrated $v_2$ (see
Figs.~\ref{hydro-RQMD},\ref{v2-vs-energy}),
it also serves to emphasize that quantitative ambiguities of scale
comparable to the EOS sensitivity remain to be understood.

\begin{figure} [thb]
\begin{center}
 \includegraphics[width=0.60\textwidth]{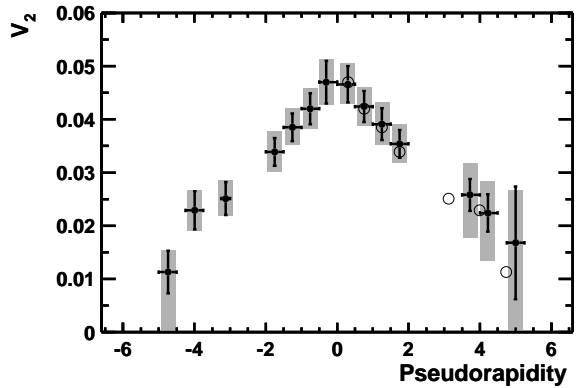}

\caption{{\it  Azimuthal anisotropies $v_2$ measured by the PHOBOS
collaboration \cite{phobos:boost-dependence} for Au+Au collisions
at \sqrtsNN=130 GeV, as a function of pseudorapidity.  Within each
pseudorapidity bin, the results are averaged over all charged
particles, over all centralities and over all \pt. The black error
bars are statistical and the grey bands systematic uncertainties.
The points on the negative side are reflected about $\eta = 0$ and
plotted as open circles on the positive side, for comparison.
Figure taken from Ref.~\cite{phobos:boost-dependence}. }}
\label{phobos:v2vseta}
\end{center}
\end{figure}

In addition to questions about the thermalization
time, viscosity and freezeout
treatment, one also needs to address the robustness of the
standard assumption of longitudinal boost-invariance in
hydrodynamics calculations \cite{kolbheinz}. There is growing
evidence at RHIC for significant deviations from boost-invariance.
This is illustrated by PHOBOS results for $v_2$ as a function of
pseudorapidity in Fig.~\ref{phobos:v2vseta}, where one sees no
evidence for a mid-rapidity plateau in elliptic flow strength
\cite{phobos:boost-dependence}. Thus, while the successes of
QGP-based hydrodynamics calculations at RHIC are tantalizing,
substantially greater systematic investigation of their
sensitivities -- including computationally challenging full
three-dimensional treatments -- would be needed to make a
compelling QGP claim on their basis
alone.

\begin{figure} [bht]
\begin{center}
\includegraphics[width=0.80\textwidth]{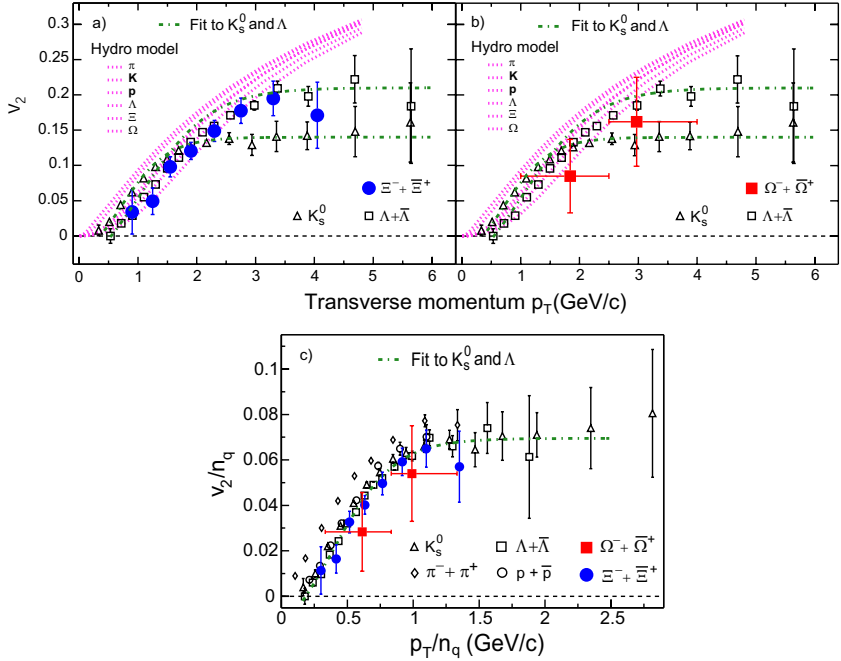}

\caption{{\it Experimental results on the transverse momentum
dependence of the event elliptic anisotropy parameter for various
hadron species produced in minimum-bias \auau\ collisions at
$\sqrt{s_{NN}}=200$ GeV.  STAR results \cite{starklv2} for $K_S^0$
and $\Lambda+\overline{\Lambda}$ are shown in all frames, together
with simple analytic function fits (dot-dashed lines) to these
data.  Additional data shown are STAR multi-strange baryon
elliptic flow \cite{star_javier} for $\Xi$ (in frames (a) and (c))
and $\Omega$ ((b) and (c)), and PHENIX results
\cite{phenix:pidflow200} for $\pi$ and $p+\overline{p}$ (frame
(c)). Hydrodynamics calculations are indicated by dotted curves in
frames (a) and (b). In (c), the flow results for all of the above
hadrons are combined by scaling both $v_2$ and $p_T$ by the number
of valence quarks ($n_q$) in each hadron. The figure is an update
of one in \cite{xin04}. }} \label{v21}
\end{center}
\end{figure}

At higher \pt values, as shown by experimental results from 200
GeV \auau minimum bias collisions in Fig.~\ref{v21}, the observed
values of $v_2$ saturate and the level of the saturation differs
substantially between mesons and baryons. Hydrodynamics
calculations overpredict the flow in this region. The dot-dashed
curves in Fig.~\ref{v21}(a)-(c) represent simple analytical
function fits to the measured $K_S^0$ and $\Lambda +
\bar{\Lambda}$ $v_2$ distributions \cite{starklv2,xin04}. It is
seen in Fig.~\ref{v21} (a) and (b) that STAR's most recent $v_2$
results for the multi-strange baryons $\Xi$ and $\Omega$
\cite{star_javier} are consistent with that of $\Lambda$'s, but
within still sizable statistical uncertainties.

%--=================================================================

%--===============================================================

In Fig.~\ref{v21} (c), particle-identified elliptic flow
measurements for the 200 GeV \auau minimum-bias sample are
combined by dividing both $v_2$ and $p_T$ by the number of valence
quarks ($n_q$) in the hadron of interest. The apparent scaling
behavior seen in this figure for $p_T/n_q > 1$ GeV/c is
intriguing, as the data themselves seem to be pointing to
constituent quarks (or at least to valence quarks sharing the full
hadron momentum, see Sec. 2.6) as the most effective degree of
freedom in determining hadron flow at intermediate $p_T$ values.
The data need to be improved in statistical precision and \pt
extent for more identified mesons and baryons in order to
establish this scaling more definitively. Within error bars the
size of those for $p_T/n_q > 1$ GeV/c, the low \pt data would also
look as though they scale with the number of constituent quarks,
whereas we already have seen in Fig.~\ref{v2low} that there is
rather a clear hydrodynamic mass-dependence in the low \pt region.
(Note that the pion data barely extend into the scaling region at
$p_T/n_q > 1$ GeV/c.)

\begin{figure} [bht]
\begin{center}
\includegraphics[width=0.9\textwidth]{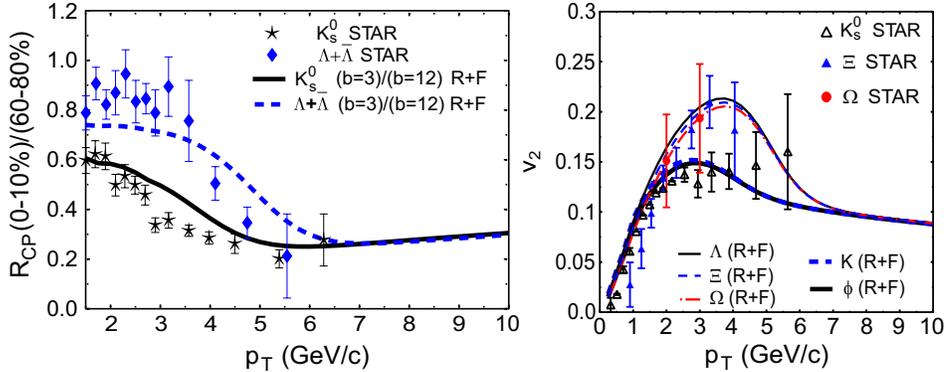}

\caption{{\it Comparisons of calculations in the Duke quark
recombination model \cite{recom03,Duke} with STAR measurements
\cite{starklv2,star_javier} of (a) $R_{CP}$ and (b) $v_2$ for
strange mesons and baryons. ``R+F'' denotes the sum of
recombination and fragmentation contributions.  Comparison of the
solid and broken curves in (b) reveals a weak mass-dependence in
the calculations, superimposed on the predominant meson-baryon
differences.  The figures are taken from Ref.~\cite{MuellerRBRC},
and they include preliminary STAR data for multi-strange baryons
that differ slightly from the values shown in Fig.~\ref{v21}. }}.
\label{recom-fits}
\end{center}
\end{figure}

If the scaling behavior at intermediate $p_T$ is confirmed with
improved data, it will provide a very important clue to the origin
of the meson-baryon differences (see also Fig.~\ref{rcp}) that
characterize this $p_T$ range.  In particular, both the $v_2$
scaling and the meson-baryon $R_{CP}$ differences can be explained
\cite{recom03,Duke} (see Fig.~\ref{recom-fits}) by assuming that
hadron formation at moderate $p_T$ proceeds via two competing
mechanisms: the coalescence of $n_q$ constituent quarks at
transverse momenta $\sim p_T/n_q$, drawn from a thermal
(exponential) spectrum \cite{recom03}, plus more traditional
fragmentation of hard-scattered partons giving rise to a power-law
component of the spectrum.  Note that, as discussed in
section~\ref{sec:theoryReco}, these models are not expected to
apply at low $p_T$. It is not yet clear that the same models could
simultaneously account as well for another observed feature
characteristic of this intermediate \pt range, namely, a jet-like
azimuthal correlation of hadron pairs that will be discussed
further in Sec. 4.

%--=================================================================

%--===============================================================

In these coalescence models, the constituent quarks carry their
own substantial azimuthal anisotropy, which is then summed to give
the hadron $v_2$. The establishment of clearer evidence for such
pre-hadronic collective flow would be an important milestone in
elucidating the nature of the matter produced in RHIC collisions.
In interpreting such evidence, it must be kept in mind that
constituent quarks are not partons: they are
effective degrees of freedom
normally associated with chiral symmetry breaking and confinement,
rather than with the deconfinement of a QGP.
Until the mechanism for the
emergence of these effective degrees of freedom from a QCD plasma
of current quarks and gluons is clarified (see Sec. 2.6),
collectively flowing constituent quarks
should not be taken as definitive
proof of a QGP stage, as we have defined it in Sec. 1. It is
unclear, for example,
whether the characteristic time
scale for constituent quarks to coalesce from current quarks and
gluons might not be shorter than
that for the establishment of thermalization in the collision
(leading to a sort of
``constituent quark plasma", as
opposed to a QGP).  Furthermore, the constituent quark $v_2$
values needed to account for the observed hadron $v_2$ saturation
might arise in part from differential energy loss of their
progenitor partons in traversing the spatially anisotropic matter
of non-central collisions \cite{GLV}, rather than strictly from
the partonic hydrodynamic flow assumed in \cite{Duke}. The
unanticipated RHIC results in this intermediate \pt range thus
raise a number of important and
fascinating questions that should be addressed further by future
measurements and theoretical calculations.

In summary, the measured yields with respect to the reaction plane
are among the most important
results to date from RHIC: they provide critical hints of the
properties of the bulk matter at early stages. They indicate that
it behaves collectively, and is consistent with rapid
(\emph{i.e.}, very short mean free path) attainment of
at least approximate local
thermal equilibrium in a QGP phase. Hydrodynamic accounts for the
mass- and $p_T$-dependence of $v_2$ for soft hadrons appear to
favor system evolution through a soft, mixed-phase EOS. The
saturated $v_2$ values observed for identified mesons and baryons
in the range $1.5 \lesssim p_T \lesssim 6$ GeV/c suggest that
hadronization in this region occurs largely via coalescence of
collectively flowing constituent quarks. What has yet to be
demonstrated is that these interpretations are unique and robust
against improvements to both the measurements and the theory.
In particular, it must be
demonstrated more clearly that the sensitivity to the role of the
QGP outweighs that to other, more mundane, ambiguities in the
theoretical treatment.

\subsection{Quantum Correlation Analyses}

Two-hadron correlation measurements in principle should provide
valuable information on the phase structure of the system at
freezeout. From the experimentally measured momentum-space
two-particle correlation functions, a Fourier transformation is
then performed in order to extract information on the space-time
structure \cite{hbt0}. Bertsch-Pratt parameterization \cite{pratt}
is often used to decompose total momentum in such measurements
into components parallel to the beam (\emph{long}), parallel to
the pair transverse component (\emph{out}) and along the remaining
third direction (\emph{side}). In this Cartesian system,
information on the source duration time is mixed into the
\emph{out} components. Hence, the ratio of inferred emitting
source radii $R_{out}/R_{side}$ is sensitive to the time duration
of the source emission. For example, if a QGP is formed in
collisions at RHIC, a long duration time and consequently large
value of $R_{out}/R_{side}$ are anticipated \cite{rischke96}.

\begin{figure} [thb]
\begin{center}
%\hspace{-5.5cm}
\includegraphics[width=0.8\textwidth]{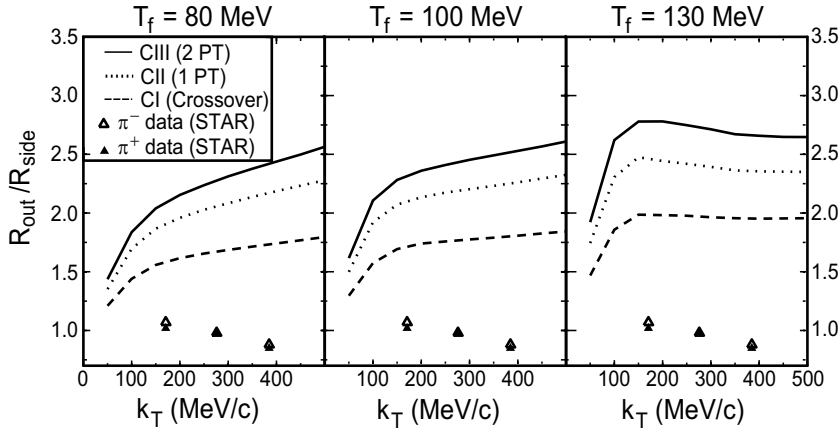}
\caption{{\it STAR measurements \cite{rhic130hbt} of
$R_{out}/R_{side}$ from pion HBT correlations for central \auau
collisions, plotted as a function of the pion pair transverse
momentum $k_T$. The experimental results are identical in the
three frames, but are compared to hydrodynamics calculations
\cite{soff01} performed for a variety of parameter values.}}
\label{hbt1}
\end{center}
\end{figure}

\begin{figure} [bht]
\begin{center}
\includegraphics[width=0.6\textwidth]{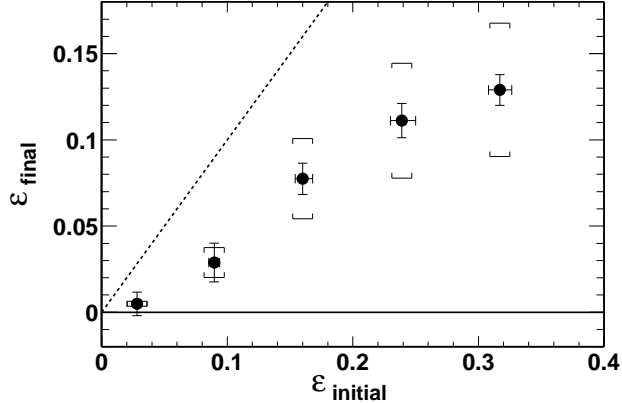}
\caption{{\it The eccentricity $\epsilon_{final}$ of the
time-integrated emitting source of soft pions, inferred from STAR
HBT correlations measured with respect to the reaction plane,
plotted versus the initial spatial eccentricity
$\epsilon_{initial}$ deduced from a Glauber calculation for five
different \auau centrality bins. The dotted line represents
$\epsilon_{final} = \epsilon_{initial}$.  See \cite{starhbt2} for
details.}} \label{hbt2}
\end{center}
\end{figure}

Measured results for Hanbury-Brown-Twiss (HBT) pion
interferometry, exploiting the boson symmetry of the two detected
particles at low relative momenta, are shown in Figs.~\ref{hbt1}
and \ref{hbt2}. A clear dependence of the `size' parameters on the
pair transverse momentum $k_T$ is characteristic of collective
expansion of the source \cite{rhic130hbt,starhbt2}, so the results
are plotted vs. $k_T$ in Fig.~\ref{hbt1}.  As indicated by the set
of curves in the figure, hydrodynamics calculations that can
account for hadron spectra and elliptic flow at RHIC
systematically over-predict $R_{out}/R_{side}$
\cite{rhic130hbt,soff01}.  One possible implication of this
discrepancy is that the collective expansion does not last as long
in reality as in the hydrodynamics accounts. However, shorter
expansion times are difficult to reconcile with the observed
magnitude of $R_{side}$, and are not supported by a recent
systematic study of HBT correlations relative to the
event-by-event reaction plane \cite{starhbt2}. The source
eccentricity at freezeout inferred from these azimuthally
sensitive measurements is shown in Fig.~\ref{hbt2} to retain a
significant fraction of the initial spatial eccentricity
characteristic of the impact parameter for each centrality bin.
The observed eccentricity retention is, in fact, quantitatively
consistent with hydrodynamics expectations for the
\emph{time-integrated} pion emission surface to which HBT is
sensitive \cite{heinz04}. Thus, the deformations in
Fig.~\ref{hbt2} tend to support the hydrodynamics view of the
expansion pressure and timeline (see Fig.~\ref{hydro-anisotropy}),
which lead to an eventual complete quenching of the initial
configuration-space anisotropy by the end of the freezeout
process.

The failure of the hydrodynamics calculations to account for the
HBT results in Fig.~\ref{hbt1}
raises another significant issue regarding the robustness of the
hydrodynamics success in reproducing $v_2$ and radial flow data.
Although the HBT interference only emerges after the freezeout of
the strong interaction, whose treatment is beyond the scope of
hydrodynamics, the measured correlation functions
receive contributions from all
times during the collision process. Furthermore, these HBT
results are extracted from the low \pt region, where soft bulk
production dominates.  It is thus reasonable to expect the correct
hydrodynamics account of the collective expansion to be consistent
with the HBT source sizes. If
improved treatment of the hadronic stage and/or the introduction
of finite viscosity during the hydrodynamic expansion
\cite{viscous} are necessary to attain this consistency, then it
is important to see how those improvements affect the agreement
with elliptic flow and spectra.

%--=================================================================

%--======================

STAR has also measured two-hadron momentum-space correlation
functions for \emph{non}-identical particles \cite{non-ident}.
These are sensitive to differences in the average emission time
and position for the different particle species.  Such differences
are very clearly revealed by the measured correlations between pions and
kaons \cite{non-ident}, and provide additional strong evidence for
a collective transverse flow of the produced hadrons.

%--===============================================================
\subsection{Correlations and fluctuations}

A system evolving near a phase boundary should develop significant
dynamical fluctuations away from the mean thermodynamic properties
of the matter. For high-energy heavy ion collisions, it has been
predicted that the general study of two-particle correlations and
event-wise fluctuations might provide evidence for the formation
of matter with partonic degrees of freedom
\cite{stephanov98,voloshin99,jeon00,asakawa00,bass00,liu03}. In
addition, nonstatistical correlations and fluctuations may be
introduced by incomplete equilibrium \cite{gazdzicki99}. With its
large acceptance and complete event-by-event reconstruction
capabilities, the STAR detector holds great potential for
fluctuation analyses of RHIC collisions.

An approach that has been used previously \cite{CERES03,PHENIX04}
to search for the presence of dynamical correlations involves
extraction of measures of the excess variance of some observable
above the statistical fluctuations that show up even in
mixed-event samples. An example shown in Fig. 24 utilizes the
square root of the covariance in \pT\  for charged-particle pairs
from collisions at SPS (CERES \cite{CERES03}) and at STAR
\cite{gary200qm04}. The presence of dynamical 2-particle
correlations is revealed by non-zero values of this quantity,
whose gross features exhibit a magnitude and a smooth
centrality-dependence that are essentially independent of
collision energy, once the variations of the inclusive mean \pT\
values ($<<\pT>>$) with centrality and energy have been divided
out.  However, the detailed nature of the dynamical correlations
is best probed by fully exploiting the impressive statistical
precision at STAR to investigate finer, multi-dimensional aspects
of the correlation densities themselves, rather than of the
integrals of correlation densities represented by excess variance
measures.

\begin{figure}[thb]
\begin{center}
\epsfig{figure=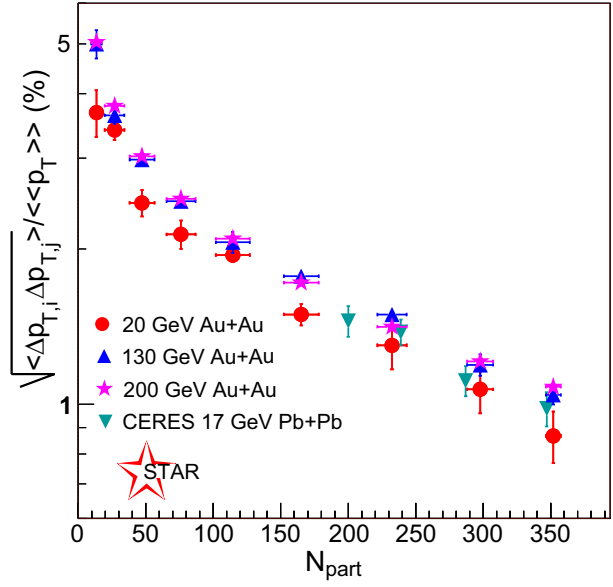,width=8.0cm}
\caption{ {\it The square root of the transverse momentum
covariance for charged particle pairs, scaled by the inclusive
mean \pT\ value for each centrality and collision energy, plotted
vs. centrality for SPS \cite{CERES03} and RHIC \cite{gary200qm04}
data at several energies.  Both the centrality-dependence (nearly
$\propto$ 1/$N_{part}$) and magnitude of this quantity are
essentially unchanged from SPS to RHIC energies, but its implicit
integration of correlation densities over the full detector
acceptance masks other interesting correlation features.} }
\label{meanpt_corr}
\end{center}
\end{figure}

\begin{figure} [bht]
\begin{center}
\includegraphics[width=0.4\textwidth]{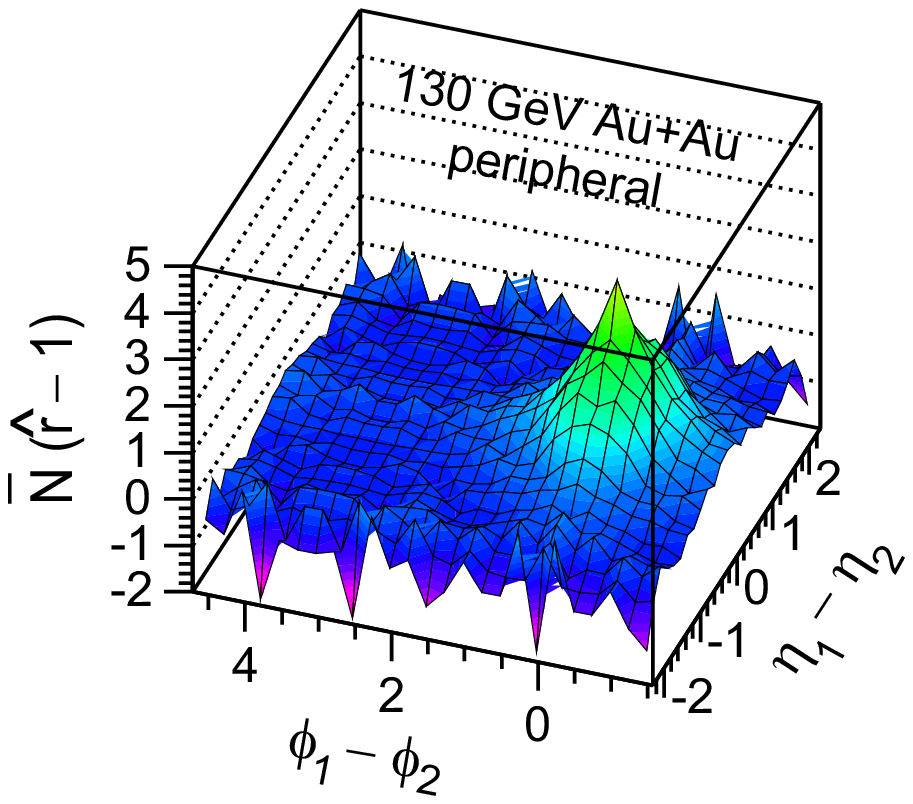}
\includegraphics[width=0.4\textwidth]{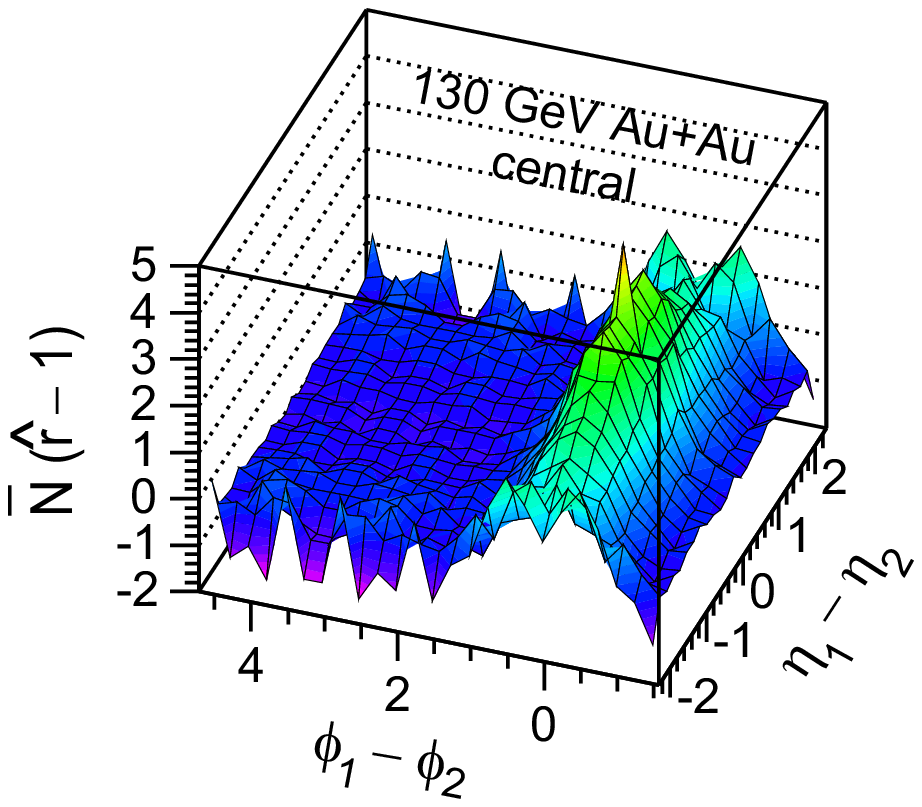}
\includegraphics[width=0.38\textwidth]{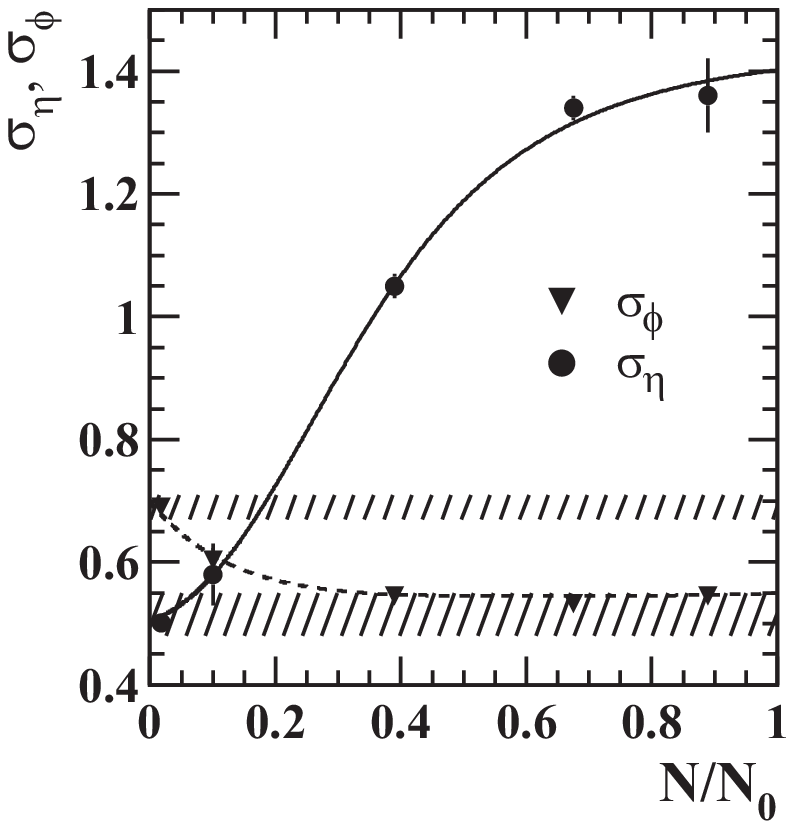}
\includegraphics[width=0.42\textwidth]{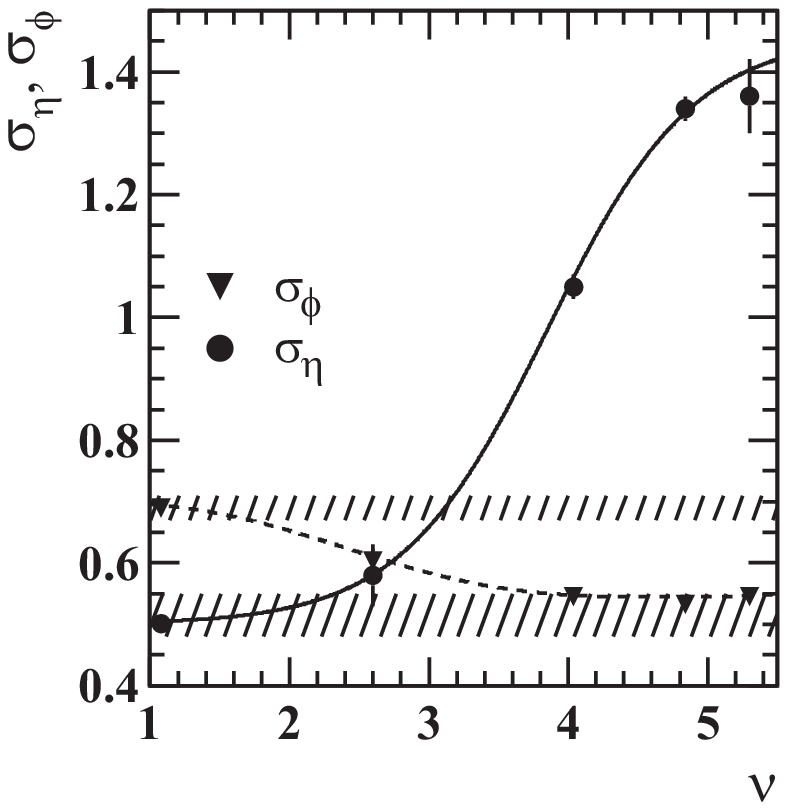}
\caption{{\it Upper frames: joint autocorrelations measured by
STAR, as a function of the hadron pair angle differences $\eta_1 -
\eta_2$ and $\phi_1 - \phi_2$, for 0.15 $\leq$ \pT\ $\leq$ 2.0
GeV/c charged hadrons detected in 130 GeV Au+Au collisions
\cite{trainor-jamaica}. The right frame contains data for central
collisions, while the left frame spans a broad range of
centralities for more peripheral collisions. The quantity
$\overline{N}(\hat{r}-1)$ plotted on the vertical axes represents
the average multiplicity for the selected centrality bin
multiplied by the relative difference in charged particle pair
yields between same events and mixed events. Elliptic flow and
momentum conservation long-range correlations have been
subtracted, as explained in \cite{trainor-jamaica}. Lower frames:
centrality-dependence of the Au+Au pseudorapidity and azimuthal
widths from two-dimensional gaussian fits to the near-side
correlation structure seen for two centrality bins in the upper
frames. The same extracted widths for Au+Au collisions are plotted
\emph{vs.} two different measures of centrality: the observed
charged-particle multiplicity divided by its maximum value (on the
left), and the mean number $\nu$ of nucleons encountered by a
typical participant nucleon (right). The hatched bands indicate
the widths observed for p+p collisions, and the curves guide the
eye. }} \label{mini-jets}
\end{center}
\end{figure}

For example, emerging STAR angular correlation results are already
suggesting that there is appreciable soft hadron emission before
the attainment of local thermal equilibrium, even in the most
central RHIC collisions. The evidence resides in remnants of
jet-like behavior observed \cite{trainor-jamaica} even in soft
($0.15 \leq p_T \leq 2.0$ GeV/c) hadron pair correlations on the
angular difference variables $\Delta\eta \equiv \eta_1 - \eta_2$
(pseudorapidity) and $\Delta\phi \equiv \phi_1 - \phi_2$
(azimuthal angle), presented for peripheral and central Au+Au
collisions in Fig.~\ref{mini-jets}. The equivalent correlations
for p+p collisions at RHIC \cite{trainor-jamaica} emphasize the
central role of parton fragmentation, even down to hadron
transverse momenta of 0.5 GeV/c, resulting in a prominent
near-side jet peak symmetric about $\Delta\eta = \Delta\phi = 0$
and a broad $\Delta\eta$-independent away-side ($\Delta\phi =
\pi$) jet ridge. One certainly anticipates some remnants of these
correlations to survive in heavy-ion collisions at sufficiently
high hadron \pt, and these correlations will be discussed in the
next chapter.  In the soft sector, however, attainment of a fully
equilibrated state of all emerging hadrons at freezeout would wash
out such initial hard-scattering dynamical correlations.

Instead, the observed soft-hadron-pair correlation for central
Au+Au collisions shown in the upper right-hand frame of
Fig.~\ref{mini-jets}, after removal of multipole components
representing elliptic flow ($v_2$) and momentum conservation
($v_1$) \cite{trainor-jamaica}, exhibits a substantially modified
remnant of the jet correlation on the near side, affecting
typically 10-30\% of the detected hadrons. Contributions to this
near-side peak from HBT correlations and Coulomb final-state
interactions between hadrons have been suppressed by cuts to
remove pairs at very low relative momentum, reducing the overall
strength of the correlation near $\Delta \eta = \Delta \phi = 0$
by $\sim$10\%~\cite{trainor-jamaica}. Simulations demonstrate that
resonance decays make no more than a few percent contribution to
the remaining near-side correlation strength. (The lack of
evidence for any remaining away-side correlation in central
collisions will be discussed further in the high-\pt context in
Sec. 4.  Its absence even for more peripheral collisions in the
upper left frame of Fig.~\ref{mini-jets} can be attributed to the
broad centrality bin used here to compensate for limited
statistics in the 130 GeV Au+Au data sample, and to the $v_1$
subtraction that removes thermalized soft hadrons balancing the
near-side jet's momentum.)

%--=================================================================

%--======================

The observed near-side correlation in central Au+Au is clearly
much broader in $\Delta\eta$ than that in p+p or peripheral Au+Au
collisions.  The pseudorapidity spread, as characterized by the
$\Delta\eta$ width of a two-dimensional gaussian function fitted
to the structure (ignoring the $\Delta\eta = \Delta\phi = 0$ bin,
where conversion electron pairs contribute), grows rapidly with
increasing collision centrality, as revealed in the lower frames
of Fig.~\ref{mini-jets}. This trend suggests that while some
parton fragments are not yet fully equilibrated in the soft
sector, they are nonetheless rather strongly coupled to the
longitudinally expanding bulk medium. The onset of this coupling
appears especially dramatic when the results are plotted (lower
right-hand frame) \emph{vs.} the alternative centrality measure
$\nu \equiv (N_{part}/2)^{1/3}$ (estimating the mean number of
nucleons encountered by a typical participant nucleon along its
path through the other nucleus), rather than the more traditional
charged-particle multiplicity (lower left frame).  The latter
comparison serves as a reminder that, as we seek evidence for a
transition in the nature of the matter produced in RHIC
collisions, it is important to consider carefully the optimal
variables to use to characterize the system.

\begin{figure} [bht]
\begin{center}
\includegraphics[width=0.8\textwidth]{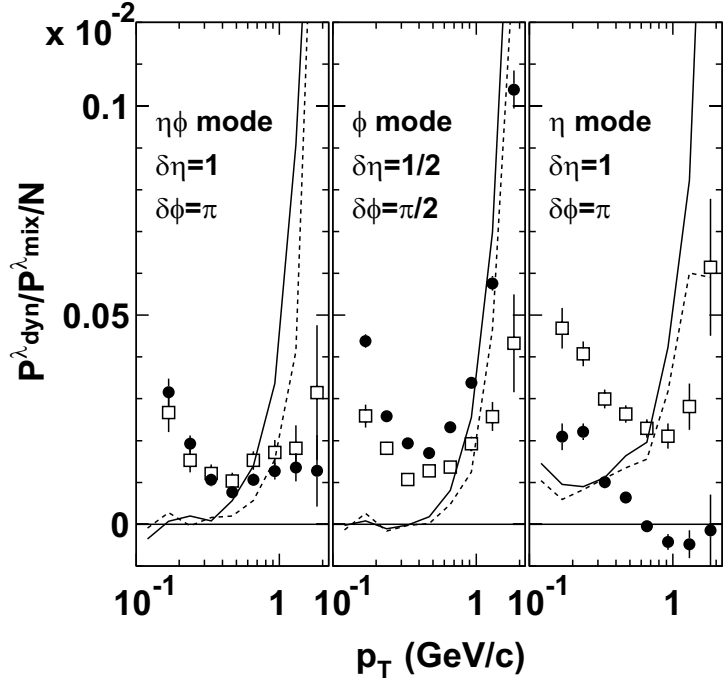}
\caption{{\it STAR measurements (filled circles) of dynamic
texture for the 4\% most central Au+Au collisions at \sqrtsNN =
200 GeV, compared to STAR peripheral (60-84\%) collision data
(boxes) renormalized for direct comparison, and to HIJING
calculations with (dashed curves) and without (solid curves)
inclusion of jet quenching.  The dynamic texture measures the
non-statistical excess in point-to-point fluctuations in the local
density of charged hadrons in an event, averaged over the event
ensemble.  Figure taken from Ref.~\cite{dynamic-texture}. }}
\label{DWT_central}
\end{center}
\end{figure}

The coupling to the longitudinal expansion can be seen more
clearly as an equilibration mechanism from measurements of the
power spectra $P^\lambda$ of local fluctuations in the density of charged
hadrons with respect to a mixed-event counterpart $P^\lambda_{mix}$.
The $\lambda$ superscript distinguishes different directions (modes)
of density variation, orthogonal in the wavelet decomposition used~\cite{dynamic-texture} for
the analysis: along $\eta$, along $\phi$, and along the diagonal $\eta\phi$.
The so-called
``dynamic texture" \cite{NA44_DWT} of the event, used to characterize
the non-statistical excess in density fluctuations, is defined as $P^\lambda_{dyn}/P^\lambda_{mix}/N$, where
$P^\lambda_{dyn} = P^\lambda - P^\lambda_{mix}$ and $N$ is the average number
of tracks in a given $p_T$ interval per event.  The dynamic texture is shown as a function
of \pt for three different modes and for both central and peripheral collisions in
Fig.~\ref{DWT_central} \cite{dynamic-texture}. HIJING simulations
\cite{HIJING} shown in the figure cannot account for the observed
fluctuations, even when jet quenching is included, although they
do suggest qualitatively that the rising trends in the data with
increasing \pt are signals of ``clumpiness" in the particle
density caused by jets. In the absence of a successful model for
these fluctuations, we can at least search for interesting
centrality dependences. The box symbols in the figure represent
what we would expect for the dynamic texture in central
collisions, based on what STAR measures for peripheral collisions,
if the correlation structure were independent of centrality.  The
strong suppression observed with respect to this expectation for
the central collision $\eta$-mode fluctuations is interpreted as
another manifestation of the coupling of parton fragments to the
longitudinally expanding bulk medium \cite{dynamic-texture}.

%--=================================================================

%--======================

\begin{figure} [thb]
\begin{center}
\includegraphics[width=0.6\textwidth]{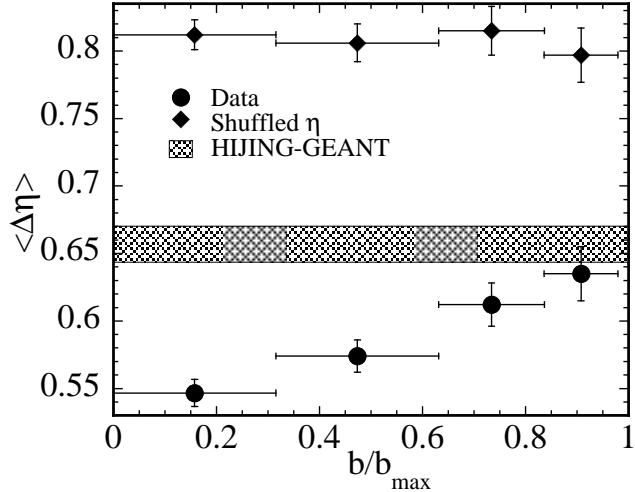}
\caption{{\it Width $\langle \Delta \eta \rangle$ of the measured
charged hadron balance function from Au+Au (filled circles)
collisions at \sqrtsNN=130 GeV, plotted as a function of the
relative impact parameter deduced from the charged particle
multiplicities for each analyzed centrality bin. The cross-hatched
band represents the centrality-independent results of HIJING
simulations \cite{HIJING} of the balance function width measured
within STAR for 130 GeV Au+Au collisions. The diamond-shaped
points illustrate constraints imposed on the balance function by
global charge conservation and the STAR detector acceptance, when
dynamical correlations are removed by randomly shuffling the
association of pseudorapidities with detected particles within
each analyzed event.  The figure is from Ref.~\cite{starbf130}. }}
\label{bf-width}
\end{center}
\end{figure}

The results in Figs.~\ref{mini-jets} and \ref{DWT_central} are
averaged over all charged hadrons without consideration of the
sign of the charge.  Detailed information on the hadronization of
the medium can be obtained from the study of
charge-\emph{dependent} (CD) correlations, \emph{e.g.}, by
examining the difference between angle-dependent correlations of
like- \emph{vs}. unlike-sign pairs.  One method focuses on the
``balance function", constructed \cite{bass00,starbf130} to
measure the excess of unlike- over like-sign pairs as a function
of their (pseudo)rapidity difference $\Delta\eta$.  The results in
Fig.~\ref{bf-width} show that the width of this function in
$\Delta\eta$ steadily decreases with increasing Au+Au centrality
\cite{gary200qm04,starbf130}, in contrast to HIJING simulations
\cite{HIJING}. A related trend is observed in the CD
two-dimensional autocorrelation \cite{trainor-jamaica} (not
plotted) analogous in format to the charge-independent results
shown in Fig.~\ref{mini-jets}.  The CD peak amplitude increases,
and its width decreases, dramatically with increasing centrality.
These trends indicate a marked change in the formation mechanism
of charged hadron pairs in central Au+Au, relative to p+p,
collisions.  The implications of that change for the nature of the
medium produced are now under intensive study with a growing array
of correlation techniques.

%--=================================================================

%--=================================================================
\subsection{Summary}

In this chapter, we have presented important results on the bulk
matter properties attained in
\auau\ collisions at RHIC.  The measured hadron spectra, yield
ratios, and low \pt $v_2$, are all consistent from all experiments
at RHIC. STAR, in
particular, has made pioneering measurements of elliptic flow, of
multi-strange baryons, and of dynamical hadron correlations that
bear directly on the matter properties critical to establishing
QGP formation.  The yield ratios are consistent with chemical equilibration
across the $u, ~d$ and $s$ sectors.  The spectra and $v_2$
clearly reveal a collective velocity field in
such collisions.  The combined evidence for near-central Au+Au
collisions at RHIC suggests that thermal equilibrium is largely,
though not quite completely, attained, and that collective flow is
established, at an early collision stage when sub-hadronic degrees
of freedom dominate the matter. However, the quality of some of
the data, and the constraints on ambiguities in some of the
theoretical models used for interpretation, are not yet sufficient
to demonstrate convincingly that thermalized, deconfined matter
has been produced.

In particular, the unprecedented
success of hydrodynamics in providing a reasonable quantitative
account for collective flow at RHIC, and of the statistical model
in reproducing hadron yields through the strange sector, together
argue for an early approach toward thermalization spanning the
$u, ~d$ and $s$ sectors.  On the
other hand, measurements of angle difference distributions for
soft hadron pairs reveal that some
(admittedly heavily modified)
remnants of jet-like dynamical correlations survive the
thermalization process, and
indicate its incompleteness.  The
fitted parameters of the statistical model analyses, combined with
inferences from the produced transverse energy per unit rapidity,
suggest attainment of temperatures and energy densities at least
comparable to the critical values for QGP formation in LQCD
calculations of bulk, static strongly interacting matter.

The data in this chapter provide
two hints of deconfinement that need to be sharpened in future
work. One is the improvement in hydrodynamics accounts for
measured low-$p_T$ flow when the calculations include a soft point
in the EOS, suggestive of a transition from partonic to hadronic
matter. It needs to be better demonstrated that comparable
improvement could not be obtained alternatively by addressing
other ambiguities in the hydrodynamics treatment.  One indication
of such other ambiguities is the failure of hydrodynamics
calculations to explain the emitting source sizes inferred from
pion interferometry. The second hint is the apparent relevance of
(constituent or valence) quark degrees of freedom in determining
the observed meson-baryon differences in flow and yield in the
intermediate-$p_T$ region. Here the data need improved precision
to establish more clearly the quark scaling behavior expected from
coalescence models, while the theory needs to establish a clearer
connection between the effective quarks that seem to coalesce and
the current quarks and gluons of QCD.

%--=================================================================

\newpage
\section{Hard Probes}

Due to the transient nature of the matter created in high energy
nuclear collisions, external probes cannot be used to study its
properties. However, the dynamical processes that produce the bulk
medium also produce energetic particles through hard scattering
processes. The interaction of these energetic particles with the
medium provides a class of unique, penetrating probes that are
analogous to the method of computed tomography (CT) in medical
science.

For $p_T \gtrsim 5$ GeV/c the observed hadron spectra in Au+Au
collisions at RHIC exhibit the power-law falloff in cross section
with increasing $p_T$ that is characteristic of perturbative QCD
hard-scattering processes \cite{star:highpt130}.  The parameters
of this power-law behavior vary systematically with collision
centrality, in ways that reveal important properties of the matter
traversed by these penetrating probes \cite{star:highpt130}. While
we focus for the most part in this chapter on hadrons of $p_T$
above 5 GeV/c, we do also consider data in the intermediate $p_T$
range down to 2 GeV/c, when those data allow more statistically
robust measurements of effects we associate with hard scattering.

\subsection{Inclusive hadron yields at high $p_T$}

\begin{figure}
\begin{center}
\includegraphics[width=.43\textwidth]{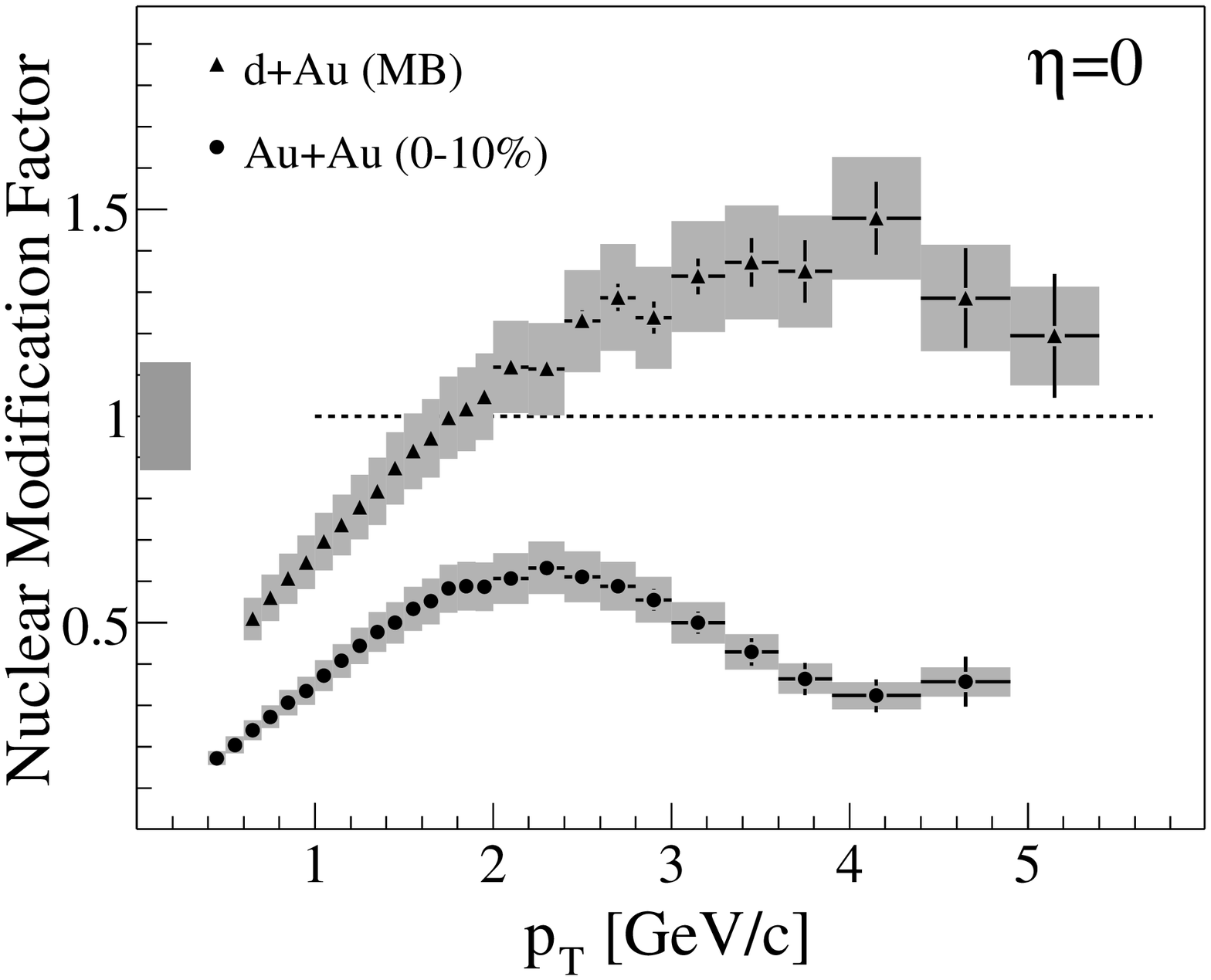}
\includegraphics[width=.52\textwidth]{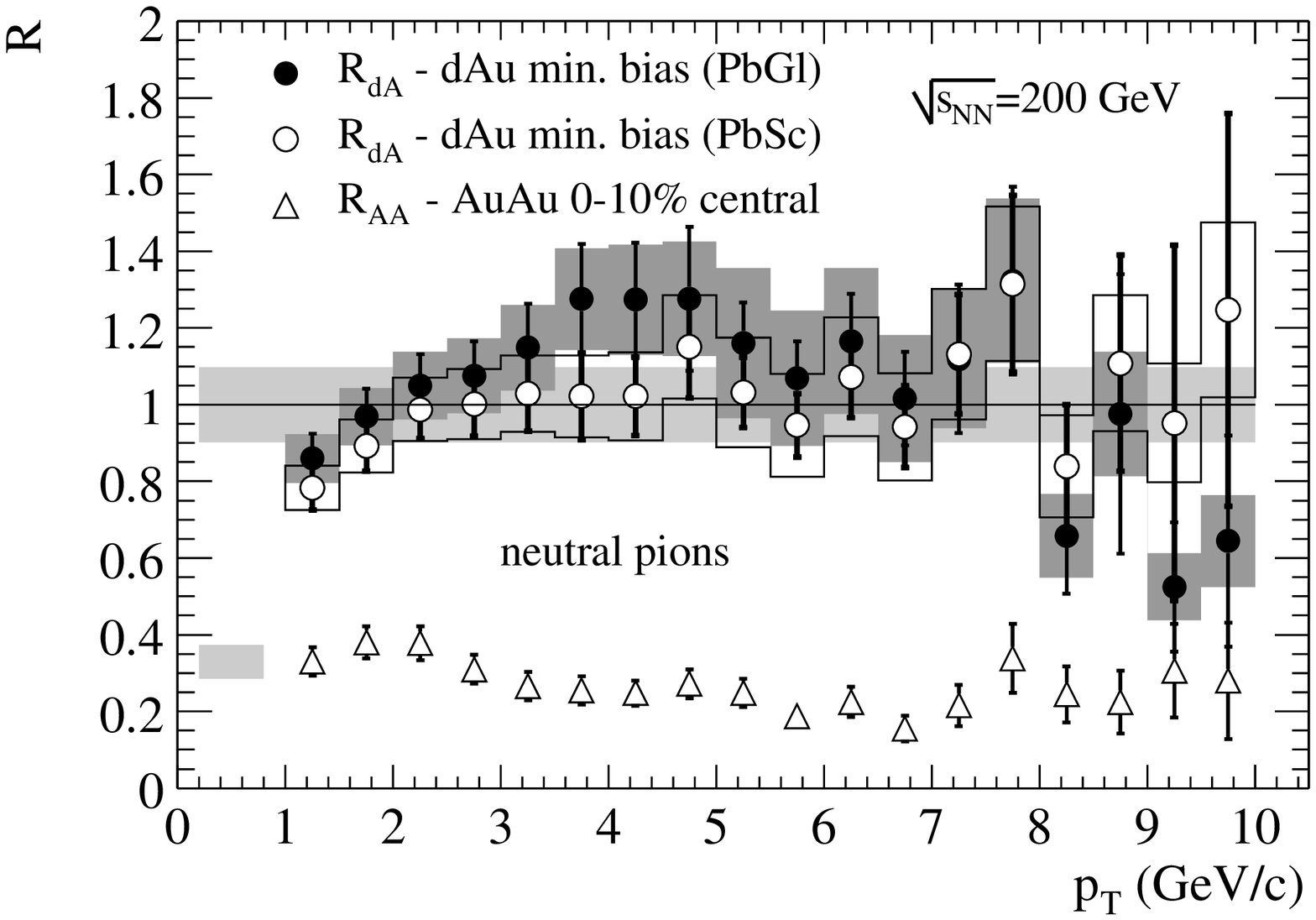}
\includegraphics[width=.43\textwidth]{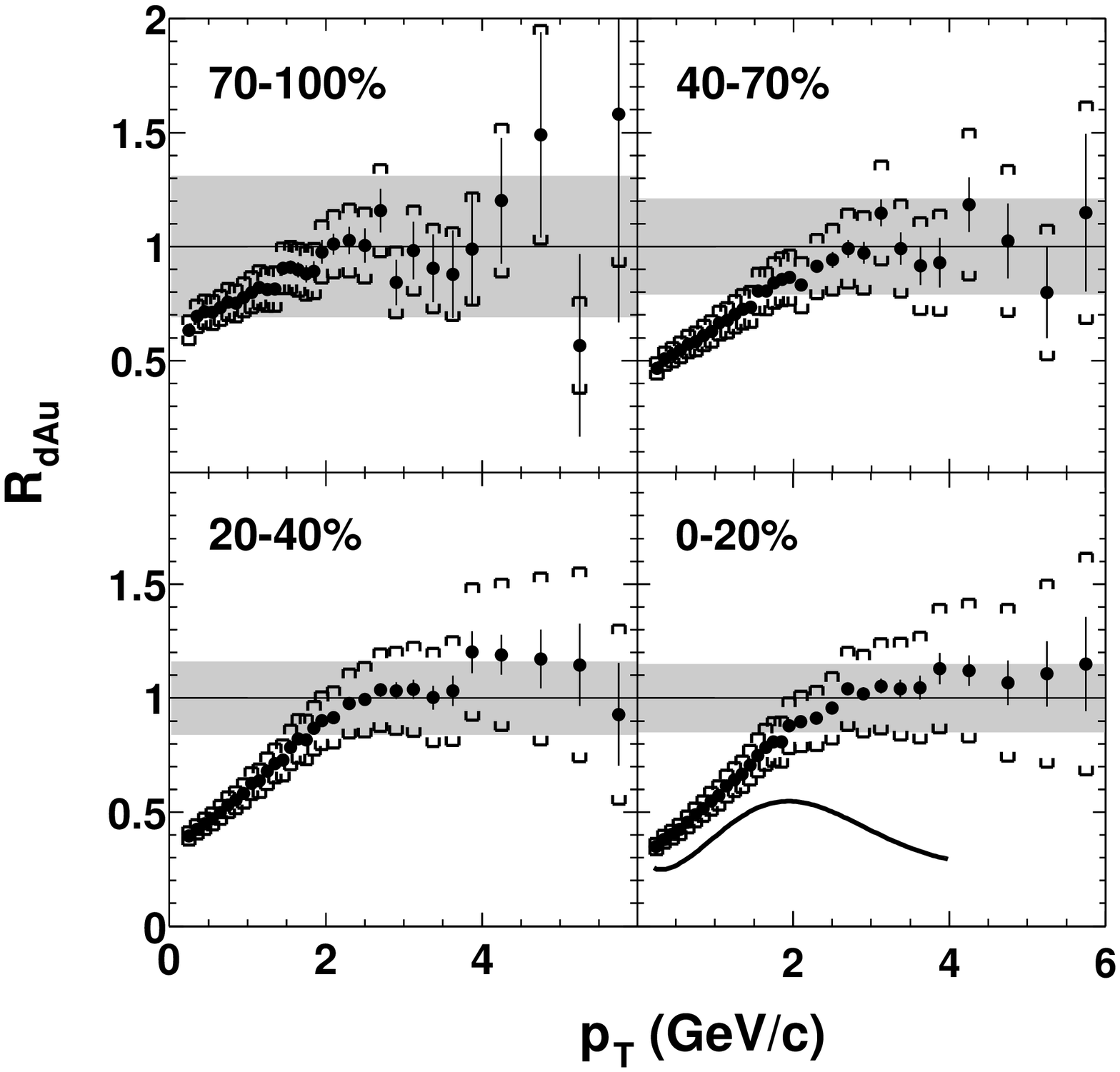}
\includegraphics[width=.52\textwidth]{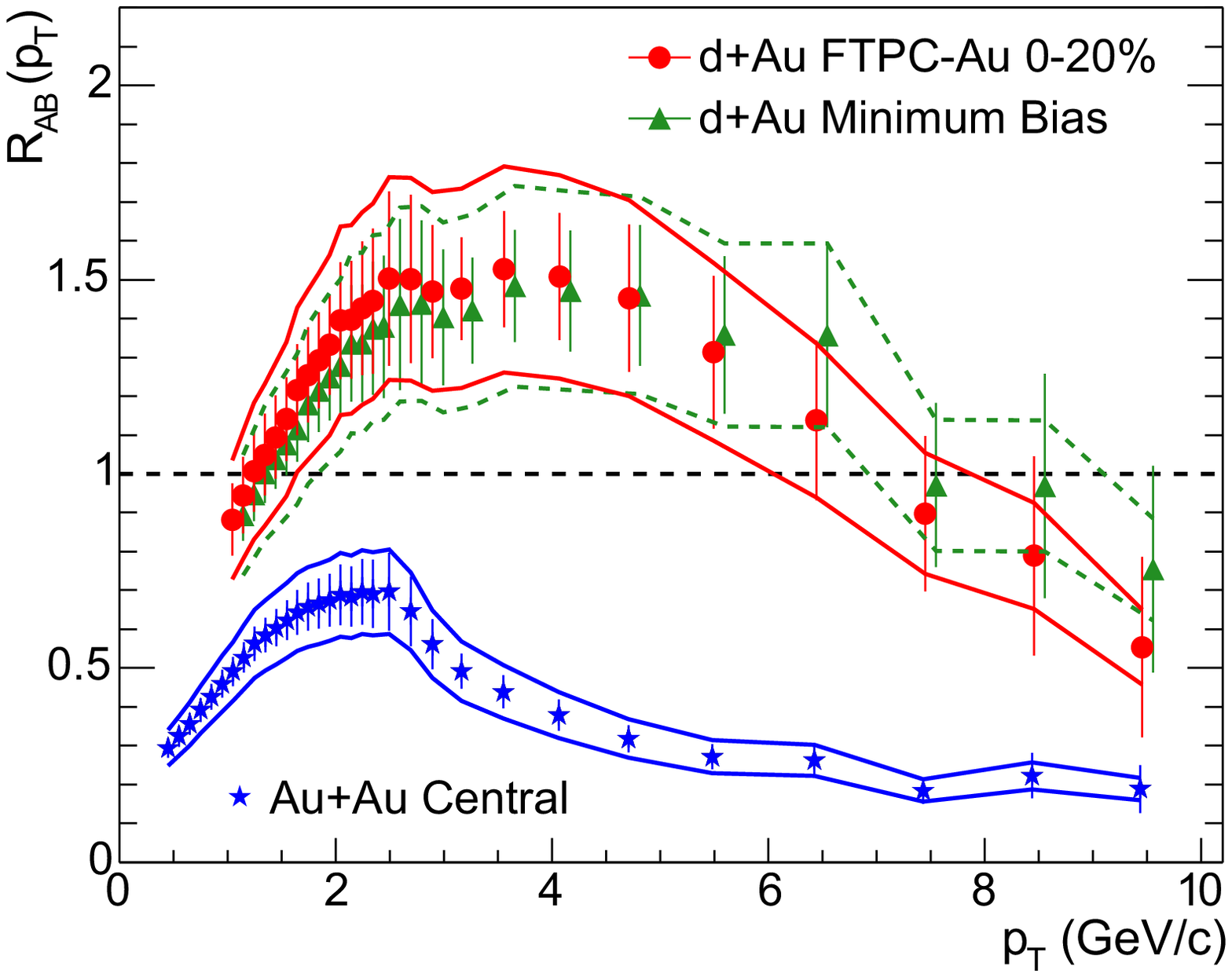}
\caption{ {\it Binary-scaled ratio \RAB\ (Eq. \ref{eq:RAB}) of
charged hadron and \pizero\ inclusive yields from 200 GeV Au+Au
and d+Au relative to that from p+p collisions, from
BRAHMS\cite{brahms:highpTdAu} (upper left),
PHENIX\cite{phenix:highpTdAu} (upper right),
PHOBOS\cite{phobos:highpTdAu} (lower left) and
STAR\cite{star:highpTdAu} (lower right). The PHOBOS data points in
the lower left frame are for d+Au,
while the solid curve represents PHOBOS central (0-6\%) Au+Au
data. The shaded horizontal bands around unity represent the
systematic uncertainties in the binary scaling corrections. } }
\label{fig:InclusiveSuppressionPR}
\end{center}
\end{figure}

There are several results to date from RHIC exhibiting large and
striking effects of the traversed matter on hard probes in central
collisions.  Figures \ref{fig:InclusiveSuppressionPR} and
\ref{fig:CorrelationsPR} show the most significant high \pT\
measurements made at RHIC thus far. Both figures incorporate
measurements of \sqrtsNN =200 GeV p+p, d+Au and
centrality-selected Au+Au collisions at RHIC, with the simpler p+p
and d+Au systems providing benchmarks for phenomena seen in the
more complex Au+Au collisions.

Figure \ref{fig:InclusiveSuppressionPR} shows \RAB, the ratio of
inclusive charged hadron yields in A+B (either Au+Au or d+Au)
collisions to p+p, corrected for trivial geometric effects via
scaling by \NbinaryMean\ , the calculated mean number of binary
nucleon-nucleon collisions contributing to each A+B centrality
bin:

\begin{equation}
\RAB=\frac{dN_{AB}/d\eta d^2\pT} {T_{AB} d\sigma_{NN}/d\eta
d^2\pT}. \label{eq:RAB}
\end{equation}

where the overlap integral $T_{AB} =
\NbinaryMean/\sigma_{inelastic}^{pp}$. A striking phenomenon is
seen: large $p_T$ hadrons in central Au+Au collisions are
suppressed by a factor $\approx 5$ relative to naive (binary
scaling) expectations. Conventional nuclear effects, such as
nuclear shadowing of the parton distribution functions and initial
state multiple scattering, cannot account for the suppression.
Furthermore, the suppression is not seen in d+Au but is unique to
Au+Au collisions, proving experimentally that it results not from
nuclear effects in the initial state (such as gluon saturation),
but rather from the final state interaction (FSI) of hard
scattered partons or their fragmentation products in the dense
medium generated in Au+Au collisions
\cite{brahms:highpTdAu,phenix:highpTdAu,phobos:highpTdAu,star:highpTdAu}.

These dominant FSI in Au+Au are presumably superimposed on a
variety of interesting initial-state effects revealed by the d+Au
results.  The enhancement seen in
Fig.~\ref{fig:InclusiveSuppressionPR} in $R_{dAu}$ for moderate
$p_T$ and mid-rapidity, known as the Cronin effect \cite{Cronin},
is generally attributed \cite{mult-scat} to the influence of
multiple parton scattering through cold nuclear matter \emph{prior
to} the hard scattering that produces the observed high-$p_T$
hadron. Other effects, revealed by the strong
\emph{rapidity}-dependence of $R_{dAu}$, will be discussed in Sec.
4.4.

\subsection{Dihadron azimuthal correlations}

\begin{figure}
\begin{center}
\includegraphics[width=.49\textwidth]{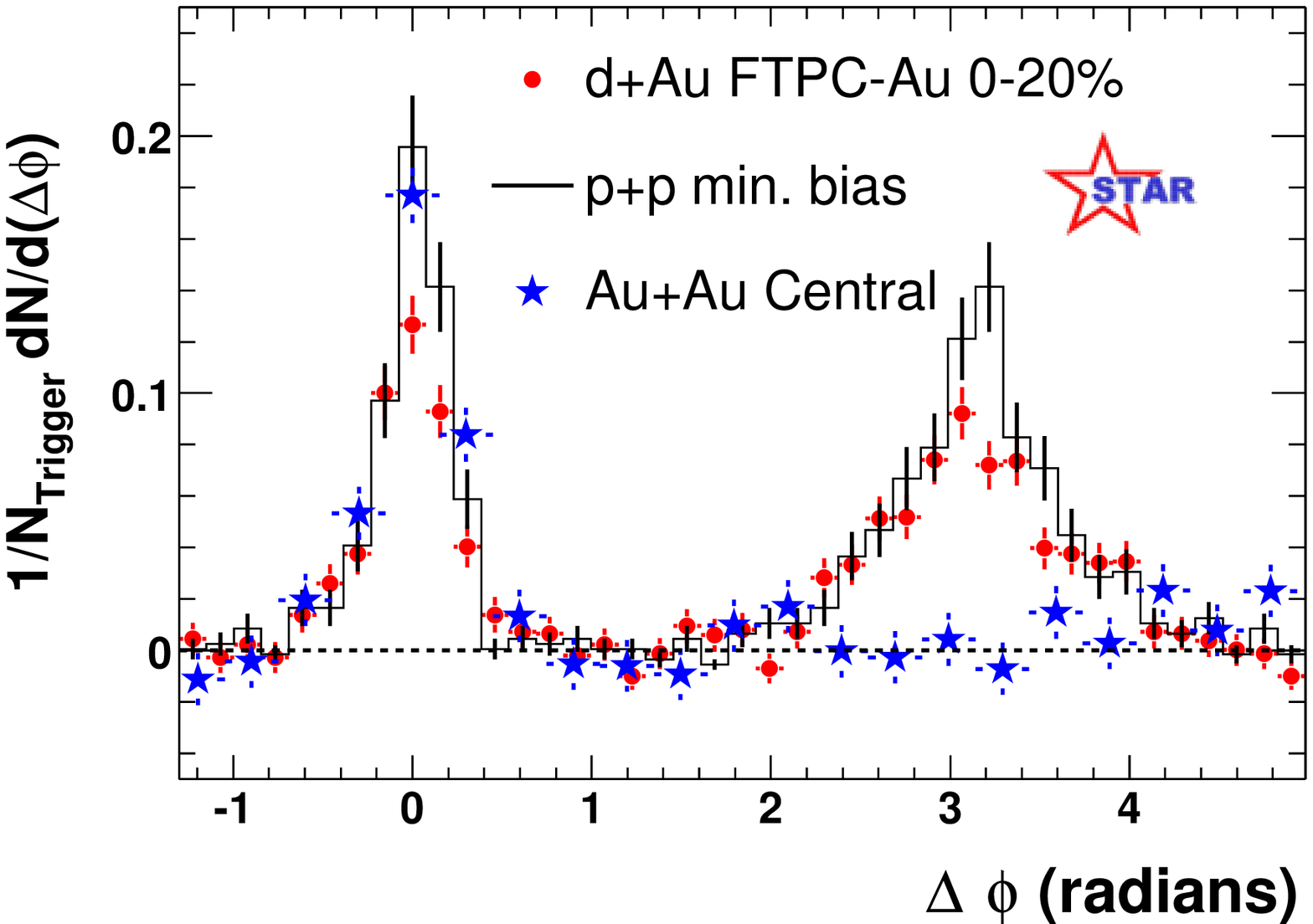}
\includegraphics[width=.48\textwidth]{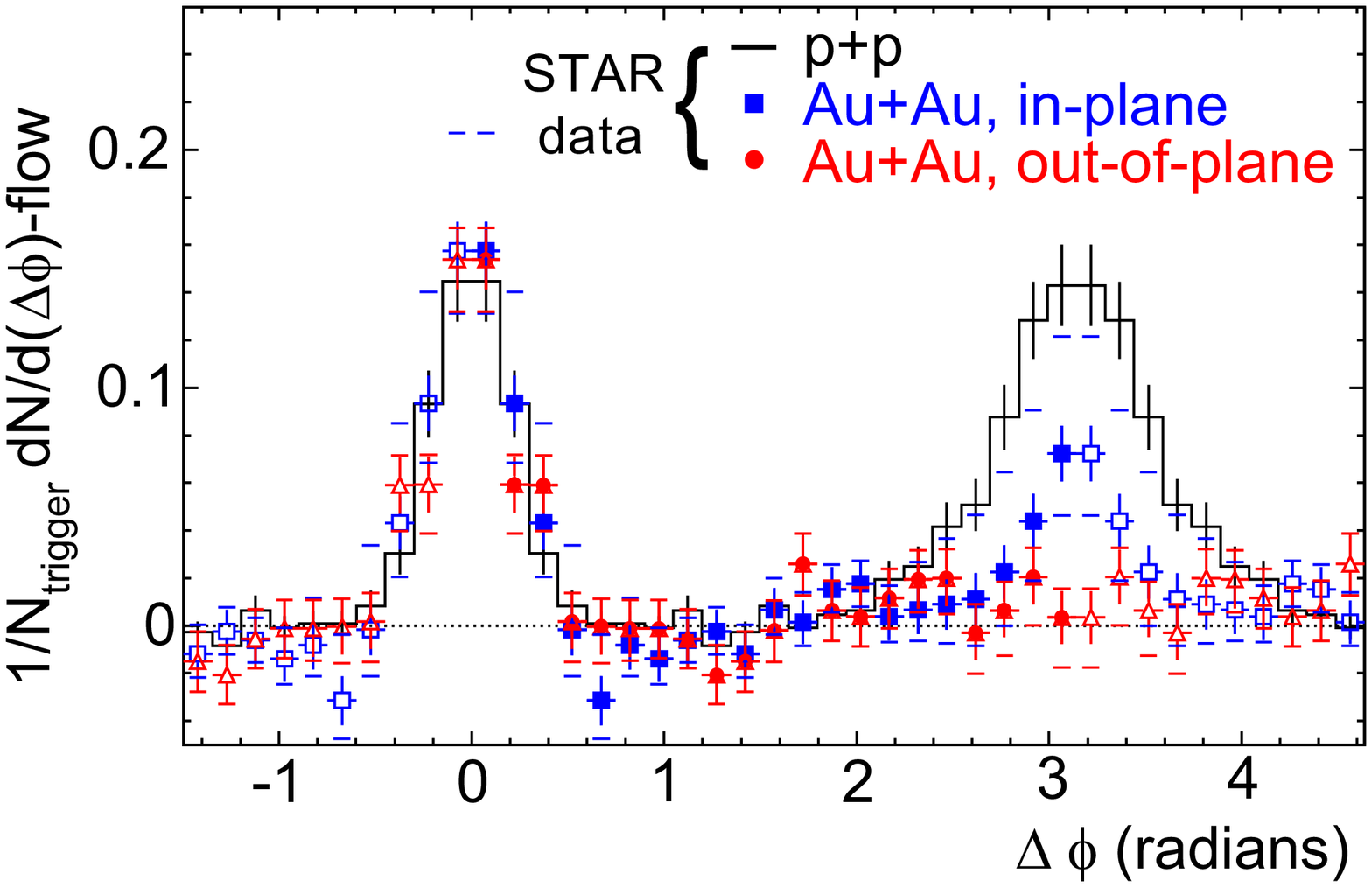}
\caption{{\it Dihadron azimuthal correlations at high \pT. Left
panel shows correlations for p+p, central d+Au and central Au+Au
collisions (background subtracted) from STAR
\cite{star:highpTdAu,star:highpTbtob}. Right panel shows the
background-subtracted high \pT\ dihadron correlation for different
orientations of the trigger hadron relative to the Au+Au reaction
plane \cite{star:highptv2}. }} \label{fig:CorrelationsPR}
\end{center}
\end{figure}

Figure \ref{fig:CorrelationsPR} shows
seminal STAR measurements of
correlations between high \pT\
hadrons. The left panel shows the azimuthal distribution of
hadrons with \pT\gt2 GeV/c relative to a trigger hadron with
$\pT^{\rm trig} \gt4$ GeV/c. A hadron pair drawn from a single jet
will generate an enhanced correlation at $\Delta\phi\approx 0$, as
observed for p+p, d+Au and Au+Au, with similar correlation
strengths, widths and (not shown) charge-sign ordering (the
correlation is stronger for oppositely charged hadron pairs
\cite{star:highpTbtob}). A hadron pair drawn from back-to-back
dijets will generate an enhanced correlation at $\Delta\phi\approx
\pi$, as observed for p+p and for d+Au with somewhat broader width
than the near-side correlation peak. However, the back-to-back
dihadron correlation is strikingly, and uniquely, absent in
central Au+Au collisions, while for peripheral Au+Au collisions
the correlation appears quite similar to that seen in p+p and
d+Au. If the correlation is indeed the result of jet
fragmentation, the suppression is again due to the FSI of
hard-scattered partons or their fragmentation products in the
dense medium generated in Au+Au collisions \cite{star:highpTdAu}.
In this environment, the hard hadrons we do see (and hence, the
near-side correlation peak) would arise preferentially from
partons scattered outward from the surface region of the collision
zone, while the away-side partons must burrow through significant
lengths of dense matter.

The qualification concerning the dominance of jet fragmentation is
needed in this case, because the correlations have been measured
to date primarily for hadrons in that intermediate $p_T$ range
(2-6 GeV/c) where sizable differences in meson vs. baryon yields
have been observed (see Fig.~\ref{rcp}), in contrast to
expectations for jets fragmenting in vacuum.  The systematics of
the meson-baryon differences in this region suggest sizable
contributions from softer mechanisms, such as quark
coalescence~\cite{recom03}. Where the azimuthal correlation
measurements have been extended to trigger particles above 6
GeV/c, they show a similar pattern to the results in
Fig.~\ref{fig:CorrelationsPR}, but with larger statistical
uncertainties~\cite{Hardtke:2002ph}. This suggests that the peak
structures in the correlations do, indeed, reflect dijet
production, and that the back-to-back suppression is indeed due to
jet quenching.  Coalescence processes in the intermediate $p_T$
range may contribute predominantly to the smooth background, with
only long-range (\emph{e.g.},
elliptic flow) correlations, that
has already been subtracted from the data in
Fig.~\ref{fig:CorrelationsPR}.

It remains an open challenge for the quark coalescence models to
account for the observed $\Delta \phi$ distributions at moderate
$p_T$ at the same time as the meson \emph{vs.} baryon yield and
elliptic flow differences discussed in Sec. 3 (see
Fig.~\ref{recom-fits} and associated discussion).  Can the size of
the jet peaks seen in Fig.~\ref{fig:CorrelationsPR} be reconciled
with the modest fragmentation contributions implied by the
coalescence fits near $p_T \approx 4$ GeV/c
(Fig.~\ref{recom-fits})? Do the jet $\Delta \phi$ peaks rather
require substantial contributions also from recombination of a
hard-scattered parton with thermal partons from the bulk matter
\cite{Hwa}?  Are models of the latter type of contributions, of
constituent quark coalescence in a thermal ensemble \cite{Duke}
and of vacuum fragmentation \cite{GVWZ} mutually compatible?  They
would appear to contain non-orthogonal contributions and to employ
incompatible degrees of freedom.  Until these questions are
successfully addressed, some ambiguity remains in physics
conclusions drawn from the intermediate-$p_T$ region, including
the dihadron correlations in Fig.~\ref{fig:CorrelationsPR}.

A more differential probe of partonic energy loss is the
measurement of high \pT\ dihadron correlations relative to the
reaction plane orientation.  The right panel of
Fig.~\ref{fig:CorrelationsPR} shows a study from STAR of the high
\pT\ dihadron correlation from 20-60\% centrality Au+Au
collisions, with the trigger hadron situated in the azimuthal
quadrants centered either in the reaction plane (``in-plane'') or
orthogonal to it (``out-of-plane'') \cite{star:highptv2}. The
same-side dihadron correlation in both cases is similar to that in
p+p collisions. In contrast, the suppression of the back-to-back
correlation depends strongly on the relative angle between the
trigger hadron and the reaction plane. This systematic dependence
is consistent with the picture of partonic energy loss: the path
length in medium for a dijet oriented out of the reaction plane is
longer than in the reaction plane, leading to correspondingly
larger energy loss. The dependence of parton energy loss on path
length is predicted \cite{GVWZ} to be substantially stronger than
linear.  The orientation-dependence of the energy loss should be
further affected by different rates of matter expansion in-plane
\emph{vs.} out-of-plane.

\begin{figure} [thb]
\begin{center}
\includegraphics[width=0.5\textwidth]{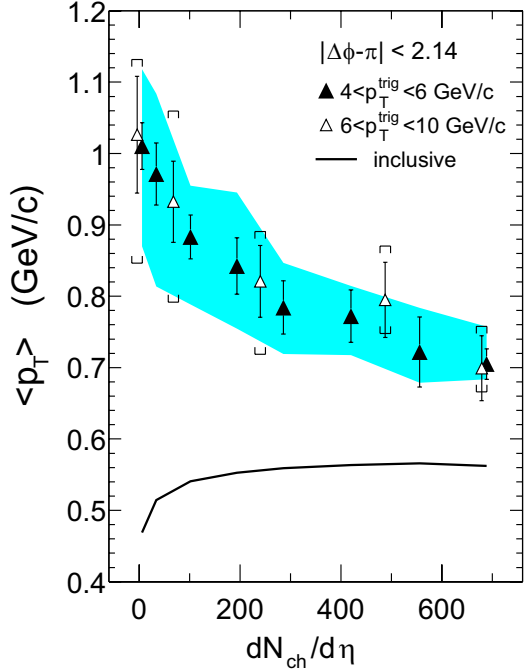}
\caption{{\it Associated charged hadron \meanpT\ from the
away-side in 200 GeV p+p (two
leftmost points) and Au+Au
collisions at various centralities, in each case opposite a
trigger hadron with $p_T$ in the 4--6 GeV/c
(filled triangles) or 6--10 GeV/c
(open triangles) range \cite{star_fqwang}. The
shaded band
and the horizontal caps represent
the systematic uncertainties for
the filled and open symbols, respectively. \meanpT\ for inclusive
hadron production in the Au+Au collisions is indicated by the
solid curve.}} \label{jetloss}
\end{center}
\end{figure}

The energy lost by away-side partons traversing the collision
matter must appear, in order to conserve transverse momentum, in
the form of an excess of softer emerging hadrons.  An analysis of
azimuthal correlations between hard and \emph{soft} hadrons has
thus been carried out for both 200 GeV p+p and Au+Au collisions
\cite{star_fqwang} in STAR, as a first attempt to trace the degree
of degradation on the away side.  With trigger hadrons still in
the range 4$<p_T^{trig}<$6 GeV/c, but the associated hadrons now
sought over 0.15$< p_T <$4 GeV/c, combinatorial coincidences
dominate these correlations, and they must be removed
statistically by a careful mixed-event subtraction, with an
elliptic flow correlation correction added by hand
\cite{star_fqwang}.  The results demonstrate that, in comparison
with p+p and peripheral Au+Au collisions, the momentum-balancing
hadrons opposite a high-$p_T$ trigger in central Au+Au are greater
in number, much more widely dispersed in azimuthal angle, and
significantly softer.  The latter point is illustrated in
Fig.~\ref{jetloss}, showing the centrality dependence of \meanpT\
of the associated away-side charged hadrons in comparison to that
of the bulk inclusive hadrons. While in peripheral collisions the
values of \meanpT\ for the away-side hadrons are significantly
larger than that of inclusive hadrons, the two values approach
each other with increasing centrality. These results are again
subject to the ambiguity arising from possible soft (\emph{e.g.},
coalescence) contributions to the observed correlations, as the
away-side strength shows little remnant of jet-like behavior
\cite{star_fqwang}.  But again,
preliminary results for higher trigger-hadron $p_T$ values, shown
in Fig.~\ref{jetloss}, appear to be consistent within larger
uncertainties.  If a hard-scattering interpretation framework
turns out to be valid, the results suggest that even a moderately
hard parton traversing a significant path length through the
collision matter makes substantial progress toward equilibration
with the bulk. The rapid attainment of thermalization via the
multitude of softer parton-parton interactions in the earliest
collision stages would then not be surprising.

\subsection{Theoretical interpretation of hadron suppression}

\begin{figure}
\begin{center}
\includegraphics[width=.6\textwidth]{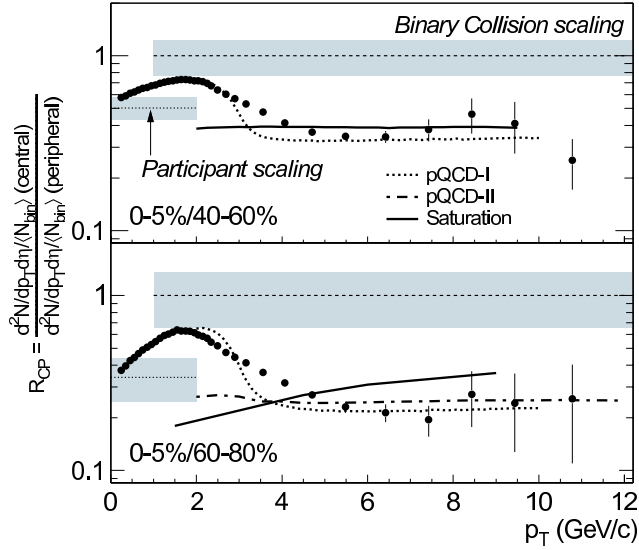}
\caption{{\it Binary-scaled yield ratio \RCP\ for central (0-5\%)
relative to peripheral (40-60\%, 60-80\%) collisions for charged
hadrons from 200 GeV Au+Au collisions \cite{star:highpTAuAu200}.
The shaded bands show the multiplicative uncertainty of the
normalization of \RCP\ relative to binary collision and
participant number scaling.}} \label{fig:HadronSuppression}
\end{center}
\end{figure}

Figure~\ref{fig:HadronSuppression} shows \RCP, the binary scaled
ratio of yields from central relative to peripheral collisions for
charged hadrons from 200 GeV Au+Au interactions. \RCP\ is closely
related to \RAB, using as reference the binary-scaled spectrum
from peripheral Au+Au collisions rather than p+p collisions. The
substitution of the reference set allows a slight extension in the
$p_T$ range for which useful ratios can be extracted.  The error
bars at the highest \pT\ are dominated by statistics and are
therefore, to a large extent, uncorrelated from point to point.
The suppression for central
collisions is again seen to be a
factor $\approx 5$ relative to the most peripheral collisions, and
for $\pT \gtrsim 6$ GeV/c it is independent of \pT\ within
experimental uncertainties. Also shown in
Fig.~\ref{fig:HadronSuppression} are results from theoretical
calculations based on pQCD incorporating partonic energy loss in
dense matter (pQCD-I \cite{Wang:2003mm}, pQCD-II
\cite{Vitev:2002pf}) and on suppression at high \pT\ due to gluon
saturation effects (Saturation \cite{Kharzeev:2002pc}, with
implications discussed further in the following subsection). The
negligible \pT-dependence of the suppression at high \pT\ is a
prediction of the pQCD models \cite{Wang:2003mm,Vitev:2002pf},
resulting from the subtle interplay of partonic energy loss,
Cronin (initial-state multiple scattering) enhancement, and
nuclear shadowing. The variation in the suppression for $p_T
\lesssim 5$ GeV/c is related to differences in suppression in this
region for mesons and baryons (see Fig.~\ref{rcp}). It is
accounted for in the pQCD-I calculation by the introduction of an
additional non-fragmentation production mechanism for kaons and
protons~\cite{Wang:2003mm}.  The
magnitude of the hadron suppression in the pQCD calculations is
adjusted to fit the measurements for central collisions, as
discussed further below.

In order to deduce the magnitude of {\it partonic} energy loss in
the medium it is essential to establish the degree to which {\it
hadronic} interactions, specifically the interaction of hadronic
jet fragments with the medium, can at least in part generate the
observed high \pT\ phenomena and contribute substantially to the
jet quenching \cite{Greiner,Falter:2002jc,Gallmeister:2002us}.
Simple considerations already argue against this scenario. The
dilated formation time of hadrons with energy $E_h$ and mass $m_h$
is $t_f=(E_h/m_h)\tau_f$, where the rest frame formation time
$\tau_f\sim0.5-0.8$ fm/c. Thus, a 10 GeV/c pion has formation time
$\sim50$ fm/c and is unlikely to interact as a fully formed pion
in the medium. Since the formation time depends on the boost, the
suppression due to hadronic absorption with constant or slowly
varying cross section should turn off with rising \pT, at variance
with observations (Fig. \ref{fig:HadronSuppression}). A detailed
hadronic transport calculation \cite{Greiner} leads to a similar
conclusion: the absorption of formed hadrons in the medium fails
by a large factor to account for the observed suppression. Rather,
this calculation attributes the suppression to \emph{ad hoc}
medium interactions of ``pre-hadrons'' with short formation time
and constant cross section.  The properties of these
``pre-hadrons" are thus similar to those of colored partons
\cite{Greiner}, and not to the expected color transparency of
hadronic matter to small color singlet particles that might evolve
into normal hadrons \cite{CT}.

Additional considerations of the available high \pT\ data
\cite{Wang} also support the conclusion that jet quenching in
heavy ion collisions at RHIC is the consequence of partonic energy
loss. In particular, large \vtwo\ values observed at high \pT\ and
the systematics of the small-angle dihadron correlations are
difficult to reconcile with the hadronic absorption scenario.
While further theoretical investigation of this question is
certainly warranted, we conclude that there is no support in the
data for {\it hadronic} absorption as the dominant mechanism
underlying the observed suppression phenomena at high \pT\, and we
consider {\it partonic} energy loss to be well established as its
primary origin.  It is conceivable that there may be minor
hadronic contributions from the fragments of soft gluons radiated
by the primary hard partons during their traversal of the
collision matter.  In any case, we emphasize that while the jet
quenching results seem to favor partons over hadrons \emph{losing}
energy, they do not allow any direct conclusion regarding whether
the energy is lost \emph{to} partonic or hadronic matter.

The magnitude of the suppression at high \pT\ in central
collisions is fit to the data in the pQCD-based models with
partonic energy loss, by adjusting the initial gluon density of
the medium. The agreement of the calculations with the
measurements at \pT\gt5 GeV/c is seen in
Fig.~\ref{fig:HadronSuppression} to be good. In order to describe
the observed suppression, these models require an initial gluon
density about a factor 50 greater than that of cold nuclear matter
\cite{Wang:2003mm,Vitev:2002pf,Eskola:2004cr}. This is the main physics result
of the high \pT\ studies carried out at RHIC to date.  It should
be kept in mind that the actual energy loss inferred for the
rapidly expanding Au+Au collision matter is not very much larger
than that inferred for static, cold nuclear matter from
semi-inclusive deep inelastic scattering data  \cite{Zhang-Wang}.
But in order to account for this slightly larger energy loss
\emph{despite} the rapid expansion, one infers the much larger
\emph{initial} gluon density at the start of the expansion
\cite{Wang:2003mm,Vitev:2002pf}.  Certainly, then, the
quantitative extraction of gluon density is subject to
uncertainties from the theoretical treatment of the expansion and
of the energy loss of partons in the entrance-channel cold nuclear
matter before they initially collide.

The gluon density derived from energy loss calculations is
consistent with estimates from the measured rapidity density of
charged hadrons \cite{Back:2001ae} using the Bjorken scenario
\cite{BjorkenHydro}, assuming isentropic expansion and duality
between the number of initial gluons and final charged hadrons.
Similar values are also deduced under the assumption that the
initial state properties in central Au+Au RHIC collisions, and
hence the measured particle multiplicities, are determined by
gluon-gluon interactions below the gluon density saturation scale
in the initial-state nuclei \cite{Kharzeev}. Additionally, the
energy density is estimated from global measurements of transverse
energy (see Sec. 3.1) to be of order 50-100 times that in cold
nuclear matter, consistent with
the values inferred from hydrodynamics accounts of measured hadron
spectra and flow.  The consistency among all these estimates,
though only semi-quantitative at present, is quite significant.
These inferred densities fall well into the regime where LQCD
calculations predict equilibrated matter to reside in the QGP
phase.

\subsection{Rapidity-dependence of high $p_T$ hadron yields in
d+Au collisions}

It had been proposed recently \cite{Kharzeev:2002pc} that gluon
saturation effects can extend well beyond the saturation momentum
scale \Qs, resulting in hadron suppression relative to binary
scaling (\RAB\lt1) for $\pT\sim5-10$ GeV/c mid-rapidity hadron
production at RHIC energies, in apparent agreement with the data
in Fig.~\ref{fig:HadronSuppression}. However, since this predicted
suppression originates in the properties of the incoming nuclear
wave function, hadron production in d+Au collisions should also be
suppressed by this mechanism \cite{Kharzeev:2002pc}.
Experimentally, an \emph{enhancement} in mid-rapidity hadron
production in d+Au is seen instead
(Fig.~\ref{fig:InclusiveSuppressionPR}
\cite{brahms:highpTdAu,phenix:highpTdAu,phobos:highpTdAu,star:highpTdAu}),
even in central d+Au collisions \cite{star:highpTdAu} where
saturation effects should be most pronounced. The observed
enhancement is at variance with saturation model expectations at
high \pT\ \cite{Kharzeev:2002pc}.

%--=================================================================
\begin{figure} [bht]
\begin{center}
\includegraphics[width=0.85\textwidth]{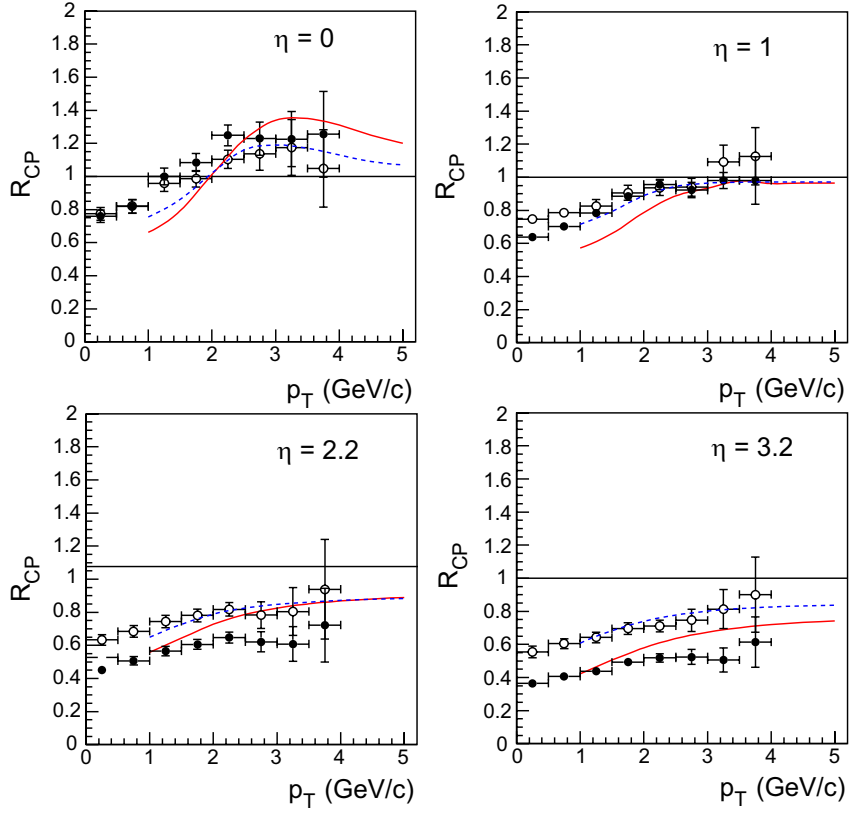}
\caption{{\it The ratio $R_{CP}$ of binary-scaled central to
peripheral hadron yields for d+Au collisions at \sqrtsNN=200 GeV,
plotted as a function of $p_T$ for four different pseudorapidity
bins, centered at $\eta = 0$, $\eta = 1$, $\eta = 2.2$ and $\eta =
3.2$.  The measurements are from the Brahms Collaboration
\cite{brahms:dAuSuppression}, for all charged hadrons (negative
hadrons only) in the case of the former (latter) two $\eta$ bins.
The curves represent gluon saturation model fits from
\cite{kharzeev04}.  The filled circles and solid curves compare
yields in the 0-20\% to 60-80\% centrality bins, while the open
circles and dashed curves compare 30-50\% to 60-80\%. The figure
is taken from Ref.~\cite{kharzeev04}. }}
\label{Brahms:dAuForwardSuppression}
\end{center}
\end{figure}

%--======================

However, at large rapidities in the deuteron direction, a
suppression of the highest $p_T$ hadrons studied is indeed
observed in d+Au collisions, as revealed by the results from the
Brahms experiment in Fig.~\ref{Brahms:dAuForwardSuppression}
\cite{brahms:dAuSuppression}. This is not true of large rapidities
in the Au direction
\cite{Frawley_QM04,stardAu_eta_asym}.
%(Fig.~\ref{Phenix:dAuRapidityDependence})
This distinct behavior is consistent with gluon saturation models,
as seen by the fits \cite{kharzeev04} to these results in
Fig.~\ref{Brahms:dAuForwardSuppression}. High-$p_T$ hadrons
produced at small angles with respect to the deuteron beam arise
preferentially from asymmetric partonic collisions involving
gluons at low Bjorken $x$ in the Au nucleus.  (For example, in a
next-to-leading order leading-twist perturbative QCD calculation,
the mean $x$-value of partons probed in the Au nucleus has been
found to be 0.03-0.05 when selecting on hadrons at $\eta=3.2$ and
$p_T = 1.5$ GeV/c~\cite{Guzey:2004zp}. Note, however, that such a
calculation may have limited validity in the regime of strong
gluonic fields.) It is precisely at low $x$ in heavy nuclei that
gluon saturation, and the resultant suppression in high-$p_T$
hadron production, should set in.  Thus, gluon saturation models
predicted the qualitative behavior of increasing suppression with
increasing rapidity in the deuteron direction before the
experimental results became available
\cite{kharzeev_dAu_prediction}, although parameter values had to
be tuned after the fact \cite{kharzeev04} to adjust the saturation
scale to obtain the fits shown in
Fig.~\ref{Brahms:dAuForwardSuppression}.

%--=================================================================
%\begin{figure} [bht]
%\begin{center}
%\includegraphics[width=0.80\textwidth]{phenix_rcp_eta_QM04.eps}
%\caption{{\it PHENIX measurements \cite{Frawley_QM04} of the ratio
%$R_{CP}$ of binary-scaled hadron yields for three centrality bins
%to peripheral yields for \sqrtsNN=200 GeV d+Au collisions. The
%measurements include all charged hadrons detected in three
%different pseudorapidity regions within the range $1 < p_T < 3$
%GeV/c.  The results show the onset of central hadron suppression
%in the deuteron direction, but significant enhancement in the Au
%direction. }} \label{Phenix:dAuRapidityDependence}
%\end{center}
%\end{figure}
%--======================

\begin{figure} [bht]
\begin{center}
\includegraphics[width=0.65\textwidth]{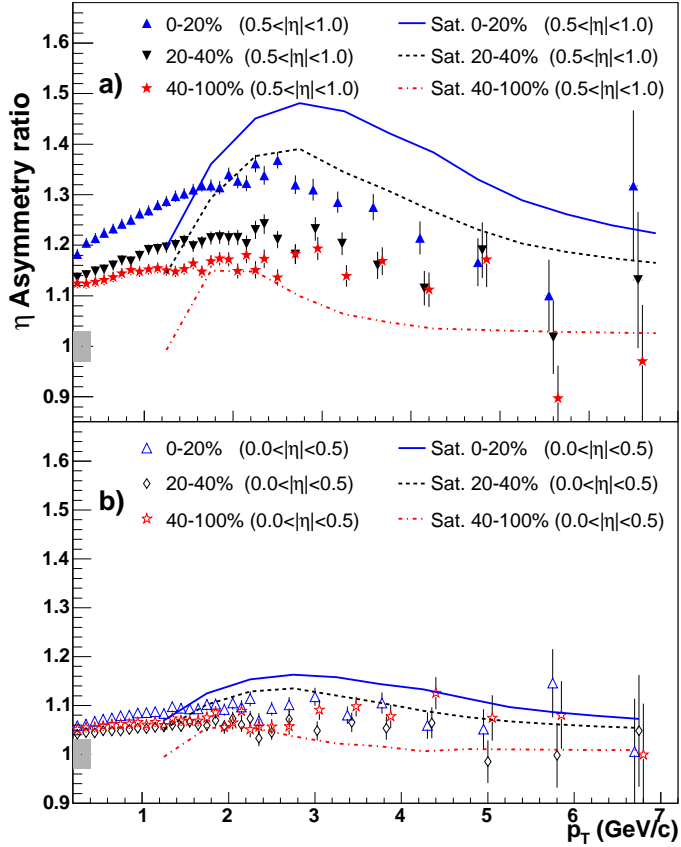}
\caption{{\it Comparison of STAR
measurements \cite{stardAu_eta_asym} with gluon saturation model
calculations \cite{kharzeev04} for backward (Au side) to forward
(d side) charged hadron yield ratios at mid-rapidity in d+Au
collisions at \sqrtsNN=200 GeV. Results are shown as a function of
hadron $p_T$ for two different pseudorapidity ranges, (a) $0.5 <
|\eta| < 1.0$ and (b) $0.0 < |\eta| < 0.5$, and three different
centrality ranges.  Centrality is determined experimentally from
the measured charged particle multiplicity in the forward Au
direction, for $-3.8 < \eta < -2.8$. Figure is taken from
Ref.~\cite{stardAu_eta_asym}. }}
\label{STAR:dAuRapidityRatios}
\end{center}
\end{figure}

At the moderate $p_T$ values
kinematically accessible at large pseudorapidity, one may worry
legitimately that softer hadron production mechanisms
(\emph{e.g.}, quark recombination) and initial-state multiple
scattering of partons before hard collisions complicate the
interpretation of the d+Au results.  The same basic suppression of
hadrons in the deuteron, relative to the Au, direction can be seen
extending to higher $p_T$ in the mid-rapidity backward/forward
yield ratios from STAR \cite{stardAu_eta_asym}, shown in
Fig.~\ref{STAR:dAuRapidityRatios}.  The same gluon saturation
model calculations \cite{kharzeev04} shown in
Fig.~\ref{Brahms:dAuForwardSuppression} are seen in
Fig.~\ref{STAR:dAuRapidityRatios} to be qualitatively, but not
quantitatively, consistent with the observed dependences of the
hadron yields on pseudorapidity, $p_T$ and centrality.  In
particular, both measurements and calculations suggest that the
mid-rapidity suppression fades away at transverse momenta above
5-6 GeV/c, as one probes higher-$x$ partons in the Au nucleus.

%--=================================================================

%--======================

The results in Fig.~\ref{Brahms:dAuForwardSuppression} and
\ref{STAR:dAuRapidityRatios} represent the strongest evidence yet
available for the applicability of Color Glass Condensate concepts
within the kinematic range spanned by RHIC collisions.
Nonetheless, more mundane origins of this forward hadron
suppression in d+Au have not been ruled out. Di-hadron correlation
measurements involving these forward hadrons in d+Au collisions
may help to distinguish between CGC and other explanations
\cite{kharzeev-cgc-correlations}.
A critical characteristic of the
CGC is that it can be treated as a classical gluon field. Forward
hadrons that result from the
interaction of a quark from the deuteron beam with this gluon
field may have their transverse
momentum balanced not by a single
recoiling parton (and therefore a jet), but rather by a number of
relatively soft hadrons with a much more smeared angular
correlation than is characteristic of di-jet processes. Such a
``mono-jet" signature would not be expected from more conventional
sources of shadowing of gluon densities in the Au nucleus
\cite{Vogt}, which still allow individual quark-gluon, rather than
quark-gluon field, scattering. On the other hand, kinematic limits
on the accessible $p_T$ values for forward hadrons imply that one
is dealing, even in a di-jet framework, with unconventional
away-side jets of only a few GeV/c \cite{Akio_DIS04}. In this
regime, a suitable reference is needed, using p+p or d+A with a
sufficiently light nucleus A to place the contributing parton x-range
above the anticipated gluon saturation regime.  The discriminating
power of di-hadron correlations for CGC
mono-jets must be demonstrated by modifications in d+Au with
respect to this reference.

\subsection{Outlook}

While large effects have been observed and the phenomenon of jet
quenching in dense matter has been firmly established, precision
data in a larger $p_T$ range are needed to fully explore the jet
quenching phenomena and their connection to properties of the
dense matter. The region $2\lt\pT\lt6$ GeV/c has significant
contributions from non-perturbative processes other than vacuum
fragmentation of partons, perhaps revealing novel hadronization
mechanisms. Most studies to date of azimuthal anisotropies and
correlations of ``jets'' have by necessity been constrained to
this region, with only the inclusive spectra extending to the
range where hard scattering is expected to dominate the inclusive
yield. High statistics data sets for much higher \pT\ hadrons are
needed to fully exploit azimuthal asymmetries and correlations as
measurements of partonic energy loss. Dihadron measurements
probing the details of the fragmentation process may be sensitive
to the {\it energy} density, in addition to the gluon density that
is probed with the present measurements. Heavy quark suppression
is theoretically better controlled, and measurement of it will
provide a critical check on the understanding of partonic energy
loss. The {\it differential} measurement of energy loss through
measurement of the emerging away-side jet and the recovery of the
energy radiated in soft hadrons is still in its initial phase of
study. A complete mapping of the modified fragmentation with
larger initial jet energy and with a direct photon trigger will
crosscheck the energy dependence of energy loss extracted from
single inclusive hadron suppression.  Experiments at different
colliding energies are also
useful to map the variation of
jet quenching with initial energy density and the lifetime of the
dense system.

At the same time as we extend the $p_T$ range for jet quenching
studies on the high side, it is crucial also to pursue further
(particle-identified) hadron correlation measurements in the soft
sector, in order to understand better how jets are modified by
interactions with the dense bulk matter.  Measurements such as
those presented in Figs.~\ref{mini-jets} and \ref{jetloss} are
just beginning to illuminate the processes leading to
thermalization of parton energy.  The properties of particles that
have been substantially degraded, but not completely thermalized,
by passage through the bulk may provide particularly fertile
ground for exposing possible fundamental modifications
(\emph{e.g.}, symmetry violations or restoration) of strong
interactions in RHIC collision matter.

\newpage

%%\documentclass[12pt]{iopart}
% Uncomment next line if AMS fonts required
%\usepackage{iopams}

%%\begin{document}

\section{Some Open Issues}

It should be clear from the detailed discussions of experimental
and theoretical results in the preceding chapters that
some open questions need to be
addressed before we can judge the evidence in favor of QGP
formation at RHIC to be
compelling. In this chapter we collect
a number of such open questions
for both experiment and theory. Convincing answers to even a few
of these questions might tip the balance in favor of a QGP
discovery claim. But even then,
it will be important to address the remaining questions to
solidify our understanding of the properties of the matter
produced in RHIC collisions.

Lattice QCD calculations suggest
that a confined state is impossible
in bulk, thermodynamically
equilibrated matter at the energy densities apparently achieved
at RHIC.  Indeed, several
experimental observations are \emph{consistent} with the creation
of deconfined matter. However, a
discovery as important as the observation of a fundamentally new
state of matter surely demands proof beyond circumstantial
evidence for deconfinement. Can we do better?

One response that has been offered is that the EOS
of strongly interacting matter is
already known from lattice QCD calculations, so that only the
conditions initially attained in
heavy-ion collisions, and the degree of thermalization
in the matter produced, are open
to doubt. Such a view tends to trivialize the QGP search by
presuming the answer. Indeed, an
important aspect of the original motivation for the experimental
program at RHIC was to explore the Equation of State of strongly
interacting matter under these extreme conditions of energy
density.  Lattice QCD, in addition to its technical difficulties
and attendant numerical uncertainties,
attempts to treat bulk, static,
thermodynamically equilibrated quark-gluon systems.  The
relationship of such idealized matter to the finite, rapidly
evolving systems produced in relativistic heavy-ion collisions is
not \emph{a priori} clear. One would prefer, then, to take LQCD
calculations as guideposts to the transition properties to search
for experimentally, but not as unassailable truth. On the other
hand, there are sufficient complexities in the theoretical
treatment of heavy-ion collisions that one would like to apply all
credible constraints in parameterizing the problem.
This dichotomy leads to our first
question:

\vspace*{1mm} \begin{itemize}

\item[$\bullet$] {\bf To what extent should LQCD results be used
to constrain the Equations of State considered in model treatments
of RHIC collisions?  How does one allow for independent checks of
the applicability of LQCD to the dynamic environment of a RHIC
collision? }

\end{itemize}

Experimentally, to verify the
creation of a fundamentally new state of matter
at RHIC one would like crosschecks
demonstrating that the matter behaves qualitatively {\it
differently} than ``normal'' (hadronic) matter
in a system known or believed to
be in a confined state.  Although
such a demonstration might be straightforward in bulk matter, it
becomes an enormous challenge with the limited experimental
control one has over thermodynamic variables in heavy-ion
collisions.  The finite size and lifetime of the matter produced
in the early collision stages, coupled with the absence of global
thermal equilibrium and of measurements (to date) of local
temperature, all work to obscure the hallmark of QGP formation
predicted by lattice QCD: a rapid transition around a critical
temperature leading to deconfinement and, quite possibly, chiral
symmetry restoration (the latter considered here as a sufficient,
but not necessary, QGP manifestation).
Given these complications, the
underlying challenge to theory and experiment is:

\vspace*{1mm}
\begin{itemize}

\item[$\bullet$] {\bf Can we make a convincing QGP discovery claim
without clear evidence of a rapid transition in the behavior of
the matter produced? Can we devise probes with sufficient
sensitivity to early, local system temperature to facilitate
observation of such an onset at RHIC? Can we predict, based on
what we now know from SPS and RHIC collisions, at what energies or
under what conditions we might produce matter below the critical
temperature, and which observables from those collisions should
not match smoothly to SPS and RHIC results? }

\end{itemize}
\vspace*{1mm}

At the most basic level, it is
conceivable that there is no rapid
deconfinement transition in
nature (or at least in the matter formed fleetingly in heavy-ion
collisions), but rather a gradual evolution from dominance of
hadronic toward dominance of partonic degrees of freedom.
It is not yet clear that we could
distinguish such behavior of QCD matter from the blurring of a
well-defined QGP transition by the use of tools with insufficient
resolution or control.

\subsection{What experimental crosschecks can be performed on
apparent QGP signatures at RHIC?}

Below we briefly discuss some of the key observations that underlie
theoretical claims
\cite{Gyulassy,McLerran-Gyulassy,RBRC}
that deconfined matter
has been produced at RHIC, and ask
what crosschecks might be
carried out to test this hypothesis.

\subsubsection{Jet quenching}

As discussed in Sec. 4, inclusive hadron spectra
%~\cite{RHIC_RAA}
and two-particle azimuthal correlations at moderate and high $p_T$
%~\cite{STAR_Hardtke}
clearly demonstrate that jets are suppressed in central RHIC Au+Au
collisions, relative to scaled NN collisions. The lack of
suppression (indeed, the enhancement, due to the Cronin effect) in
d+Au collisions at RHIC
%~\cite{dAuRHIC}
provides a critical crosscheck that the quenching is not an
initial-state effect.  Measurements with respect to the event
reaction plane orientation (see Fig.~\ref{fig:CorrelationsPR})
provide another important crosscheck, demonstrating that the
magnitude of the suppression depends strongly on the amount of
matter traversed.  Such jet
quenching was first predicted \cite{Bjorken} within the framework
of parton energy loss in traversing a QGP.  However, more recent
theoretical work \cite{Baier} casts doubt that deconfinement of
the medium is essential to the phenomenon, or would be manifested
clearly in the energy-dependence of quenching.  Nonetheless,
experimental hints of a possibly interesting energy dependence to
quenching phenomena should be pursued as a potential crosscheck on
formation of a new state of matter.

Moderate-$p_T$ (up to 4 GeV/c) yields from Pb-Pb collisions at the
SPS \cite{WA98highpT} appear to show an enhancement over a scaled
{\it parameterized} p-p reference spectrum.  However, questions
raised about the p-p parameterization \cite{D'Enterria}, combined
with the unavailability of measurements constraining the
initial-state (Cronin) enhancement at these energies, leave open
the possibility that even at SPS, jets in central A+A collisions
may turn out to be suppressed \emph{relative to expectations}.
Indeed, the data in \cite{WA98highpT} do demonstrate hadron
suppression in central relative to semi-peripheral collisions.
Also, it is unclear whether the suppression of away-side
two-particle correlations out of the reaction plane, observed at
RHIC (see Fig.~\ref{fig:CorrelationsPR}), might be
%~\cite{STAR_jetsWrtRP}
of similar origin as the away-side out-of-plane broadening
observed at the SPS~\cite{CERESAwaySide}. These ambiguities are
amplified by the limited $p_T$ range covered in SPS measurements,
spanning only a region where RHIC results suggest that hard parton
scattering and fragmentation may not yet be the dominant
contributing hadron production mechanism.
These observations lead to the
following question:

\vspace*{1mm}
\begin{itemize}

\item[$\bullet$] {\bf Is there a
qualitative change in the yield of high-$p_T$ hadrons in A+A
collisions between SPS and RHIC energies?  Or does hadron
suppression rather evolve smoothly with energy, reflecting a
gradual growth in initial gluon density and parton energy loss? Is
it feasible to make meaningful measurements of hard probes at
sufficiently low collision energy to test for the absence or gross
reduction of jet quenching in matter believed to be in a hot
hadronic (\emph{i.e.}, confined) gas state?}

\end{itemize}
\vspace*{1mm}

\subsubsection{Constituent-quark scaling of yields and anisotropies}

The baryon \emph{vs}. meson systematics of $R_{CP}$
(Fig.~\ref{rcp}) and the apparent scaling of elliptic flow with
the number of constituent quarks (Fig.~\ref{v21}) in the
intermediate $p_T$ range strongly suggest collective behavior at a
pre-hadronic level, a necessary aspect of QGP formation and
thermalization in heavy-ion collisions.  Once again, one would
like to observe the {\it absence} of this behavior for systems in
which QGP is not formed. High-quality, particle-identified
elliptic flow data do not yet exist at SPS (or lower) energies in
this $p_T$ region.

\vspace*{1mm}
\begin{itemize}

\item[$\bullet$] {\bf Should constituent-quark scaling of $v_2$ in
the intermediate $p_T$ sector be broken if a QGP is {\it not}
formed? If so, is an appropriate statistically meaningful,
particle-identified measurement of $v_2$ at intermediate $p_T$
feasible at $\sqrt{s_{NN}}$ below the QGP formation threshold? }

\end{itemize}
\vspace*{1mm}

Alternatively, we could seek to
establish the role of constituent quarks more convincingly by
additional predictions of
the quark coalescence models introduced to characterize this
intermediate $p_T$ region.  For
this purpose it may be helpful to integrate the coalescence models
with other (e.g., gluon saturation or hydrodynamics) models that
might serve to constrain the anticipated initial conditions and
coalescence parameters as a function of centrality or collision
energy.

\vspace*{1mm}
\begin{itemize}

\item[$\bullet$] {\bf Coalescence models have provided a simple
ansatz to recognize the possible importance of constituent quark
degrees of freedom in the hadronization process in A+A collisions
at RHIC, and to suggest that these constituent quarks exhibit
collective flow.  Once model parameters have been adjusted to
account for the observed ratios of yields and elliptic flow
strengths for baryons vs. mesons,
can integration of key features
from other models enhance predictive power?
For example, can the
centrality-dependence of these ratios, or meson vs. baryon
correlations (angular or otherwise) at moderate $p_T$ be
predicted?}

\end{itemize}
\vspace*{1mm}

\subsubsection{Strong elliptic flow in agreement with hydrodynamics}

In contrast to the above signatures, which require access to
moderate-to-high $p_T$ values, observables in the soft sector have
already been extensively explored, even from Bevalac energies. The
only soft-sector observable selected as a ``pillar''
%~\cite{Gyulassy_3QGPpillars}
of the QGP claim at RHIC, in Ref.~\cite{Gyulassy}, is the strong
elliptic flow, whose magnitude, mass and $p_T$-dependence for
mid-central collisions are in reasonable agreement with
expectations based on ideal hydrodynamic flow (see
Fig.~\ref{v2low}).
%~\cite{HeinzKolbHydroReview}.
Furthermore, the agreement appears better for an Equation of State
that includes passage through a phase transition from partonic to
hadronic matter.
%~\cite{HeinzKolbHydroReview}.

\begin{figure}[thb]
\begin{center}
\epsfig{figure=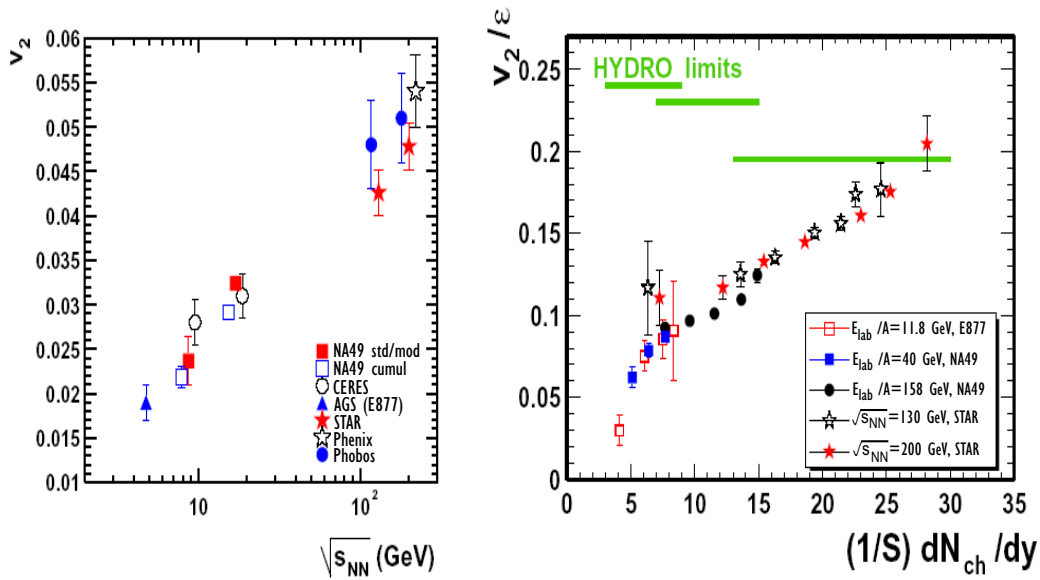,width=15.0cm} \caption{
{\it (a) Energy dependence of elliptic flow measured near
mid-rapidity for mid-central collisions ($\sim 12-34\%$ of the
geometric cross section) of $A \sim 200$ nuclei at the AGS, SPS
and RHIC. (b) Mid-rapidity elliptic flow measurements from various
energies and centralities combined in a single plot of $v_2$
divided by relevant initial spatial eccentricity vs.
charged-particle rapidity density per unit transverse area in the
$A+A$ overlap region.  The figures, taken from
Ref.~\cite{NA49flow}, highlight the smooth behavior of flow vs.
energy and centrality. The rightmost points represent near-central
STAR results, where the observed $v_2/\epsilon$ ratio becomes
consistent with limiting hydrodynamic expectations for an ideal
relativistic fluid.  The hydrodynamic limits are represented by
horizontal lines \cite{NA49flow} drawn for AGS, SPS and RHIC
energies (from left to right), for one particular choice of EOS
that assumes no phase transition in the matter produced.}}
\label{v2-vs-energy}
\end{center}
\end{figure}

This success leads to the claim \cite{kolbheinz,Gyulassy} that the
elliptic flow has finally, in near-central collisions at RHIC
energies, reached the ideal hydrodynamic ``limit,'' suggesting
creation of equilibrated,
low-viscosity matter at an early
stage in the collision (when geometric anisotropy is still large).
However, the results from many experiments clearly indicate a
smoothly rising $v_2( \sqrt{s_{NN}})$, while the hydrodynamic
limit for given initial spatial eccentricity
and fixed EOS is falling with
increasing energy (see Fig.~\ref{v2-vs-energy}).  It is thus
unclear from the available data whether we are observing at RHIC
the interesting onset of saturation of a simple physical limit
particularly relevant to QGP
matter, or rather an accidental crossing point of experiment with
a necessarily somewhat simplified
theory.  It is of major
significance that ideal hydrodynamics appears to work at RHIC for
the first time. This conclusion -- and in particular the evidence
for an Equation of State containing a phase change -- would be
much strengthened if the hydrodynamic limit were demonstrated to
be relevant as well under conditions far removed from those in
RHIC measurements to date. Future measurements in central
collisions of heavier and highly deformed nuclei (\emph{e.g.}, U+U
\cite{kolbheinz}) possible after a planned upgrade of the ion
source for RHIC, or at significantly lower or higher energy (the
latter awaiting LHC turn-on) will provide the possibility of
additional crosschecks of this important conclusion.

\vspace*{1mm}
\begin{itemize}

\item[$\bullet$] {\bf  Is the ideal hydrodynamic limit for
elliptic flow relevant to heavy-ion collisions over a broad range
of conditions, within which near-central Au+Au collisions at full
RHIC energy represent merely a first ``sighting"? Will $v_2$ at
LHC energies surpass the hydrodynamic limit?  Is thermalization
likely to be sufficiently established in collisions below
\sqrtsNN$\approx 100$ GeV to permit meaningful tests of
hydrodynamics?  If so, will measurements at lower RHIC energies
reveal a non-trivial energy dependence of $v_2$, such as that
predicted in Fig.~\ref{hydro-excitation} by
ideal hydrodynamics
incorporating a phase transition?
Can one vary the initial spatial eccentricity of the bulk matter
independently of centrality and degree of thermalization, via
controlled changes in the relative alignment of deformed colliding
nuclei such as uranium?}

\end{itemize}

\vspace*{1mm}

\subsubsection{Dependence of observables on system size}

The above questions focused on excitation function measurements,
which traditionally have played a crucial role in heavy-ion
physics. It is also desirable to explore the appearance and
disappearance of possible QGP signatures as a function of system
size.  To date, system size variations have been examined at RHIC
primarily via the centrality dependence of many observables.  A
number of variables have been observed to change rapidly from the
most peripheral to mid-peripheral collisions, and then to saturate
for mid-central and central collisions.  Examples of this type of
behavior include:  the strength ($I_{AA}$ in
Ref.~\cite{star:highpTbtob}) and $\Delta\eta$ width
(Fig.~\ref{mini-jets}) of near-side di-hadron correlations; the
ratio of measured $v_2$ to the hydrodynamic limit for relevant
impact parameter \cite{star130flow}; the strangeness saturation
parameter $\gamma_s$ deduced from statistical model fits to
measured hadron yield ratios (inset in Fig.~\ref{ratio})
\cite{barranikQM04}. Do these changes reflect a (QGP) transition
with increasing centrality in the nature of the matter first
produced, or merely the gradual growth in importance of hadronic
initial- and final-state interactions, and in the degree of
thermalization achieved, as the number of nucleon participants
increases?  One's answer to this question may depend on how rapid
the variation with centrality appears, but this in turn depends on
what measure one uses for centrality, as emphasized in the lower
frames of Fig.~\ref{mini-jets}.

As the centrality changes for given colliding nuclei, so,
unavoidably, does the initial shape of the overlap region.  In
order to unravel the influence of different initial conditions on
the evolution of the matter formed in heavy-ion collisions, it
will be important to measure as well the dependence of observables
such as those above on the size of the colliding nuclei.

\vspace*{1mm}
\begin{itemize}

\item[$\bullet$] {\bf  Do RHIC measurements as a function of
centrality already contain hints
of the onset of QGP formation in relatively peripheral regions?
Will future measurements for lighter colliding nuclei permit more
definitive delineation of these
apparently rapid changes with system size?}

\end{itemize}

\vspace*{1mm}

\subsection{Do the observed consistencies with QGP formation demand
a QGP-based explanation?}

Because it is difficult to control the degree of thermalization
achieved in heavy-ion collisions, and to measure directly the
temperature at which it is initially achieved, it is possible that
none of the crosschecks discussed in the preceding subsection for
RHIC energies and below may provide definitive experimental
resolution concerning QGP formation. In this case, our reliance on
the comparison with theory would be significantly increased, and
the questions posed below become especially important.  Here, we
question the {\it uniqueness} of a QGP-based explanation. In other
words, do the data {\it demand} a scenario characterized by
thermalized, deconfined matter?

\subsubsection{Strong elliptic flow}
\label{sec:HydroConsistent}

The hydrodynamic overestimate of elliptic flow at energies below
RHIC has been attributed either
to a failure to achieve complete thermalization in those
collisions \cite{kolbheinz} or to
their earlier transition to a viscous hadronic phase
\cite{Gyulassy}.  These
interpretations suggest that the observed energy-dependence of
flow (Fig.~\ref{v2-vs-energy}) is dominated by the
complex dynamics of early
thermalization and late hadronic
interactions.  While application of hydrodynamics relies on local
thermal equilibrium, it is not obvious that agreement with data
after parameter adjustment necessarily proves thermalization. The
following question is posed in this light.

\vspace*{1mm}
\begin{itemize}

\item[$\bullet$] {\bf The unprecedented success of hydrodynamics
calculations assuming ideal relativistic fluid behavior in
accounting for RHIC elliptic flow results has been interpreted as
evidence for both early attainment of local thermal equilibrium
and an Equation of State with a soft point, characteristic of the
predicted phase transition.  How do we know that the observed
elliptic flow can't result, alternatively, from a harder EOS
coupled with incomplete or late thermalization
and/or significant viscosity in
the produced matter? }

\end{itemize}
\vspace*{1mm}

Even if we {\it assume} thermalization (and hence the
applicability of hydrodynamics), it is clear that a complete
evaluation of the ``theoretical error bars'' has yet to be
performed. When parameters are adjusted to reproduce spectra,
agreement with $v_2$ measurements in different centrality bins is
typically at the 20-30\% level. The continuing systematic
discrepancies from HBT results, and from the energy dependence of
elliptic flow when simplified
freezeout parameterizations are applied, suggest some level of
additional ambiguity from the
treatment of late-stage hadronic
interactions and from possibly faulty assumptions of the usual
hydrodynamics calculations (see Sec.~2.2). When theoretical
uncertainties within hydrodynamics are fairly treated, does a
convincing signal for an EOS with a soft point survive?

\vspace*{1mm}
\begin{itemize}

\item[$\bullet$] {\bf The indirect evidence for a phase transition
of some sort in the elliptic flow results comes primarily from the
sensitivity in hydrodynamics calculations of the magnitude and
hadron mass-dependence of $v_2$ to the EOS.  How does the level of
this EOS sensitivity compare quantitatively to that of
uncertainties in the calculations, gleaned from the range of
parameter adjustments, from the observed deviations from the
combination of elliptic flow, spectra and HBT correlations, and
from the sensitivity to the freezeout treatment and to such
normally neglected effects as viscosity and boost non-invariance?
}

\end{itemize}
\vspace*{1mm}

\subsubsection{Jet quenching and high gluon density}

The parton energy loss treatments do not directly distinguish
passage through confined vs. deconfined systems. Although effects
of deconfinement must exist at some level, \emph{e.g.}, on the
propagation of radiated soft gluons, their inclusion in the energy
loss models might well be quantitatively masked by other
uncertainties in the calculations. Evidence of deconfinement must
then be indirect, via comparison of the magnitude of inferred
gluon or energy densities early in the collision to those
suggested by independent partonic treatments such as gluon
saturation models. The actual energy loss inferred from fits to
RHIC data, through the rapidly expanding collision matter, is only
slightly larger than that indicated through static cold nuclei by
fits to semi-inclusive deep inelastic scattering data.  The
significance of the results is then greatly magnified by the
correction to go from the expanding collision matter to an
equivalent static system at the time of the initial hard
scattering.  The quantitative uncertainties listed in the question
below will then be similarly magnified.  What, then, is a
reasonable guess of the range of initial gluon or energy densities
that can be accommodated, and how does one demonstrate that those
densities can only be reached in a deconfined medium?

\vspace*{1mm}
\begin{itemize}

\item[$\bullet$] {\bf
    Does the magnitude of the parton energy loss inferred from RHIC hadron
    suppression observations \emph{demand} an explanation in terms of traversal
    through deconfined matter?  The answer must take into account quantitative
    uncertainties in the energy loss treatment arising, for example, from the
    uncertain applicability of factorization in-medium, from potential differences
    (other than those due to energy loss) between in-medium and vacuum
    fragmentation, and from effects of the expanding matter and of energy loss
    of the partons through cold matter preceding the hard scattering.
}

\end{itemize}
\vspace*{1mm}

Gluon saturation models set a QCD scale for anticipated gluon
densities, that can then be compared to values inferred from
parton energy loss treatments, modulo the questions asked above
and below.  An important question,
given that RHIC multiplicity data are used as input to the models
(\emph{e.g.}, to fix the proportionality between gluon density and
hadron yields) is whether they provide information truly
independent from the initial energy density inferred via the
simple Bjorken hydrodynamic expansion scenario (Eq.~\ref{epsilon})
from measured rapidity densities of transverse energy.

\vspace*{1mm}
\begin{itemize}

\item[$\bullet$] {\bf If there is a truly universal gluon density
saturation scale, determined already from HERA e-p deep inelastic
scattering measurements, why can't RHIC A+A particle
multiplicities be predicted a
priori without input from RHIC experimental data?  Is not the A-
(or $N_{part}$-) dependence of the gluon densities at the relevant
Bjorken x-ranges predicted in gluon saturation treatments? Can
saturated entrance-channel gluon densities in overlapping cold
nuclei be directly compared to the early gluon densities in
thermalized hot matter, inferred from parton energy loss
treatments of jet quenching?}

\end{itemize}
\vspace*{1mm}

% next, add the contents of the file ``Remainder.tex''

%%\end{document}

\newpage
\section{Overview and Outlook}

\subsection{What have we learned from the first three years of RHIC measurements?}

Already in their first three years, all four RHIC experiments have
been enormously successful in producing
a broad array of
high-quality data illuminating the dynamics of heavy-ion
collisions in a new regime of very high energy densities.
STAR, in particular, has
established a number of seminal, striking results highlighted in
Secs. 3 and 4 of this document. In parallel, there have been
significant advances in the theoretical treatment of these
collisions.  The theory-experiment
comparison indicates that central Au+Au collisions at RHIC produce
a unique form of strongly interacting matter, with some dramatic
and surprisingly simple properties. A number of the most striking
experimental results have been
described to a reasonable quantitative level, and in some cases
even predicted beforehand, using theoretical treatments inspired
by QCD and based on QGP formation in the early stages of the
collisions.

The observed hadron spectra and correlations at RHIC reveal three
transverse momentum ranges with distinct behavior: a soft range
($p_T \lesssim$1.5 GeV/c) containing the vast majority of produced
hadrons, representing most of the remnants of the bulk collision
matter; a hard-scattering range ($p_T \gtrsim$ 6 GeV/c), providing
partonic probes of the early collision matter; and an intermediate
range ($1.5 \lesssim p_T \lesssim 6$ GeV/c) where hard processes
coexist with softer ones. The behavior in each of these ranges is
quite different than would be expected from an incoherent sum of
independent nucleon-nucleon collisions; for the hard sector, in
particular, this is one of the most important new observations at
RHIC. Below we summarize the major findings described in earlier
chapters within each of these three ranges, in each case listing
them in approximate decreasing order of what we judge to be their
level of robustness with respect to current experimental and
theoretical ambiguities. This is not intended necessarily to
represent order of importance, as some of the presently
model-dependent conclusions are among the strongest arguments in
favor of QGP formation.

\subsubsection{Soft sector}

\begin{figure}[thb]
\begin{center}
\epsfig{figure=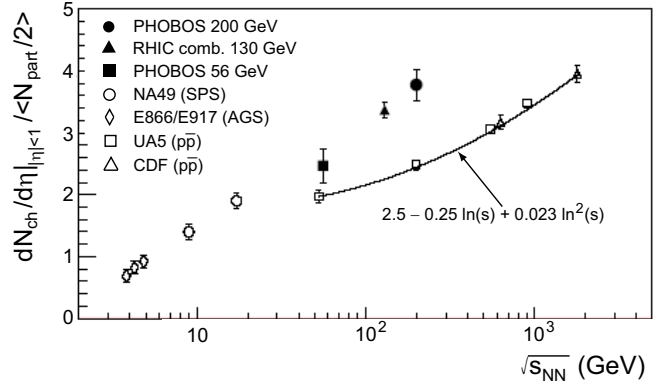,width=10.0cm}
\caption{ {\it Measured mid-rapidity charged particle densities,
scaled by the calculated number of participant nucleons, for
central collisions of $A \sim 200$ nuclei at AGS, SPS and RHIC,
plotted as a function of the center-of-mass energy.  Results for
$\overline{p} + p$ collisions are shown for comparison.  Figure
from \cite{phobos03}. } } \label{dNdeta}
\end{center}
\end{figure}

\begin{figure}[thb]
\begin{center}
\epsfig{figure=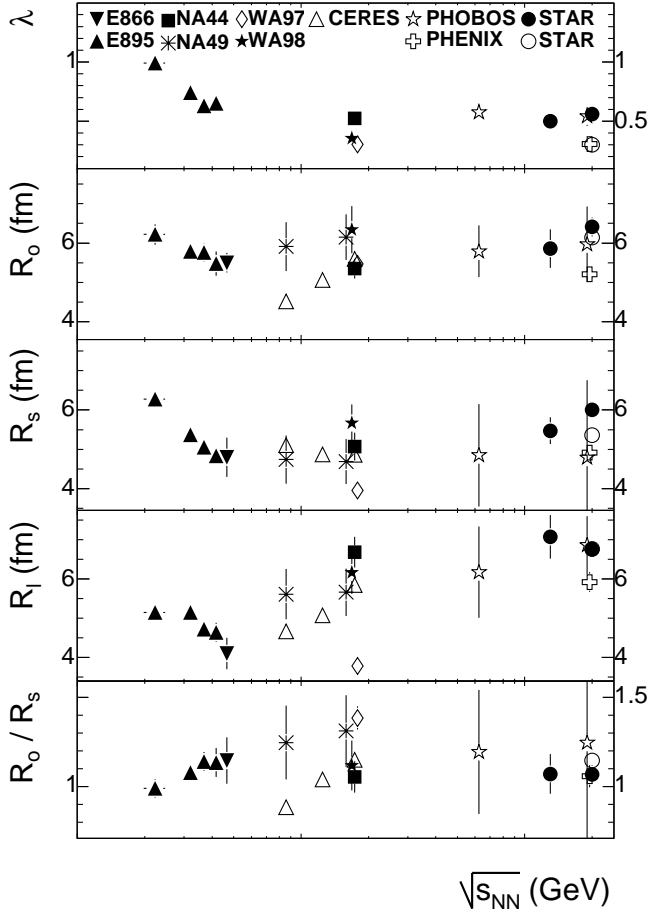,width=9.0cm} \caption{ {\it
Energy dependence of HBT parameters extracted from pion pair
correlations in central $A+A$ ($A \sim 200$) collisions at
mid-rapidity and pair $k_T \approx 0.2$ GeV/c.  The data span the
AGS, SPS and RHIC.  Figure from \cite{HBT_PRC}.} }
\label{HBT-vs-energy}
\end{center}
\end{figure}

\begin{itemize}

\item[$\bullet$] The matter produced exhibits \textbf{strong
collective flow}: most hadrons at low $p_T$ reflect a communal
transverse velocity field resulting from conditions early in the
collision, when the matter was clearly expanding rapidly under
high, azimuthally anisotropic, pressure gradients and frequent
interactions among the constituents. The commonality of the
velocity is clearest from the systematic dependence of elliptic
flow strength on hadron mass at low $p_T$ (see Fig.~\ref{v2low}),
from the common radial flow velocities extracted by fitting
observed spectra (Fig.~\ref{figtbeta}), and from the measurements
of HBT and non-identical particle correlations~\cite{non-ident}.
All of these features fit naturally, at least in a qualitative
way, within a hydrodynamic description of the system evolution.

\item[$\bullet$] Most bulk properties measured appear to fall on
quite smooth curves with similar results from lower-energy
collisions. Examples shown include
features of integrated two-hadron
$p_T$ correlations (Fig.~\ref{meanpt_corr}), elliptic flow
(Fig.~\ref{v2-vs-energy}), charged particle density
(Fig.~\ref{dNdeta})  and emitting
source radii inferred from HBT analyses
(Fig.~\ref{HBT-vs-energy}). Similarly, the centrality-dependences
observed at RHIC are generally smooth (but see
Fig.~\ref{mini-jets} for a possible exception).
These experimental results
contrast with theoretical speculations and predictions made before
RHIC start-up, which often
\cite{Harris,kolbsollfrank,gyulassy-hbt} suggested strong
energy dependences accompanying the hadron-to-QGP transition.  The
observed smooth general behavior has been primarily attributed to
the formation of matter over a range of initial local conditions,
even at a given collision energy or centrality, and to the absence
of any direct experimental determination of early temperature. In
any case, the results clearly highlight \textbf{the difficulty of
observing any rapid ``smoking-gun" onset of a transition to a new
form of matter}.

\item[$\bullet$] Despite the smoothness of the energy and
centrality dependences, two important milestones related to the
attainment of thermal equilibrium appear to be reached for the
first time in near-central RHIC collisions at or near full energy.
The first is that \textbf{the yields of different hadron species,
\emph{up to and including multi-strange hadrons}, become
consistent with a Grand Canonical statistical distribution} at a
chemical freezeout temperature of $160 \pm 10$ MeV and a baryon
chemical potential $\approx 25$ MeV (see Fig.~\ref{ratio}).  This
result places an effective lower limit on the temperatures
attained if thermal equilibration is reached during the collision
stages preceding this freezeout.  This lower limit is
\textbf{essentially equal to the QGP transition temperature
predicted by lattice QCD calculations} (see
Fig.~\ref{LQCD-pressure}).

\item[$\bullet$] At the same time (\emph{i.e.}, for near-central
RHIC collisions) the mass- and $p_T$-dependence of the observed
hadron spectra and of the strong elliptic flow in the soft sector
become \textbf{consistent, at the $\pm 20-30\%$ level, with
hydrodynamic expectations for an \emph{ideal} relativistic fluid}
formed with an initial eccentricity characteristic of the impact
parameter. These hydrodynamic calculations have not yet succeeded
in also quantitatively explaining the emitting
hadron source size inferred from
measured HBT correlations (see Fig.~\ref{hbt1}). Nonetheless,
their overall success suggests that the interactions among
constituents in the initial stages of these near-central
collisions are characterized by very short mean free paths,
leading to \textbf{quite rapid ($\tau \lesssim 1$ fm/c) attainment
of at least approximate local thermal equilibrium}. The short mean
free path in turn suggests a very dense initial system.

\item[$\bullet$] Based on the rapid attainment of thermal
equilibrium, and making the assumption of longitudinal
boost-invariant expansion, one can extract \cite{BjorkenHydro} a
rough
lower bound on the initial energy density from measured
rapidity densities \cite{phenix:et130,star:et200} of the total transverse
energy ($dE_T/dy$) produced in the collisions. These estimates
suggest that in central Au+Au collisions at RHIC, \textbf{matter
is formed at an initial energy density well above the critical
density} ($\sim 1.0$ GeV/fm$^3$) predicted by LQCD for a
transition to the QGP.

\item[$\bullet$] Measurements of two-hadron angular correlations
and of the power spectrum of local charged-particle density
fluctuations reveal strong near-side
correlations surviving in the soft sector, reminiscent of jet-like
behavior in some aspects, but with a strong pseudorapidity broadening
introduced by the presence of the collision matter.
The observed structure (see Fig.~\ref{mini-jets}) suggests that
\textbf{soft jet fragments are not fully thermalized with the bulk
matter, but nonetheless show the effects of substantial coupling
to that matter} in a considerable broadening of the jet ``peak" in
pseudorapidity difference between two hadrons.

\item[$\bullet$] Hydrodynamics calculations are best able to
reproduce RHIC results for hadron spectra and the magnitude and
mass-dependence of elliptic flow (Fig.~\ref{v2low}) by utilizing
\textbf{an Equation of State incorporating a soft LQCD-inspired
phase transition from QGP to hadronic matter}. However, the
calculations also exhibit
comparable sensitivity to other
\emph{a priori} unknown features, \emph{e.g.}, the details of the
hadronic final-state interactions and the time at which thermal
equilibrium is first attained. In light of these competing
sensitivities, it is not yet clear if the experimental results
truly \emph{demand} an EOS with a soft point.

\end{itemize}

\subsubsection{Intermediate sector}

\begin{itemize}

\item[$\bullet$] In the intermediate $p_T$ range, the elliptic
flow strength $v_2$ saturates and we see systematic meson vs.
baryon differences (rather than a systematic mass-dependence) in
both yield (see Fig.~\ref{rcp}) and $v_2$ value (Figs.~\ref{v21}).
In the same region we also observe clear jet-like angular
correlation peaks in the near-side azimuthal difference
distributions between pairs of hadrons (see
Fig.~\ref{fig:CorrelationsPR}). The most natural interpretation
for this combination of characteristics is that \textbf{the
intermediate-$p_T$ yield arises from a mixture of partonic
hard-scattering (responsible for the jet-like correlations) and
softer processes (responsible for the meson-baryon differences) }.

\item[$\bullet$] \textbf{The saturated $v_2$ values appear to
scale with the number of constituent
(or valence) quarks $n$ in the
hadron studied}, \emph{i.e.}, $v_2/n$ vs. $p_T/n$ falls on a
common curve for mesons and baryons (see Fig.~\ref{v21}). If this
trend persists as the particle-identified intermediate-$p_T$ data
are improved in statistical precision for a suitable variety of
hadron types, it would provide direct experimental evidence for
the relevance of sub-hadronic
degrees of freedom in determining flow for hadrons produced at
moderate $p_T$ in RHIC collisions.

\item[$\bullet$] Quark recombination models are able to provide a
reasonable account of the observed meson and baryon spectra, as
well as the $v_2$ systematics, in the intermediate sector by
\textbf{a sum of contributions from coalescence of thermalized
constituent quarks following an exponential $p_T$ spectrum and
from fragmentation of initially hard-scattered partons with a
power-law spectrum} \cite{MuellerRBRC}.  It is not yet clear if
the same mixture can also account quantitatively for the azimuthal
dihadron correlation (including background under the jet-like
peaks) results as a function of $p_T$. Other models
\cite{recom03,Hwa} mix the above contributions by also invoking
recombination of hard-scattered with thermal partons.

\end{itemize}

\subsubsection{Hard sector}

\begin{figure}[thb]
\begin{center}
\epsfig{figure=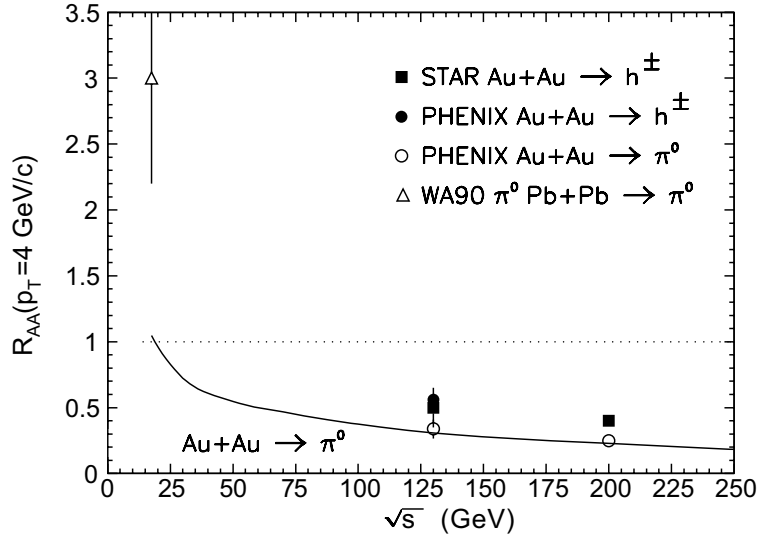,width=10.0cm} \caption{ {\it
The nuclear modification factor measured for 4 GeV/c hadrons in
central $A+A$ ($A \sim 200$) collisions at SPS and two RHIC
energies, showing (Cronin) enhancement at the lower energy and
clear jet-quenching suppression at RHIC.  The small difference
between RHIC charged hadron and identified $\pi^0$ results
reflects meson vs. baryon differences in this $p_T$ range.  The
solid curve represents a parton energy loss calculation under
simplifying assumptions concerning the energy-dependence, as
described in \cite{Wang:2003mm}.}} \label{RAA-vs-energy}
\end{center}
\end{figure}

\begin{itemize}

\item[$\bullet$] The dominant characteristic of the hard regime is
\textbf{the strong suppression of hadron yields in central Au+Au
collisions}, in comparison to expectations from p+p or peripheral
Au+Au collisions, scaled by the number of contributing binary
(nucleon-nucleon) collisions (see
Fig.~\ref{fig:HadronSuppression}). Such suppression sets in
already in the intermediate sector, but saturates and remains
constant as a function of $p_T$ throughout the hard region
explored to date. Such suppression was not seen in d+Au collisions
at RHIC, indicating that it is \textbf{a final-state effect
associated with the collision matter produced in Au+Au}. It is
consistent with effects of parton energy loss in traversing dense
matter, predicted before the data were available
\cite{Wang:2003mm,Vitev:2002pf}.

\item[$\bullet$] Azimuthal correlations of moderate- (see
Fig.~\ref{fig:CorrelationsPR}) and high-$p_T$
\cite{Hardtke:2002ph} hadrons exhibit clear jet-like peaks on the
near side. However, \textbf{the anticipated away-side peak
associated with dijet production is suppressed} by progressively
larger factors as the Au+Au centrality is increased, and for given
centrality, as the amount of (azimuthally anisotropic) matter
traversed is increased (see Fig.~\ref{fig:CorrelationsPR}). Again,
no such suppression is observed in d+Au collisions.  The
suppression of hadron yields and back-to-back correlations firmly
establish that \textbf{jets are quenched by very strong
interactions with the matter produced in central Au+Au
collisions}.  The jet-like near-side correlations survive
presumably because one observes preferentially hard fragments of
partons scattered outward from the surface region of the collision
zone. Effects of interaction with the bulk matter are nonetheless
still seen on the near side, primarily by the broadened
distribution in pseudorapidity of softer correlated fragments (see
Fig.~\ref{mini-jets} and Ref.~\cite{star_fqwang}).

\item[$\bullet$] Many features of the observed suppression of
high-$p_T$ hadrons, including the centrality-dependence and the
$p_T$-independence, can be described efficiently by
\textbf{perturbative QCD calculations incorporating parton energy
loss} in a thin, dense medium (see
Fig.~\ref{fig:HadronSuppression}). To reproduce the magnitude of
the observed suppression, despite the rapid expansion of the
collision matter the partons traverse, these treatments need to
assume that \textbf{the initial gluon density when the collective
expansion begins is more than an order of magnitude greater than
that characteristic of cold, confined nuclear matter}
\cite{Wang:2003mm}. The inferred gluon density is consistent, at a
factor $\sim 2$ level, with the saturated densities needed to
account for RHIC particle multiplicity results in gluon saturation
models (see Fig.~\ref{mult}).

\item[$\bullet$] The yields of hadrons at moderate-to-high $p_T$
in central d+Au collisions exhibit a systematic dependence on
pseudorapidity, marked by \textbf{substantial suppression, with
respect to binary scaling expectations, of products near the
deuteron beam direction, in contrast to substantial enhancement of
products at mid-rapidity and near the Au beam direction}
(see
Figs.~\ref{Brahms:dAuForwardSuppression} and
\ref{STAR:dAuRapidityRatios}).  This pattern suggests a depletion
of gluon densities at low Bjorken $x$ in the colliding Au nucleus,
and is \textbf{qualitatively consistent with predictions of gluon
saturation models}. Measurements to date cannot yet distinguish
interactions with a classical gluon field
(Color Glass Condensate) from
interactions with a more conventionally shadowed density of
individual gluons.

\item[$\bullet$] Angular correlations between moderate-$p_T$ and
soft hadrons have been used to explore how transverse momentum
balance is achieved, in light of jet quenching, opposite a
high-$p_T$ hadron in central Au+Au collisions.  The results show
the balancing hadrons to be significantly larger in number, softer
(see Fig.~\ref{jetloss}) and more widely dispersed in angle
compared to p+p or peripheral Au+Au collisions, with
\textbf{little remnant of away-side jet-like behavior}. To the
extent that hard scattering dominates these correlations
at moderate and low $p_T$, the
results could signal an approach of the away-side parton toward
thermal equilibrium with the bulk matter it traverses.  As
mentioned earlier, progress toward thermalization of jet fragments
on the near-side is also suggested
by soft-hadron correlations.

\item[$\bullet$] The hard sector was not accessed in SPS
experiments, so any possible energy dependence of jet quenching
can only be explored via the hadron nuclear modification factor in
the intermediate-$p_T$ range.
While the results (see
Fig.~\ref{RAA-vs-energy}) leave open the possibility of a rapid
transition \cite{Wang}, one is not
expected on the basis of theoretical studies of parton energy loss
\cite{Baier}.  Furthermore, serious questions have been raised
\cite{D'Enterria} about the validity of the p+p reference data
used to determine the SPS result in the figure.

\end{itemize}

In summary, the RHIC program has enabled
dramatic advances in the study of
hot strongly interacting matter, for two basic reasons. With the
extended reach in initial energy density, the matter produced in
the most central RHIC collisions appears to have attained
conditions that considerably simplify its theoretical treatment:
essentially ideal fluid expansion, and approximate local thermal
equilibrium beyond the LQCD-predicted threshold for QGP formation.
With the extended reach in particle momentum, the RHIC experiments
have developed probes for behavior that was difficult to access at
lower collision energies: jet quenching and apparent constituent
quark scaling of elliptic flow. These results indicate, with
fairly modest reliance on theory, that RHIC collisions produce
highly opaque and dense matter
that behaves collectively. The magnitude of the density inferred
from parton energy loss treatments, together with the hints of
constituent quark collective flow, argue against the effectiveness
of a purely hadronic treatment of this
unique strongly interacting
matter.  It appears from the most
robust signals to evolve for a significant fraction of its
lifetime as a low-viscosity, pre-hadronic liquid.

If one takes seriously all of the theoretical successes mentioned
above, they suggest the following
more detailed overall picture of
RHIC collisions: Interactions of very short mean free path within
the gluon density saturation regime lead to a rapidly thermalized
partonic system at energy densities and temperatures above the
LQCD critical values. This thermalized matter expands collectively
and cools as an ideal fluid, until the phase transition back to
hadronic matter begins, leading to a significant pause in the
build-up of elliptic flow.  During the phase transition,
constituent quarks emerge as the effective degrees of freedom in
describing hadron formation at medium $p_T$ out of this initially
partonic matter. Initially hard-scattered partons (with lower
color interaction cross sections than the bulk partons) traversing
this matter lose substantial energy to the medium via gluon
radiation, and thereby approach, but do not quite reach,
equilibration with the bulk matter.  Thus, some evidence of
degraded jets survives (\emph{e.g.}, see
Fig.~\ref{fig:CorrelationsPR}), depending on the amount of matter
traversed.  Any claim of QGP discovery based on RHIC results to
date requires an assessment of the robustness, internal
consistency, quantitative success and predictive power of this
emerging picture.

\subsection{Are we there yet?}

The consistency noted above of many RHIC results with a QGP-based
theoretical framework is an important and highly non-trivial
statement! Indeed, it is the basis of some claims
\cite{Gyulassy,McLerran-Gyulassy,RBRC} that the Quark-Gluon Plasma
has already been discovered at
RHIC.  However, these claims are
associated with QGP definitions \cite{Gyulassy,McLerran-Gyulassy}
that do not specifically highlight deconfinement as an essential
property to be demonstrated. In our judgment,
for reasons mentioned below, and
also reflected in the list of open questions provided in Chap. 5
of this document, it is premature
to conclude definitively that the matter produced in central RHIC
collisions is a Quark-Gluon Plasma, as this term has been
understood by the scientific community for the past 20 years (see
Appendix B).

\begin{itemize}

\item[$\bullet$] The RHIC experiments have not yet produced
\emph{direct} evidence for deconfinement, or indeed for any
clear transition in thermodynamic
properties of the matter produced.
It may be unreasonable to expect a
clear onset of deconfinement in heavy-ion systems as a function of
collision energy, because the matter, even if locally thermalized,
is presumably formed over a range of initial temperatures at any
given collision energy.  Thus, in the emerging theoretical
picture, the matter produced in heavy-ion collisions at SPS was
probably also formed in part above the critical energy density,
but over a smaller fraction of the volume and with shorter-lived
(or perhaps never attained) thermal equilibrium, in comparison
with RHIC collisions.  At still lower collision energies, where
the critical conditions might never be reached, various aspects of
the theoretical framework applied at RHIC become inapplicable,
precluding a simple theory-experiment comparison over a range from
purely hadronic to allegedly QGP-dominated matter.

\item[$\bullet$] The indirect evidence for a
thermodynamic transition and for
attainment of local thermal equilibrium in the matter
produced at RHIC are intertwined
in the hydrodynamics account for
observed hadron spectra and
elliptic flow results.  The uniqueness of the solution involving
early thermalization and an EOS with a soft mixed phase is not yet
demonstrated.  Nor is its robustness against changes in the
treatment of the late hadronic
stage of the evolution, including the introduction of viscosity
and other modifications that
might be needed to reduce discrepancies from HBT measurements.

\item[$\bullet$] The indirect evidence for deconfinement rests
primarily on the large initial gluon densities inferred from
parton energy loss fits to the observed hadron suppression at high
$p_T$, and on the supposition that
such high densities could only be achieved in deconfined matter.
The latter supposition has yet to be demonstrated in a compelling
theoretical argument.  The agreement with initial gluon densities
suggested by Color Glass Condensate approaches is encouraging, but
is still at a basically qualitative level.
The measurements suggest that
matter is formed at initial temperatures and energy densities at
or above the critical values predicted by LQCD for a deconfinement
transition. But they do not establish the detailed relevance of
the lattice calculations to the fleeting dynamic matter produced
in heavy-ion collisions.

\item[$\bullet$] The role of collectively flowing constituent
quarks in hadron formation at
intermediate $p_T$ is not yet well established experimentally. If
it becomes so established by subsequent measurements and analyses,
this will hint at the existence of a collective, thermalized
partonic stage in the system evolution.
However, that hint will fall short
of a conclusive QGP demonstration until some interpretational
ambiguities are resolved: Is it really \emph{constituent}, rather
than \emph{current} (valence) quarks that coalesce?  If the
former, do the constituent quarks then merely represent the
effective degrees of freedom for hadronization of a QGP, or do
they indicate an intermediate, pre-hadronic evolutionary stage,
after the abundant gluons and current quarks have coalesced and
dynamical chiral symmetry breaking has been re-introduced? If
there is a distinct constituent quark formation stage, is
thermalization achieved before, or only during, that stage?

\item[$\bullet$] The theory remains a patchwork of different
treatments applied in succession to each stage of the
collision
evolution, without yet a clear
delineation of the different aspects as distinct limits of one
overarching, seamless theory. The theoretical claims of QGP
discovery in \cite{RBRC}, considered together, rely on five
``pillars of wisdom" for RHIC central Au+Au collisions, and each
invokes a separate model or theoretical approach for its
interpretation: (i) statistical model fits to measured hadron
yields to infer possible chemical equilibrium across the $u, ~d$
and $s$ sectors; (ii) hydrodynamics calculations of elliptic flow
to suggest early thermalization and soft EOS; (iii) quark
recombination models to highlight the role of thermalized
constituent quarks in intermediate-sector $v_2$ scaling; (iv)
parton energy loss models to infer an initial gluon density from
high-$p_T$ hadron suppression observations; (v) gluon saturation
model fits to observed hadron multiplicities and yields at large
rapidity, to suggest how high-density QCD may predetermine the
achieved initial gluon densities.  Each
movement of the
theoretical suite has its own
assumptions, technical difficulties, adjusted parameters and
quantitative uncertainties, and they fit together somewhat
uneasily.  Until they are assimilated into a
more self-consistent whole with
only a few overall parameters fitted to existing data, it may be
difficult to assess theoretical uncertainties quantitatively or to
make non-trivial quantitative predictions whose comparison with
future experimental results have the potential to prove the theory
wrong.

\end{itemize}

\textbf{The bottom line is that in the absence of a direct
``smoking gun" signal of deconfinement revealed by experiment
alone, a QGP discovery claim must rest on the comparison with a
promising, but still not yet mature, theoretical framework.  In
this circumstance, clear predictive power with quantitative
assessments of theoretical uncertainties are necessary for the
present appealing picture to survive as a lasting one.
The matter produced in RHIC
collisions is fascinating and unique.  The continuing delineation
of its properties will pose critical tests for the theoretical
treatment of non-perturbative QCD.  But we judge that a QGP
discovery claim based on RHIC measurements to date would be
premature.}  We do not propose that a comprehensive theoretical
understanding of all observed phenomena must be attained before a
discovery claim is warranted, but only that at least some of the
serious open questions posed above and in Sec. 5 be successfully
answered.

\subsection{What are the critical needs from future experiments?}

The above comments make it clear what is needed most urgently from
theory. But how can future measurements, analyses and heavy-ion
collision facilities bring us to a
clearer delineation of the fundamental properties of the unique
matter produced, and hopefully to a more definitive conclusion
regarding the formation of a Quark-Gluon Plasma?  We briefly
describe below the goals of some important anticipated programs,
separated into short-term and long-term prospects, although the
distinction in time scale is not always sharp. In the short term,
RHIC measurements should concentrate on verifying and extending
its new observations of jet quenching and $v_2$ scaling; on
testing quantitative predictions
of theoretical calculations
incorporating a QGP transition at lower energies and for
different system sizes; on measuring charmed-hadron and charmonium
yields and flow to search for other evidence of deconfinement; and
on testing more extensive predictions of gluon saturation models
for forward hadron production. Some of the relevant data have
already been acquired during the highly successful 2004 RHIC run
-- which has increased the RHIC database by an order of magnitude
-- and simply await analysis, while other measurements require
anticipated near-term upgrades of the detectors.  In the longer
term, the LHC will become available to provide crucial tests of
QGP-based theoretical extrapolations to much higher energies, and
to focus on very high $p_T$ probes of collision matter that is
likely to be formed deep into the gluon saturation regime. Over
that same period, RHIC should provide the extended integrated
luminosities and upgraded detectors needed to undertake
statistically challenging measurements to probe directly the
initial system temperature, the
pattern of production yields among various heavy quarkonium
species, the quantitative energy loss of partons traversing
the early collision matter, and
the fate of strong-interaction symmetries in
that matter.

Important short-term goals include the following:

\begin{itemize}

\item[$\bullet$] \textbf{Establish $v_2$ scaling more
definitively.} Extend the particle-identified flow measurements
for hadrons in the medium-$p_T$ region over a broader $p_T$ range,
a wider variety of hadron species, and as a function of
centrality.  Does the universal curve of $v_2/n$ vs. $p_T/n$
remain a good description of all the data?  How is the scaling
interpretation affected by anticipated hard contributions
associated with differential jet quenching through spatially
anisotropic collision matter? Can the observed di-hadron angular
correlations be quantitatively accounted for by a 2-component
model attributing hadron production in this region to quark
coalescence (with correlations reflecting only the collective
expansion) plus fragmentation (with jet-like correlations)?  Do
hadrons such as $\phi$-mesons or $\Omega$-baryons, containing no
valence $u$ or $d$ quarks, and hence with quark-exchange contributions
to hadronic interaction cross sections suppressed in normal nuclear matter,
follow the
same flow trends as other hadrons?  Do the measured $v_2$ values
for resonances reflect their constituent quark, or rather their
hadron, content?  These investigations have the potential to
establish more clearly that constituent quarks exhibiting
collective flow are the relevant degrees of freedom for
hadronization at medium $p_T$.

\item[$\bullet$] \textbf{Establish that jet quenching is an
indicator of parton, and not hadron, energy loss.} Extend the
measurements of hadron energy loss and di-hadron correlations to
higher $p_T$, including particle identification in at least some
cases.  Do the meson-baryon suppression differences seen at lower
$p_T$ truly disappear?  Does the magnitude of the suppression
remain largely independent of $p_T$, in contrast to expectations
for hadron energy loss \cite{Wang}?  Does one begin to see a
return of away-side jet behavior, via punch-through of correlated
fragments opposite a higher-$p_T$ trigger hadron? Improve the
precision of di-hadron correlations with respect to the reaction
plane, and extend jet quenching measurements to lighter colliding
nuclei, to observe the non-linear dependence on distance
traversed, expected for radiating partons \cite{GVWZ}.  Measure
the nuclear modification factors for charmed meson production, to
look for the ``dead-cone" effect predicted \cite{Djordjevic} to
reduce energy loss for heavy quarks.

\item[$\bullet$] \textbf{Extend RHIC Au+Au measurements down
toward SPS measurements in energy,
to test quantitative predictions
of the energy-dependence.}  Does
the suppression of high-$p_T$ hadron yields persist, and does it
follow the gentle energy-dependence predicted in
Fig.~\ref{RAA-vs-energy}? Do the gluon densities inferred from
parton energy loss model fits to hadron yields follow
energy-dependent trends expected from gluon saturation models?
Does elliptic flow remain in agreement with calculations that
couple expansion of an ideal partonic fluid to a late-stage,
viscous hadron cascade?  Do meson-baryon differences and
indications of constituent-quark scaling persist in hadron yields
and flow results at intermediate $p_T$?  Do quark coalescence
models remain viable, with inferred thermal quark spectra that
change sensibly with the (presumably) slowly varying initial
system temperatures?  The study of the evolution with collision energy
of differential measurements such as those
in Fig.~\ref{mini-jets} promises to yield important insight into
the dynamical processes which occur during system evolution.

\item[$\bullet$] \textbf{Measure charmonium yields and open charm
yields and flow, to search for signatures of color screening and
partonic collectivity.} Use particle yield ratios for charmed
hadrons to determine whether the apparent thermal equilibrium in
the early collision matter at RHIC extends even to quarks with
mass significantly greater than the anticipated system
temperature.  From the measured $p_T$ spectra, constrain the
relative contributions of coalescence vs. fragmentation
contributions to charmed-quark hadron production. Compare D-meson
flow to the trends established in the $u, d$ and $s$ sectors, and
try to extract the implications for flow contributions from
coalescence vs. possibly earlier partonic interaction stages of
the collision.  Look for the extra suppression of charmonium,
compared to open charm, yields expected to arise from the strong
color screening in a QGP state (see Fig.~\ref{LQCD-screening}).

\item[$\bullet$] \textbf{Measure angular correlations with far
forward high-energy hadrons in d+Au or p+Au collisions.} Search
for the mono-jet signature anticipated for quark interactions with
a classical (saturated) gluon field, as opposed to di-jets from
quark interactions with individual gluons.  Correlations among two
forward hadrons are anticipated to provide the best sensitivity to
the gluon field at sufficiently low Bjorken $x$ to probe the
possible saturation regime.

\end{itemize}

Longer-term prospects, requiring much greater integrated
luminosities (as anticipated at
RHIC II) or other substantial
facility developments, include:

\begin{itemize}

\item[$\bullet$] \textbf{Develop thermometers for the early stage
of the collisions, when thermal equilibrium is first established.}
In order to pin down experimentally where a
thermodynamic transition may
occur, it is critical to find probes with direct sensitivity to
the temperature well before chemical freezeout.  Promising
candidates include probes with
little final-state interaction: direct photons -- measured down
to low momentum, for example, via $\gamma-\gamma$ HBT, which is
insensitive to the large $\pi^0$ background -- and thermal
dileptons.  The former would
require enhanced pair production tracking and the latter the
introduction of hadron-blind detectors and techniques.

\item[$\bullet$] \textbf{Measure
the yields and spectra of various heavy quarkonium species.}
Recent LQCD calculations \cite{Asakawa} predict the onset of
charmonium melting -- which can be taken as a signature for
deconfinement -- at quite different temperatures above $T_c$ for
$J/\psi$ vs. $\psi^\prime$. Similar differences are anticipated
for the various $\Upsilon$ states.  While interpretation of the
yield for any one quarkonium species may be complicated by
competition in a QGP state between enhanced heavy quark production
rates and screened quark-antiquark interactions, comparison of a
measured hierarchy of yields with LQCD expectations would be
especially revealing. They would have to be compared to measured
yields for open charm and beauty, and to the corresponding
quarkonium production rates in p+p and p+A collisions. Clear
identification of $\psi^\prime$ and separation of $\Upsilon$
states require upgrades to detector resolution and vertexing
capabilities.

\item[$\bullet$] \textbf{Quantify parton energy loss by
measurement of mid-rapidity jet fragments tagged by a hard direct
photon, a heavy-quark hadron, or a far forward energetic hadron.}
Such luminosity-hungry coincidence measurements will elucidate the
energy loss of light quarks vs. heavy quarks vs. gluons,
respectively, through the collision matter. They should thus
provide more quantitative sensitivity to the details of parton
energy loss calculations.

\item[$\bullet$] \textbf{Test quantitative predictions for
elliptic flow in U+U collisions.} The large size and deformation
of uranium nuclei make this a considerable extrapolation away from
RHIC Au+Au conditions, and a significant test for the details of
hydrodynamics calculations that are consistent with the Au+Au
results \cite{kolbheinz}.  If the relative alignment of the
deformation axes of the two uranium nuclei can be experimentally
controlled, one would be able to vary initial spatial eccentricity
largely independently of centrality and degree of thermalization
of the matter.

\item[$\bullet$] \textbf{Measure hadron multiplicities, yields,
correlations and flow at LHC and GSI energies, and compare to
quantitative predictions based on models that work at RHIC.} By
fixing parameters and ambiguous features of gluon saturation,
hydrodynamics, parton energy loss and quark coalescence models to
fit RHIC results, and with guidance from LQCD calculations
regarding the evolution of strongly interacting matter with
initial temperature and energy density, theorists should make
quantitative predictions for these observables at LHC and GSI
before the data are collected.  The success or failure of those
predictions will represent a stringent test of the viability of
the QGP-based theoretical framework.

\item[$\bullet$] \textbf{Devise tests for the fate of fundamental
QCD symmetries in the collision matter formed at RHIC.} If the
nature of the QCD vacuum is truly modified above the critical
temperature, then chiral and U$_A$(1) symmetries may be restored,
while parity and CP may conceivably be broken \cite{kharzeevCP}.
Testing these symmetries in this unusual form of strongly
interacting matter is of great importance, even if we do not have
a crisp demonstration beforehand that the matter is fully
thermalized and deconfined. Indeed, if evidence were found for a
clear change in the degree of adherence to one of the strong
interaction symmetries, in
comparison with normal nuclear matter, this would likely provide
the most compelling ``smoking gun" for production of a new form of
matter in RHIC collisions. Approaches that have been discussed to
date include looking for meson mass shifts in dilepton spectra as
a signal of chiral symmetry restoration, and searching for CP
violation via $\Lambda - \overline{\Lambda}$ spin correlations or
electric dipole distributions of produced charge with respect to
the reaction plane \cite{kharzeevCP}. It may be especially
interesting to look for evidence among particles emerging opposite
an observed high-$p_T$ hadron tag, since the strong suppression of
away-side jets argues that the fate of the away-side particles may
reflect strong interactions with a maximal amount of early
collision matter. These tests will begin in the short term, but
may ultimately need the higher statistics available in the longer
term to distinguish subtle signals from dominant backgrounds.

\end{itemize}

\subsection{Outlook}

The programs we have outlined above for desirable advances in
theory and experiment represent a decade's worth of research, not
all of which must, or are even expected to, \emph{precede} a
discovery announcement for the Quark-Gluon Plasma.
We can imagine several possible
scenarios leading to a more definitive QGP conclusion.
Identification of a single compelling experimental signature is
still conceivable, but the most promising prospects are long-term:
establishment of a telling pattern of quarkonium suppression
\emph{vs}. species; observation of clear parity or CP violation,
or of chiral symmetry restoration, in the collision matter;
extraction of a transition signal as a function of measured early
temperature. It is also possible that a single theoretical
development could largely seal the case: \emph{e.g.}, a compelling
argument that gluon densities more than an order of magnitude
higher than those in cold nuclear matter really do \emph{demand}
deconfinement; or sufficient hydrodynamics refinement to
demonstrate that RHIC flow results really do \emph{demand} a soft
point in the EOS.  Perhaps the most likely path would involve
several additional successes in theory-experiment comparisons,
leading to a preponderance of evidence that RHIC collisions have
produced thermalized, deconfined quark-gluon matter.

In any scenario, however, RHIC has
been, and should continue to be, a tremendous success in its
broader role as an instrument for discovery of new features of QCD
matter under extreme conditions.  The properties already
delineated, with seminal contributions from STAR, point toward a
dense, opaque, non-viscous, pre-hadronic liquid state that was not
anticipated before RHIC.   Determining whether the quarks and
gluons in this matter reach thermal equilibrium with one another
before they become confined within hadrons, and eventually whether
chiral symmetry is restored, are two among many profound questions
one may ask.  Further elaboration of the properties of this
matter, with eyes open to new unanticipated features, remains a
vital research mission, independent of the answer that nature
eventually divulges to the more limited question that has been the
focus of this document.

\newpage
\section{Appendix A: Charge}

This report was prepared for the
STAR Collaboration in response to the following charge from the
Spokesperson, delivered to a drafting committee on March 18,
2004.

``Thank you very much for agreeing
to help in preparing a draft whitepaper to serve as the starting
point for a focused discussion by the STAR Collaboration of the
experimental evidence regarding the role of the Quark-Gluon Plasma
in RHIC heavy ion collisions."

``The charge to this panel is to
make a critical assessment of the presently available evidence to
judge whether it warrants a discovery announcement for the QGP,
using any and all experimental and theoretical results that
address this question.  The white paper should pay particular
attention to identifying the most crucial features of the QGP that
need to be demonstrated experimentally for a compelling claim to
be made.  It should summarize those data that may already
convincingly demonstrate some features, as well as other data that
may be suggestive but with possible model-dependence, and still
other results that raise questions about a QGP interpretation. If
the conclusion is that a discovery announcement is at present
premature, the paper should outline critical additional
measurements and analyses that would make the case stronger, and
the timeline anticipated to produce those new results."

``The white paper should be of
sufficient quality and scientific integrity that, after
incorporation of collaboration comments, it may be circulated
widely within the RHIC and larger physics communities as a
statement of STAR's present assessment of the evidence for the
QGP."

\newpage
\section{Appendix B: Definitions of the Quark-Gluon Plasma in
Nuclear Physics Planning Documents}

One's conclusion concerning the
state of the evidence in support of Quark-Gluon Plasma formation
is certainly influenced by the definition one chooses for the QGP
state.  Recent positive claims have been based on definitions
different from that chosen in this work (see Sec. 1), leaning more
toward either an operational definition based on actual RHIC
measurements \cite{Gyulassy}, or a demonstration that experiments
have reached conditions under which lattice QCD calculations
predict a QGP state \cite{McLerran-Gyulassy}.  We have rather
chosen to extract what we believe to be the consensus definition
built up in the physics community from the past 20 years' worth of
planning documents and proposals for RHIC.  In this section, we
collect relevant quotes concerning the QGP from a number of these
documents.

A relativistic heavy-ion collider
facility was first established as the highest priority for new
construction by the 1983 Nuclear Science Advisory Committee (NSAC)
Long Range Plan \cite{LRP-83}. In discussing the motivation for
such a facility, that document states:

\begin{itemize}
\item[] ``Finally, under
conditions of very elevated energy density, nuclear matter will
exist in a wholly new phase in which there are no nucleons or
hadrons composed of quarks in individual bags, but an extended
\emph{quark-gluon plasma}, within which the quarks are deconfined
and move independently. ... The production and detection of a
quark-gluon plasma in ultra-relativistic heavy ion collisions
would not only be a remarkable achievement in itself, but by
enabling one to study quantum chromodynamics (QCD) over large
distance scales it would enable one to study fundamental aspects
of QCD and confinement unattainable in few-hadron experiments. ...
A second, chiral-symmetry restoring, transition is also expected
at somewhat higher energy density, or perhaps coincident with the
deconfinement transition; such a transition would be heralded by
the quarks becoming effectively massless, and low mass pionic
excitations no longer appearing in the excitation spectrum."

\end{itemize}

The high priority of such a
collider was confirmed in the 1984-6 National Academy of Sciences
survey of Nuclear Physics \cite{NAS-84}, which stated:

\begin{itemize}
\item[] ``A major scientific
imperative for such an accelerator derives from one of the most
striking predictions of quantum chromodynamics: that under
conditions of sufficiently high temperature and density in nuclear
matter, a transition will occur from excited hadronic matter to a
quark-gluon plasma, in which the quarks, antiquarks and gluons of
which hadrons are composed become `deconfined' and are able to
move about freely.  The quark-gluon plasma is believed to have
existed in the first few microseconds after the big bang, and it
may exist today in the cores of neutron stars, but it has never
been observed on Earth. Producing it in the laboratory will thus
be a major scientific achievement, bringing together  various
elements of nuclear physics, particle physics, astrophysics, and
cosmology."

\end{itemize}

The glossary of the above document
\cite{NAS-84} defined Quark-Gluon Plasma in the following way:
\begin{itemize}
\item[] ``An extreme state of
matter in which quarks and gluons are deconfined and are free to
move about in a much larger volume than that of a single hadron
bag.  It has never been observed on earth."

\end{itemize}

In the 1984 proposal for RHIC from
Brookhaven National Laboratory \cite{RHIC_proposal}, the QGP was
described as follows:
\begin{itemize}
\item[] ``The specific motivation
from QCD is the belief that we can assemble macroscopic volumes of
nuclear matter at such extreme thermodynamic conditions as to
overcome the forces that confine constituents in normal hadrons,
creating a new form of matter in an \emph{extended confined plasma
of quarks and gluons}."
\end{itemize}

The 1989 NSAC Long Range Plan
\cite{LRP-89}, in reconfirming the high priority of RHIC, states:

\begin{itemize}
\item[] ``The most outstanding
prediction based on the theory of the strong interaction, QCD, is
that the properties of matter should undergo a profound and
fundamental change at an energy density only about one order of
magnitude higher than that found in the center of ordinary nuclei.
This change is expected to involve a transition from the confined
phase of QCD, in which the degrees of freedom are the familiar
nucleons and mesons and in which a quark is able to move around
only inside its parent nucleon, to a new deconfined phase, called
the quark-gluon plasma, in which hadrons dissolve into a plasma of
quarks and gluons, which are then free to move over a large
volume."

\end{itemize}

The 1994 NSAC Assessment of
Nuclear Science \cite{NSAC-94} states:

\begin{itemize}
\item[] ``When nuclear matter is
heated to extremely high temperatures or compressed to very large
densities we expect it to respond with a drastic transformation,
in which the quarks and gluons, that are normally confined within
individual neutrons and protons, are able to move over large
distances.  A new phase of matter, called Quark-Gluon-Plasma
(QGP), is formed.  At the same time chiral symmetry is restored
making particles massless at the scale of quark masses.  Quantum
Chromodynamics (QCD) of massless quarks is chirally (or
left-right) symmetric, but in the normal world this symmetry is
spontaneously broken giving dynamical masses to quarks and the
particles composed of quarks."

\end{itemize}

The 1996 NSAC Long Range Plan
\cite{LRP-96} repeats the emphasis on chiral symmetry restoration
in addition to deconfinement:
\begin{itemize}
\item[] ``At temperatures in
excess of $T_c$ nuclear matter is predicted to consist of
unconfined, nearly massless quarks and gluons, a state called the
\emph{quark-gluon plasma}. The study of deconfinement and chiral
symmetry restoration is the primary motivation for the
construction of the Relativistic Heavy Ion Collider (RHIC) at
Brookhaven National Laboratory."

\end{itemize}

The most recent National Academy
of Sciences survey of Nuclear Physics \cite{NAS-99} puts it this
way:
\begin{itemize}
\item[] ``At RHIC such high energy
densities will be created that the quarks and gluons are expected
to become deconfined across a volume that is large compared to
that of a hadron.  By determining the conditions for
deconfinement, experiments at RHIC will play a crucial role in
understanding the basic nature of confinement and shed light on
how QCD describes the matter of the real world. ...Although the
connection between chiral symmetry and quark deconfinement is not
well understood at present, chiral symmetry is expected to hold in
the quark-gluon plasma."

\end{itemize}

Finally, the 2004 NuPECC (Nuclear
Physics European Collaboration Committee) Long Range Plan for
nuclear physics research in Europe \cite{NuPECC-04} states:

\begin{itemize}
\item[] ``The focus of the
research in the ultra-relativistic energy regime is to study and
understand how collective phenomena and macroscopic properties,
involving many degrees of freedom, emerge from the microscopic
laws of elementary particle-physics. ... The most striking case of
a collective bulk phenomenon predicted by QCD is the occurrence of
a phase transition to a deconfined chirally symmetric state, the
quark gluon plasma (QGP)."

\end{itemize}

In short, every statement
concerning the QGP in planning documents since the conception of
RHIC has pointed to deconfinement of quarks and gluons from
hadrons as the primary characteristic of the new phase.  More
recent definitions have tended to include chiral symmetry
restoration as well.  Based on the above survey, we believe that
the definition used in this paper would be very widely accepted
within the worldwide physics community as a ``minimal" requirement
for demonstrating formation of a Quark-Gluon Plasma.

\newpage
\textbf{Acknowledgements:}
We thank the RHIC Operations Group and RCF at BNL, and the NERSC Center at LBNL for their support. This work was supported in part by the HENP Divisions of the Office of Science of the U.S. DOE; the U.S. NSF; the BMBF of Germany; IN2P3, RA, RPL, and EMN of France; EPSRC of the United Kingdom; FAPESP of Brazil; the Russian Ministry of Science and Technology; the Ministry of Education and the NNSFC of China; Grant Agency of the Czech Republic, FOM of the Netherlands, DAE, DST, and CSIR of the Government of India; Swiss NSF; the Polish State Committee for Scientific Research; and the STAA of Slovakia.


\newpage
\begin{thebibliography}{00}


% notes:

% \bibitem{label2}
% If there is a note, it should come last:
% \bibitem{label3} \note
% \end{subbibitems}

\bibitem{Jacobs-Wang} P. Jacobs and X.N. Wang, hep-ph/0405125.

\bibitem{Rischke} D.H. Rischke, Prog. Part. Nucl. Phys. 52 (2004) 197.

\bibitem{kolbheinz} P. F. Kolb and U. Heinz, in
Quark Gluon Plasma 3, eds. R.C. Hwa and X.N. Wang (World
Scientific, Singapore, 2003); nucl-th/0305084.

\bibitem{GVWZ} M. Gyulassy, I. Vitev, X.N. Wang and B.W. Zhang, in
Quark Gluon Plasma 3, eds. R.C. Hwa and X.N. Wang (World
Scientific, Singapore, 2003), nucl-th/0302077;
R. Baier, D. Schiff, B.G. Zakharov, Ann. Rev. Nucl. Part. Sci. 50 (2000) 37.

\bibitem{tomasik-wiedemann} B. Tom\'a$\breve{s}$ik and U.A.
Wiedemann, in Quark Gluon Plasma 3, eds. R.C. Hwa and X.N.
Wang (World Scientific, Singapore, 2003);
hep-ph/0210250.


\bibitem{Gyulassy} M. Gyulassy, nucl-th/0403032.

\bibitem{McLerran-Gyulassy} M. Gyulassy and L. McLerran, Nucl. Phys. A750 (2005) 30.

\bibitem{RBRC} RIKEN-Brookhaven Research Center Scientific Articles,
Vol. 9: New Discoveries at RHIC:  the current case for the
Strongly Interactive QGP (May, 2004).

\bibitem{NYT} J. Glanz, Like Particles, 2 Houses of Physics
Collide, New York Times, January 20, 2004 (Section F, page 1); K.
Davidson, Universe-Shaking Discovery, or More Hot Air?, San
Francisco Chronicle, January 19, 2004.

\bibitem{Nature} G. Brumfiel, What's In a Name?, Nature,
July 26, 2004.

\bibitem{Harris} J.W. Harris and B. M\"uller, Ann. Rev. Nucl.
Part. Sci. 46 (1996) 71.

\bibitem{CERN} U. Heinz and M. Jacob, nucl-th/0002042.

\bibitem{star:nim} K.H. Ackermann et al., Nucl. Instrum. Methods A499 (2003) 624.

\bibitem{Karsch} F. Karsch, Lecture Notes in Physics
583 (2002) 209.

\bibitem{Kaczmarek} O. Kaczmarek, F. Karsch, E. Laermann and M.
L\"utgemeier, Phys. Rev. D62 (2000) 034021.

\bibitem{Matsui} T. Matsui and H. Satz, Phys. Lett. B178 (1986) 416;
F. Karsch, M.T. Mehr and H. Satz, Z. Phys. C37 (1988) 617.

\bibitem{Asakawa} M. Asakawa and T. Hatsuda, Phys. Rev. Lett. 92 (2004) 012001;
S. Datta et al., J. Phys. G 30 (2004) S1347.

\bibitem{Laermann} E. Laermann and O. Philipsen,
Ann. Rev. Nucl. Part. Sci. 53 (2003) 163.

\bibitem{Fodor-2} Z. Fodor and S.D. Katz, JHEP 0404 (2004) 050.

\bibitem{Fodor-1} Z. Fodor and S.D. Katz, Phys. Lett. B534 (2002) 87;
 Z. Fodor and S.D. Katz, JHEP 0203 (2002) 014.

\bibitem{landau}L.D. Landau, Izv. Akad. Nauk Ser, Fiz. 17 (1953) 51;
L.D. Landau and E.M. Lifshitz, Fluid Mechanics.

\bibitem{shuryak0312} E. Shuryak, Prog. Part. Nucl. Phys. 53 (2004) 273.

\bibitem{stoeck} R. Stock, J. Phys. G 30 (2004) S633.

\bibitem{biro04} T. Biro and B. M\"uller, Phys.
Lett. B 578 (2004) 78, and references therein.

\bibitem{hagedorn}R. Hagedorn, Nuovo Cim. Suppl. 3 (1965) 147.

% Sorge v2 early
\bibitem{sorge97} H. Sorge, Phys. Lett. B402 (1997) 251.

\bibitem{sorge99} H. Sorge, Phys. Rev. Lett. 82 (1999) 2048.

\bibitem{bass} S.A. Bass and A. Dumitru, Phys. Rev. C61 (2000) 064909.

\bibitem{teaney} D. Teaney, J. Lauret, and E. Shuryak,
Phys. Rev. Lett. 86 (2001) 4783;
nucl-th/0110037.

\bibitem{kolbsollfrank} P.F. Kolb, J. Sollfrank and U. Heinz,
Phys. Rev. C62 (2000) 054909.

\bibitem{RQMD} H. Sorge, Phys. Rev. C52 (1995) 3291.

\bibitem{Huovinen} P. Huovinen, P.F. Kolb, U. Heinz, P.V.
Ruuskanen and S.A. Voloshin, Phys. Lett. B503 (2001) 58.

\bibitem{sorge} H. van Hecke, H. Sorge and N. Xu,
Phys. Rev. Lett. 81 (1998) 5764.

\bibitem{xu2} N. Xu and Z. Xu, Nucl. Phys. A715 (2003) 587c.

\bibitem{viscous}D. Teaney, Phys. Rev. C68 (2003) 034913.

\bibitem{3Dhydro} K. Morita, S. Muroya, C. Nonaka and T. Hirano,
Phys. Rev. C66 (2004) 054904, and references therein.

\bibitem{pbm} P. Braun-Munzinger, K. Redlich and J. Stachel,
in Quark Gluon Plasma 3, eds. R.C. Hwa and X.N. Wang (World
Scientific, Singapore, 2003); nucl-th/0304013.

\bibitem{huang} K. Huang,
Statistical Mechanics (John Wiley and Sons, 1988).

\bibitem{fermi}E. Fermi, Prog. Theor. Phys. 5 (1950) 570.

\bibitem{redlich} E.V. Shuryak, Phys. Lett. B42 (1972) 357;
J. Rafelski and M. Danos, Phys. Lett. B97 (1980) 279; R.
Hagedorn and K. Redlich Z. Phys. C 27 (1985) 541.

\bibitem{nxu} P. Braun-Munzinger, J. Stachel, J.P. Wessels and N. Xu,
Phys. Lett. B365 (1996) 1.

\bibitem{koch}V. Koch, Nucl. Phys. A715 (2003) 108c.

\bibitem{heinz-qm99} U. Heinz, Nucl. Phys. A661 (1999) 140c.

\bibitem{art} A.M. Poskanzer and S.A. Voloshin, Phys. Rev. C58 (1998)1671.

\bibitem{rafelski1} G. Torrieri and J. Rafelski,
Phys. Lett. B509 (2001) 239; Z. Xu,
J. Phys. G 30 (2004) S325; M. Bleicher and H. Stocker,
J. Phys. G 30 (2004) S111.

\bibitem{Bjorken} J.D. Bjorken, FERMILAB-PUB-82-59-THY and erratum
(unpublished).

\bibitem{GLV} M. Gyulassy, P. Levai and I. Vitev, Nucl. Phys. A661 (1999) 637;
M. Gyulassy, P. Levai and I. Vitev, Nucl. Phys. B571 (2000) 197;
M. Gyulassy, P. Levai and I. Vitev, Phys. Rev. Lett. 85 (2000) 5535;
M. Gyulassy, P. Levai and I. Vitev, Nucl. Phys. B594 (2001) 371.

\bibitem{WW} E. Wang and X.N. Wang, Phys. Rev. Lett. 87 (2001) 142301.

\bibitem{BDMS} R. Baier, Y.L. Dokshitzer, A.H. Mueller, S. Peigne and D. Schiff, Nucl. Phys. B483 (1997) 291;
R. Baier, Y.L. Dokshitzer, A.H. Mueller and D.
Schiff, Nucl. Phys. B531 (1998) 403;
R. Baier, Y.L. Dokshitzer, A.H. Mueller and D. Schiff,  Phys. Rev. C60 (1999) 064902;
R. Baier, Y.L. Dokshitzer, A.H. Mueller and D. Schiff, JHEP 0109 (2001) 033.

\bibitem{Wied} U.A. Wiedemann, Nucl. Phys. B588 (2000) 303; U.A. Wiedemann, Nucl. Phys. A690 (2001) 731; C.A. Salgado and U.A. Wiedemann, Phys. Rev. Lett. 89 (2002) 092303;
C.A. Salgado, U.A. Wiedemann, Phys. Rev. D68 (2003) 014008.

\bibitem{Greiner} W. Cassing, K. Gallmeister and C. Greiner,
Nucl. Phys. A735 (2004) 277.

\bibitem{CT} G.R. Farrar, H. Liu, L.L. Frankfurt and M.I.
Strikman, Phys. Rev. Lett. 61 (1998) 686;
S.J. Brodsky and A.H. Mueller, Phys. Lett. B206 (1988) 685;
B.K. Jennings and G.A. Miller, Phys. Lett. B236 (1990) 209;
B.K. Jennings and G.A. Miller, Phys. Rev. D44 (1991) 692;
B.K. Jennings and G.A. Miller, Phys. Rev. Lett. 69 (1992) 3619;
B.K. Jennings and G.A. Miller, Phys. Lett. B274 (1992) 442.

\bibitem{Wang} X.N. Wang, Phys. Lett. B579 (2004) 299.

\bibitem{Djordjevic} M. Djordjevic and M. Gyulassy,
Phys. Lett. B560 (2003) 37.

\bibitem{HERMES} A. Airapetian et al. [HERMES
Collaboration], Eur. Phys. J. C 20 (2001) 479; V.
Muccifora [HERMES Collaboration], Nucl. Phys. A715 (2003) 506.

\bibitem{E866} J.M. Moss et al., hep-ex/0109014.

\bibitem{Zhang-Wang} X.N. Wang and X.F. Guo, Nucl. Phys. A
696 (2001) 788; B.W. Zhang and X.N. Wang,
Nucl. Phys. A720 (2003) 429.

\bibitem{Baier} R. Baier, Nucl. Phys. A715 (2003) 209c.

\bibitem{HERA-glue} J. Breitweg et al., Eur. Phys. J. C 7 (1999) 609.

\bibitem{mueller} A.H. Mueller, Nucl. Phys. B335 (1990) 115;
A.H. Mueller,  Nucl. Phys. B572 (2002) 227.

\bibitem{McLerran1}L.D. McLerran and R. Venugopalan,
Phys. Rev. D49 (1994) 2233.

\bibitem{McLerran2}L.D. McLerran, hep-ph/0311028.

\bibitem{raju} E. Iancu and R. Venugopalan, in
Quark Gluon Plasma 3, eds. R.C. Hwa and X.N. Wang (World
Scientific, Singapore, 2003); hep-ph/0303204.

\bibitem{DGLAP} V.N. Gribov and L.N. Lipatov, Sov. J. Nucl. Phys. 15 (1972) 438
and 675;
                L.N. Lipatov, Sov. J. Nucl. Phys. 20 (1975) 94;
                G. Altarelli and G. Parisi, Nucl. Phys. B126 (1977) 298;
                Yu. L. Dokshitzer, Sov. Phys. JETP 46 (1977) 641.

\bibitem{BFKL}  E.A. Kuraev, L.N. Lipatov and V.S. Fadin, Phys. Lett. B60 (1975) 50;
                Sov. Phys. JETP 44 (1976) 443;  Sov. Phys. JETP 45
(1977) 199;
             L.N. Lipatov, Sov. J. Nucl. Phys. 23 (1976), 338;
             Ya. Ya. Balitsky and L.N. Lipatov, Sov. J. Nucl. Phys. 28
(1978) 822;
                 Sov. Phys. JETP Lett. 30 (1979) 355.

\bibitem{Kharzeev} D. Kharzeev and M. Nardi, Phys. Lett. B507 (2001) 121;
D. Kharzeev and E. Levin, Phys. Lett. B523 (2001) 79.

\bibitem{recom77} V.V. Anisovich and V.M. Shekhter,
Nucl. Phys. B55 (1973) 455;
J.D. Bjorken and G.E. Farrar, Phys. Rev. D9 (1974) 1449;
K.P. Das and R.C. Hwa, Phys. Lett. B68 (1977) 459;
Erratum ibid. 73 (1978) 504;
R.G. Roberts, R.C. Hwa and S. Matsuda, J. Phys. G 5 (1979) 1043.

\bibitem{coalescence95}C. Gupt, R.K. Shivpuri, N.S. Verma and A.P. Sharma,
Nuovo Cim. A75 (1983) 408;
T. Ochiai, Prog. Theor. Phys. 75 (1986) 1184;
T.S. Biro, P. Levai and J. Zimanyi, Phys. Lett. B347 (1995) 6;
T.S. Biro, P. Levai and J. Zimanyi, J. Phys. G 28 (2002) 1561.

\bibitem{recom03} S.A. Voloshin, Nucl. Phys. A. 715 (2003) 379c;
D. Molnar and S.A. Voloshin, Phys. Rev. Lett. 91 (2003) 092301;
R.J. Fries, B. M\"uller, C. Nonaka and S.A. Bass,
Phys. Rev. Lett. 90 (2003) 202303;
V. Greco, C.M. Ko and P. Levai, Phys. Rev. Lett. 90 (2003) 202302;
Z.W. Lin and C.M. Ko, Phys. Rev. Lett. 89 (2002) 202302;
Z.W. Lin and D. Molnar, Phys. Rev. C68 (2003) 044901.

\bibitem{MuellerRBRC} B. M\"uller, nucl-th/0404015.

\bibitem{star:highpTbtob} C. Adler et al. [STAR Collaboration],
Phys. Rev. Lett. 90 (2003) 082302.

\bibitem{Hwa} R.C. Hwa and C.B. Yang, Phys. Rev. C66 (2002) 025205;
R.C. Hwa and C.B. Yang, Phys. Rev. C67 (2003) 034902;
R.C. Hwa and C.B. Yang, Phys. Rev. C70 (2004) 024905.

\bibitem{ritter97} W. Reisdorf and H.G. Ritter,
Ann. Rev. Nucl. Part. Sci. 47 (1997) 663.

\bibitem{ollitrault92} J.-Y. Ollitrault, Phys. Rev. D46 (1992) 229.

\bibitem{Wang-Gyulassy} X.N. Wang and M. Gyulassy,
Phys. Rev. Lett. 86 (2001) 3496.

\bibitem{phobos02} B.B. Back et al. [PHOBOS Collaboration],
Phys. Rev. C65 (2002) 061901R.

\bibitem{Laue-spectra} J. Adams et al. [STAR Collaboration],
nucl-ex/0311017.

\bibitem{EKRT} K.J. Eskola, K. Kajantie, P.V. Ruuskanen and K. Tuominen,
Nucl. Phys. B570 (2000) 379.

\bibitem{Czyz} W. Czyz and L.C. Maximon, Annals Phys. 52 (1969) 59.

\bibitem{phobos03} B.B. Back et al. [PHOBOS Collaboration],
nucl-ex/0301017.

\bibitem{phenix:et130} K. Adcox et al. [PHENIX Collaboration],
Phys. Rev. Lett. 87 (2001) 052301.

\bibitem{WA98:dETdeta} M.M. Aggarwal et al. [WA98 Collaboration],
 Eur. Phys. J. C 18 (2001) 651.

\bibitem{star:et200} C. Adler et al. [STAR Collaboration],
Phys. Rev. C70 (2004) 054907.

\bibitem{BjorkenHydro} J.D. Bjorken, Phys. Rev. D27 (1983) 140.


\bibitem{star:multi-strange} J. Adams et al. [STAR Collaboration],
Phys. Rev. Lett. 92, (2004) 182301.

\bibitem{barranikQM04} O. Barannikova et al. [STAR Collaboration],
nucl-ex/0403014.

\bibitem{star_pikp} C. Adler et al. [STAR Collaboration],
Phys. Rev. Lett. 89 (2002) 092301;
J. Adams et al. [STAR Collaboration], Phys. Lett. B595 (2004) 143;
J. Adams et al. [STAR Collaboration], Phys. Rev. C70 (2004) 041901;
J. Adams et al. [STAR Collaboration], nucl-ex/0406003.

\bibitem{Xu-Kaneta} N. Xu and M. Kaneta, Nucl. Phys. A698 (2002) 306c.

\bibitem{Hagedorn:1984hz} R. Hagedorn, CERN-TH-3918/84.

\bibitem{heinz93} E. Schnedermann, J. Sollfrank, and
U. Heinz, Phys. Rev. C48 (1993) 2462.

\bibitem{pbm9596} P. Braun-Munzinger, J. Stachel, J. Wessels, and N.
  Xu, Phys. Lett. B344 (1995) 43;
P. Braun-Munzinger, I. Heppe, and J. Stachel, Phys. Lett. B465 (1999) 15.

\bibitem{bass99} S.A. Bass et al., Phys. Rev. C60 (1999) 021902.

\bibitem{dumitru} A. Dumitru and S.A. Bass, Phys. Lett. B460 (1999) 411.

\bibitem{phenixspectra} S.S. Adler et al. [PHENIX Collaboration],
Phys. Rev. C69 (2004) 034909.

\bibitem{starphi} E. Yamamoto et al. [STAR Collaboration],
Nucl. Phys. A715 (2003) 466c.

\bibitem{kaiqm04} K. Schweda et al. [STAR Collaboration],
J. Phys. G 30 (2004) S693.

\bibitem{phenix:meson-baryon} K. Adcox et al. [PHENIX Collaboration],
 Phys. Rev. Lett. 88 (2002) 242301.

\bibitem{Poskanzer} A.M. Poskanzer and S.A. Voloshin,
Phys. Rev. C58 (1998) 1671.

\bibitem{Borghini} N. Borghini, P.M. Dinh and J.Y. Ollitrault,
Phys. Rev. C64 (2001) 054901.

\bibitem{NA49flow} C. Alt et al. [NA49 Collaboration],
Phys. Rev. C68 (2003) 034903.

\bibitem{star:flow4part130} C. Adler et al. [STAR Collaboration],
Phys. Rev. C66 (2002) 034904.

\bibitem{Tang:2003kz} A. Tang et al. [STAR Collaboration],
AIP Conf. Proc. 698 (2004) 701; J. Adams et al. [STAR
Collaboration], nucl-ex/0409033.

\bibitem{CERESAwaySide} G. Agakichiev et al. [CERES/NA45 Collaboration],
Phys. Rev. Lett. 92 (2004) 032301.

\bibitem{starklv2} J. Adams et al. [STAR Collaboration],
Phys. Rev. Lett.  92 (2004) 052302.

\bibitem{star130flow} C. Adler et al. [STAR Collaboration],
Phys. Rev. Lett. 87 (2001) 182301.

\bibitem{pasi03} P. Huovinen, private communications (2003).

\bibitem{Shuryak} E. Shuryak, Prog. Part. Nucl. Phys.53 (2004) 273.

\bibitem{phobos:boost-dependence} B.B. Back, et al. [PHOBOS Collaboration],
Phys. Rev. Lett. 89 (2002) 222301.

\bibitem{xin04} X. Dong, S. Esumi, P. Sorensen, N. Xu, and Z. Xu,
Phys. Lett. B597 (2004) 328.

\bibitem{star_javier} J. Castillo et al. [STAR Collaboration],
J. Phys. G 30 (2004) S1207.

\bibitem{phenix:pidflow200} S.S. Adler et al. [PHENIX Collaboration],
Phys. Rev. Lett. 91 (2003) 182301.

\bibitem{Duke} C. Nonaka, R.J. Fries and S.A. Bass,
Phys. Lett. B583 (2004) 73;
C. Nonaka et al., Phys. Rev. C69 (2004) 031902.

\bibitem{hbt0} For general introduction of two-particle correlation
studies, see U.A. Wiedemann and U. Heinz, Phys. Rept. 319 (1999) 145;
B. Jacak and U. Heinz, Ann. Rev. Nucl. Part. Sci. 49 (1999) 529.

\bibitem{pratt} G. Bertsch, Nucl. Phys. A498 (1989) 173c.

\bibitem{rischke96} D. Rischke and M. Gyulassy,
Nucl. Phys. A597 (1996) 701;
D. Rischke and M. Gyulassy, Nucl. Phys. A608 (1996) 479.

\bibitem{rhic130hbt} C. Adler et al. [STAR Collaboration],
Phys. Rev. Lett. 87 (2001) 082301;
 K. Adcox et al. [PHENIX Collaboration], Phys. Rev. Lett. 88 (2002) 242301.

\bibitem{starhbt2} J. Adams et al. [STAR Collaboration],
Phys. Rev. Lett. 92 (2003) 062301.

\bibitem{soff01} S. Soff, hep-ph/0202240;
D. Zschiesche, S. Schramm, H. Stocker, and W. Greiner,
Phys. Rev. C65 (2001) 064902;
S. Soff, S.A. Bass, and A. Dumitru,
Phys. Rev. Lett. 86 (2001) 3981.

\bibitem{heinz04} U. Heinz, hep-ph/0407360.

\bibitem{wong04} C.Y. Wong, J. Phys. G 30 (2004) S1053.

\bibitem{non-ident} J. Adams et. al. [STAR Collaboration],
Phys. Rev. Lett. 91 (2003) 262302.

\bibitem{stephanov98} M. Stephanov, K. Rajagopal, and E. Shuryak,
Phys. Rev. Lett. 81 (1998) 4816.

\bibitem{voloshin99} S. Voloshin, V. Koch, and H.G. Ritter,
Phys. Rev. C60 (1999) 024901.

\bibitem{jeon00} S. Jeon and V. Koch, Phys. Rev. Lett. 85 (2000) 2076.

\bibitem{asakawa00} M. Asakawa, U. Heinz, and B. M\"{u}ller,
Phys. Rev. Lett. 85 (2000) 2072.

\bibitem{bass00} S.A. Bass, P. Danielewicz, and S. Pratt,
Phys. Rev. Lett. 85 (2000) 2689.

\bibitem{liu03} Q. Liu and T. Trainor, Phys. Lett. B567 (2003) 184.

\bibitem{gazdzicki99} M. Gaz\'dzicki and St. Mr\'owczy\'nski,
Z. Phys. C54 (1999) 127.

\bibitem{CERES03} D. Adamava et al. [CERES Collaboration], Nucl. Phys. A727 (2003) 97.

\bibitem{PHENIX04} S.S. Adler et al. [PHENIX Collaboration], Phys. Rev. Lett. 93 (2004) 092301.

\bibitem{gary200qm04} G. Westfall et al. [STAR Collaboration],
J. Phys. G 30 (2004) S1389.

\bibitem{trainor-jamaica} J. Adams et al. [STAR Collaboration], nucl-ex/0411003;T.A. Trainor et al. [STAR Collaboration],
hep-ph/0406116.

\bibitem{NA44_DWT} I. Bearden et al. [NA44 Collaboration],
Phys. Rev. C65 (2002) 044903.

\bibitem{dynamic-texture} J. Adams et al. [STAR Collaboration],
Phys. Rev. C 71 (2005) 031901(R).

\bibitem{starbf130} J. Adams et al. [STAR Collaboration],
Phys. Rev. Lett. 90 (2003) 172301.

%\bibitem{bialas03} A. Bialas, Phys. Lett. {\bf B579, 31(2004).

%\bibitem{starnchebye03} J. Adams, {\it et al., (STAR Collaboration),
%Phys. Rev. {\bf C68, 044905(2003).

%\bibitem{starptebye04} J. Adams, {\it et al., (STAR Collaboration),
%nucl-ex/0308033.

\bibitem{HIJING} X.N. Wang and M. Gyulassy,
Phys. Rev. D44 (1991) 3501;
X.N. Wang and M. Gyulassy, Comput. Phys. Commun. 83 (1994) 307.


\bibitem{star:highpt130} C. Adler et al. [STAR Collaboration],
Phys. Rev. Lett. 89 (2002) 202301.

\bibitem{brahms:highpTdAu} I. Arsene et al. [BRAHMS Collaboration],
Phys. Rev. Lett. 91 (2003) 072305.

\bibitem{phenix:highpTdAu} S.S. Adler et al. [PHENIX Collaboration],
Phys. Rev. Lett. 91 (2003) 072303.

\bibitem{phobos:highpTdAu} B.B. Back et al. [PHOBOS Collaboration],
Phys. Rev. Lett. 91 (2003) 072302.

\bibitem{star:highpTdAu} J. Adams et al. [STAR Collaboration],
Phys. Rev. Lett. 91 (2003) 072304.

\bibitem{Cronin} D. Antreasyan et al.,
Phys. Rev. D19 (1979) 764.

\bibitem{mult-scat} X.N. Wang, Phys. Rept. 280 (1997) 287;
M. Lev and B. Petersson, Z. Phys. C21 (1987) 155;
T. Ochiai et al., Prog. Theor. Phys. 75 (1986) 288.

\bibitem{star:highptv2} J. Adams et al. [STAR Collaboration],
Phys. Rev. Lett. 93 (2004) 252301.

%\bibitem{Tang:2004vc} A.H. Tang, nucl-ex}/0403018.

%\bibitem{Filimonov:2004qz} K. Filimonov, nucl-ex}/0403060.

\bibitem{Hardtke:2002ph} D. Hardtke et al. [STAR Collaboration],
Nucl. Phys. A715 (2003) 272.

\bibitem{star_fqwang} F. Wang et al. [STAR Collaboration],
J. Phys. G. 30 (2004) S1299; J. Adams et al. [STAR Collaboration],
nucl-ex/0501016.

\bibitem{Wang:2003mm} X.N. Wang, Phys. Lett. B 595 (2004) 165.

\bibitem{Vitev:2002pf} I. Vitev and M. Gyulassy,
Phys. Rev. Lett. 89 (2002) 252301.

\bibitem{Eskola:2004cr} K.J. Eskola, H. Honkanen, C.A. Salgado, and U.A. Wiedemann,
Nucl. Phys. A747 (2005) 511.

\bibitem{Kharzeev:2002pc} D. Kharzeev, E. Levin and L. McLerran,
Phys. Lett. B561 (2003) 93.

\bibitem{star:highpTAuAu200} J. Adams et al. [STAR Collaboration],
Phys. Rev. Lett. 91 (2003) 172302.

\bibitem{Falter:2002jc} T. Falter and U. Mosel,
Phys. Rev. C66 (2002) 024608.

\bibitem{Gallmeister:2002us} K. Gallmeister, C. Greiner and Z. Xu,
Phys. Rev. C67 (2003) 044905.

\bibitem{Back:2001ae} B.B. Back et al. [PHOBOS Collaboration],
Phys. Rev. Lett. 88 (2002) 022302.

\bibitem{brahms:dAuSuppression} I. Arsene et al. [BRAHMS Collaboration],
Phys. Rev. Lett. 93 (2004) 242303.

\bibitem{Guzey:2004zp} V. Guzey, M. Strikman, and W. Vogelsang,
Phys. Lett. B603 (2004) 173.

\bibitem{kharzeev04} D. Kharzeev, Y.V. Kovchegov and K. Tuchin,
Phys Lett. B599 (2004) 23.

\bibitem{Frawley_QM04} A.D. Frawley et al. [PHENIX Collaboration],
J. Phys. G30 (2004) S675.

\bibitem{stardAu_eta_asym} J. Adams et al. [STAR Collaboration],
Phys. Rev. C 70 (2004) 064907.

\bibitem{kharzeev_dAu_prediction} D. Kharzeev, Y.V. Kovchegov and
K. Tuchin, Phys. Rev. D68 (2003) 094013.

\bibitem{kharzeev-cgc-correlations} D. Kharzeev, E. Levin and L.
McLerran, hep-ph/0403271.

\bibitem{Vogt} R. Vogt, hep-ph/0405060.

\bibitem{Akio_DIS04} A. Ogawa et al. [STAR Collaboration],
nucl-ex/0408004.

\bibitem{WA98highpT} M.M. Aggarwal et al. [WA98 Collaboration],
Phys. Rev. Lett. 81 (1998) 4087;
M.M. Aggarwal et al. [WA98 Collaboration], Phys. Rev. Lett. 84 (2000) 578;
M.M. Aggarwal et al. [WA98 Collaboration], Eur. Phys. J. C 23 (2002) 225.

\bibitem{D'Enterria} D. d'Enterria, J. Phys. G 30 (2004), S767.

%\bibitem{CERESAwaySide} G. Agakichiev et al. [CERES/NA45 Collaboration],
%Phys. Rev. Lett. 92 (2004) 032301.

\bibitem{gyulassy-hbt} D.H. Rischke and M. Gyulassy, Nucl. Phys.
A608 (1996) 479.

\bibitem{HBT_PRC} J. Adams et al. [STAR Collaboration],
nucl-ex/0411036.

\bibitem{kharzeevCP} D. Kharzeev et al.,
 Phys. Rev. Lett. 81 (1998) 512 (1998);
D. Kharzeev, hep-ph/0406125.

\bibitem{LRP-83} A Long Range Plan for Nuclear Science
(NSAC Report, 1983).

\bibitem{NAS-84} Physics Through the 1990s: Nuclear
Physics, J. Cerny et al. (National Academy Press,
Washington, D.C., 1986).

\bibitem{RHIC_proposal} RHIC and Quark Matter: Proposal for a
Relativistic Heavy Ion Collider at Brookhaven National Laboratory
(BNL Report 51801, August 1984).

\bibitem{LRP-89} Nuclei, Nucleons, Quarks: Nuclear Science in the 1990's
(NSAC Long Range Plan Report, December 1989).

\bibitem{NSAC-94} Nuclear Science: Assessment and Promise
(NSAC Report, May 1994).

\bibitem{LRP-96} Nuclear Science: A Long Range Plan (NSAC
Report, February 1996).

\bibitem{NAS-99} Nuclear Physics: The Core of Matter, The Fuel of
Stars, J.P. Schiffer, et al. (National Academy Press,
Washington, D.C., 1999).

\bibitem{NuPECC-04} NuPECC Long Range Plan 2004: Perspectives for
Nuclear Physics Research in Europe in the Coming Decade and
Beyond, ed. M. Harakeh, et al. (April 2004).

%\bibitem{kolbv4}P.F. Kolb, \Journal{\PRC}{68}{031902}{2003}.

%\bibitem{becattini} F. Becattini, hep-ph/0202071; F. Becattini,
%\Journal{\ZPC}{69}{485}{1996}; F. Becattini and U. Heinz,
%\Journal{\ZPC}{76}{269}{1996}.

\end{thebibliography}
\end{document}